\documentclass[onecolumn,10pt]{IEEEtran}
\usepackage[mathscr]{eucal}
\usepackage[cmex10]{amsmath}
\usepackage{epsfig,epsf,psfrag}
\usepackage{amssymb,amsmath,amsthm,amsfonts,latexsym}
\usepackage{amsmath,graphicx,bm,xcolor,url,overpic}
\usepackage{array}
\usepackage{verbatim}
\usepackage{bm}
\usepackage{algorithmic}
\usepackage{algorithm}
\usepackage{verbatim}
\usepackage{textcomp}
\usepackage{mathrsfs}
\usepackage{epstopdf}

\newcommand{\openone}{\leavevmode\hbox{\small1\normalsize\kern-.33em1}}

\catcode`~=11 \def\UrlSpecials{\do\~{\kern -.15em\lower .7ex\hbox{~}\kern .04em}} \catcode`~=13 

\allowdisplaybreaks[4]

\newcommand{\mmd}{\mathrm{MMD}}
\newcommand{\nn}{\nonumber}

\newcommand{\calA}{\mathcal{A}}
\newcommand{\calB}{\mathcal{B}}

\newcommand{\calE}{\mathcal{E}}
\newcommand{\calF}{\mathcal{F}}

\newcommand{\calM}{\mathcal{M}}
\newcommand{\calN}{\mathcal{N}}

\newcommand{\calP}{\mathcal{P}}
\newcommand{\calQ}{\mathcal{Q}}
\newcommand{\calR}{\mathcal{R}}

\newcommand{\calT}{\mathcal{T}}

\newcommand{\calX}{\mathcal{X}}

\newcommand{\hatcalA}{\hat{\calA}}

\newcommand{\bX}{\mathbf{X}}

\newcommand{\bY}{\mathbf{Y}}

\newcommand{\rmc}{\mathrm{c}}

\newcommand{\rmf}{\mathrm{f}}

\newcommand{\rms}{\mathrm{s}}

\newcommand{\rmu}{\mathrm{u}}
\newcommand{\rmU}{\mathrm{U}}

\newcommand{\bbE}{\mathsf{E}}

\newcommand{\bbN}{\mathbb{N}}

\newcommand{\bbP}{\mathbb{P}}

\newcommand{\bbR}{\mathbb{R}}

\DeclareMathAlphabet{\mathbsf}{OT1}{cmss}{bx}{n}
\DeclareMathAlphabet{\mathssf}{OT1}{cmss}{m}{sl}

\DeclareSymbolFont{bsfletters}{OT1}{cmss}{bx}{n}  
\DeclareSymbolFont{ssfletters}{OT1}{cmss}{m}{n}
\DeclareMathSymbol{\bsfGamma}{0}{bsfletters}{'000}
\DeclareMathSymbol{\ssfGamma}{0}{ssfletters}{'000}
\DeclareMathSymbol{\bsfDelta}{0}{bsfletters}{'001}
\DeclareMathSymbol{\ssfDelta}{0}{ssfletters}{'001}
\DeclareMathSymbol{\bsfTheta}{0}{bsfletters}{'002}
\DeclareMathSymbol{\ssfTheta}{0}{ssfletters}{'002}
\DeclareMathSymbol{\bsfLambda}{0}{bsfletters}{'003}
\DeclareMathSymbol{\ssfLambda}{0}{ssfletters}{'003}
\DeclareMathSymbol{\bsfXi}{0}{bsfletters}{'004}
\DeclareMathSymbol{\ssfXi}{0}{ssfletters}{'004}
\DeclareMathSymbol{\bsfPi}{0}{bsfletters}{'005}
\DeclareMathSymbol{\ssfPi}{0}{ssfletters}{'005}
\DeclareMathSymbol{\bsfSigma}{0}{bsfletters}{'006}
\DeclareMathSymbol{\ssfSigma}{0}{ssfletters}{'006}
\DeclareMathSymbol{\bsfUpsilon}{0}{bsfletters}{'007}
\DeclareMathSymbol{\ssfUpsilon}{0}{ssfletters}{'007}
\DeclareMathSymbol{\bsfPhi}{0}{bsfletters}{'010}
\DeclareMathSymbol{\ssfPhi}{0}{ssfletters}{'010}
\DeclareMathSymbol{\bsfPsi}{0}{bsfletters}{'011}
\DeclareMathSymbol{\ssfPsi}{0}{ssfletters}{'011}
\DeclareMathSymbol{\bsfOmega}{0}{bsfletters}{'012}
\DeclareMathSymbol{\ssfOmega}{0}{ssfletters}{'012}

\newcommand{\hatK}{\hat{K}}

\newcommand{\tilP}{\tilde{P}}

\newcommand{\tilQ}{\tilde{Q}}

\newcommand{\hatT}{\hat{T}}

\newcommand{\bari}{\bar{i}}
\newcommand{\barj}{\bar{j}}

\newcommand{\barQ}{\bar{Q}}

\newtheorem{theorem}{Theorem}

\usepackage{subfigure,epstopdf,bbm} 
\usepackage[utf8]{inputenc} 
\usepackage[T1]{fontenc}    
\usepackage{url,bbm}        
\usepackage{booktabs}       
\usepackage{amsfonts}       
\usepackage{nicefrac}       
\usepackage{microtype}      
\usepackage{cite}
\usepackage{subfigure,transparent,color,graphicx} 

\newcommand{\gjs}{\mathrm{GJS}}
\usepackage[ colorlinks = true,
linkcolor = blue,
urlcolor  = blue,
citecolor = red,
anchorcolor = green]{hyperref}
\allowdisplaybreaks[2]
\begin{document}
\title{Exponentially Consistent Low Complexity Tests for Statistical Sequence Matching}
\author{Lin Zhou

\thanks{L. Zhou is the School of Automation and Intelligent Manufacturing, Southern University of Science and Technology (Email: zhoul9@sustech.edu.cn).}
}

\maketitle

\begin{abstract}
Towards practical applications of statistical sequence matching, we propose low complexity tests that are exponentially consistent and bound the exponential decay rates for error probabilities of our proposed tests. In statistical sequence matching, one is given two databases of multiple sequences, where in each database, each sequence is generated i.i.d. from a  distinct distribution. A pair of sequences, one from each database, is said matched if they are generated from the same distribution. The number of matched pairs of sequences, a.k.a, the number of matches, can be either zero or positive. We consider both cases of known and unknown number of matches. When the number of matches is known and positive, the task is to identify all matched pairs of sequences. To construct fixed-length tests, we calculate the scoring function values for all pairs of sequences across the two databases and identify the matched pairs as those having small enough pairwise scoring function values. We show that our proposed test is exponentially consistent, which strikes a better tradeoff between complexity and performance than existing fixed-length tests that use exhaustive search. To construct sequential tests, we need an additional positive threshold to determine the stopping time, and we show that our sequential test achieves strictly better performance than our fixed-length test. Subsequently, we generalize our results to the case with unknown number of matches. In this case, one needs to simultaneously estimate the number of matches and identify all matched pairs if the number of matches is positive. For both cases of known and unknown number of matches, we propose exponentially consistent low-complexity fixed-length and sequential tests for discrete and continuous observed sequences, characterize the exponential tradeoff among three error probabilities, and show that our proposed sequential test achieves better performance than our proposed fixed-length test.
\end{abstract}


\section{Introduction}
Motivated by database de-anonymization, statistical sequence matching (SSM) was pioneered by Unnikrishnan~\cite{unnikrishnan2015asymptotically}. In this problem, one is given two databases, where each database consists of multiple sequences that are generated from distinct distributions.  Each sequence can be either discrete, generated i.i.d. from a probability mass function (pmf), or continuous, generated i.i.d. from a probability density function (pdf). The generating distribution of each sequence in each database is assumed unknown. A pair of sequences, one from each database, is said matched if they are generated from the same distribution. The number of matched pairs of sequences, a.k.a. the number of matches, can be zero or positive. The task of SSM is to determine whether there are matched pairs of sequences and identify all matched pairs when the number of matches is positive.  

The number of matches can be either known~\cite{unnikrishnan2015asymptotically} or unknown~\cite{zhou2024tit}. When the number of matches is known, the performance criterion is the mismatch probability, which quantifies the probability that the test fails to identify the set of matched pairs of sequences. When the number of matches is unknown, there are two additional performance criteria: the false reject and the false alarm probabilities. Specifically, the false reject probability quantities the probability that the test incorrectly claims that there are no matched pairs of sequences  when there exists matched pairs, while the false alarm probability quantifies the probability that the test incorrectly claims there are matched pairs of sequences when there is no. The test can be either fixed-length~\cite{unnikrishnan2015asymptotically} or sequential~\cite{zhou2025seq}. A fixed-length test~\cite{blahut1974hypothesis} corresponds to offline decision, where the length of each sequence is fixed, while a sequential test~\cite{wald1945sequential,wald1948optimum} corresponds to online decision, where the lengths of sequences are random, which are determined by the test designer to make reliable decisions. The sequential test usually yields superior performance~\cite{wald1945sequential,wald1948optimum,zhou2025seq}.

When the number of matches is known and the observed sequences are discrete, Unnikrishnan~\cite[Section IV]{unnikrishnan2015asymptotically} proposed a fixed-length test using empirical distributions of all sequences, derived the exponential decay rates of the mismatch probability and proved optimality of the test in the generalized Neyman-Pearson sense when the test is allowed to output an additional reject decision, indicating that a reliable decision cannot be made. Zhou \emph{et al.}~\cite{zhou2024tit} generalized the results in~\cite[Section IV]{unnikrishnan2015asymptotically} to the case with both known and unknown number of matches, studied the non-asymptotic performance of optimal tests in the generalized Neyman-Pearson sense and characterized the ignored false reject probability. When the reject decision is not allowed with known number of matches, the results in \cite{zhou2024tit} have been generalized to sequential tests~\cite[Section III]{zhou2025seq}, where it was shown that sequential tests achieve better performance. Furthermore, the benefit of sequentiality with unknown number of matches was also revealed~\cite[Section IV]{zhou2025seq}. When the observed sequences are continuous, using the maximum mean discrepancy (MMD) metric~\cite{gretton2012jmlr}, regardless whether the number of matches is known or unknown, Zhou \emph{et al.}~\cite{zhou2025csm} proposed fixed-length and sequential tests and derived the exponential decay rates of error probabilities, generalizing the results in \cite{zhou2024tit,zhou2025seq} to continuous sequences.

The above results have comprehensively characterized the performance of tests for SSM, with optimality guarantees in certain cases. However, in all above information theoretic studies, the crux is to ensure good performance while the computational complexity is not considered, in the spirit of Shannon~\cite{shannon1948mathematical,shannon1959coding}. In particular, to run the test in \cite{unnikrishnan2015asymptotically,zhou2024tit,zhou2025seq,zhou2025csm}, the computational complexity scales exponentially with respect to the number of matches. When there are $100$ sequences in the first database, $50$ sequences in the second database and $10$ matches between the two databases, one needs to check roughly $6.4526\times 10^{29}$ possibilities to identify the set of matched pairs of sequences even when the number of matches is known. Such high computational complexity is unaffordable, rendering the above theoretically guaranteed tests impractical.

To address the above problem, in this paper, we propose low complexity tests, characterize the performance of our tests and show that our tests strike a good balance between performance and computational complexity. We consider cases where the number of matches is known and unknown, and study the performance of both fixed-length and sequential tests. Specifically, we propose four tests for both discrete and continuous observed sequences. For discrete sequences, we propose a scoring function that uses empirical distributions of observed sequences while for continuous sequences, we propose another scoring function that uses observed sequences in MMD estimators. Our main contributions are summarized below.

\subsection{Main Contributions}
We first consider the case where the number of matches is known. In this case, the task is to identify all matched pairs of sequences. When the first database has $M_1$ sequences, the second database has $M_2$ sequences, and there are $K$ matches, the total number of possible sets of matched sequences is $T_K:={M_1\choose K}{M_2\choose K}K!$, which increases exponentially with respect to the number of matches $K$. The tests in~\cite{unnikrishnan2015asymptotically,zhou2024tit,zhou2025csm,zhou2025seq} use exhaustive search over all possibilities and thus suffer from intractable computational complexity. To address this problem, we propose a fixed-length test that has polynomial complexity with respect to the number of sequences $(M_1,M_2)$ of two databases, regardless of the number of matches $K$. The key idea is to use a scoring function that quantifies the closeness of generating distributions of two sequences. Asymptotically, if two sequences are matched, the scoring function value vanishes; otherwise, the value is non-vanishing. Using this idea, we calculate the values of all $M_1M_2$  pairwise scoring functions and identify the matched pairs as those having $K$ smallest scoring function values. Our test is shown to be exponentially consistent and we lower bound the exponential decay rate of the mismatch probability, a.k.a., the mismatch exponent. Although the exponent is smaller than the test in \cite{zhou2024tit,zhou2025csm}, our test enjoys low-complexity, enabling practical applications and achieving a good tradeoff between detecting performance and running times. We also generalize our results to sequential tests and reveal the benefit of sequentiality.

We next consider the case where the number of matches is unknown. In this case, one needs to estimate the number of matches and identify all matched pairs of sequences if the estimated number of matches is positive. Each test should trade off three error probabilities: mismatch, false reject and false alarm. Since the number of matches is unknown, the total number of possibilities increases to $T:=\sum_{K\in[1,\min\{M_1,M_2\}]}T_K+1$. To address the forbidden high computational complexity of the tests in \cite{unnikrishnan2015asymptotically,zhou2024tit,zhou2025csm,zhou2025seq}, we first calculate the values of all $M_1M_2$  pairwise scoring functions. Since the number of matches is unknown, we introduce a positive threshold $\lambda$ to determine the null case with no match and to identify the matched pairs of sequences. Specifically, we output the set of matched pairs as all pairs of sequences that have pairwise scoring function no greater than $\lambda$. If the set is empty, then the null hypothesis is claimed. Under mild conditions on the tuple of unknown generating distributions, we show that our fixed-length test ensures exponentially small probabilities of mismatch, false reject and false alarm. Furthermore, we generalize the above results to sequential tests, show that our sequential tests have bounded expected stopping time under the same mild conditions for our fixed-length tests to have positive exponents and reveal the benefit of sequentiality. For both fixed-length and sequential tests, we reveal the penalty on the detection performance of not knowing the number of matches.

\subsection{Related Studies}
The cornerstone of our test design and performance analysis is binary classification, dating back to Gutman~\cite{gutman1989asymptotically}. In this problem, one is giving a testing sequence and two training sequences that are generated i.i.d. from two distinct distributions. The task of binary classification is to infer the generating distribution of the testing sequence. Gutman~\cite[Theorem 1]{gutman1989asymptotically} proposed a non-parametric test that checks the similarity between the generating distributions of the testing sequence and the first training sequence. If the similarity level is high, the testing sequence is believed to be generated from the same distribution as the first training sequence; otherwise, the other case is believed to be true. Such a test is shown to be asymptotically optimal in the generalized Neyman-Pearson sense. Subsequently, Zhou, Tan and Motani~\cite{zhou2018binary} proved that such a test is optimal in the second-order asymptotic regime that approximates the performance of optimal tests with finite sample sizes. The above results for fixed-length tests have also been generalized to sequential tests. The sequential case was pioneered by Haghifam, Khisti and Tan ~\cite{haghifam2019sequential}, who proposed a semi-sequential test and demonstrated the benefit of sequentiality by showing that the Bayesian exponent of their sequential test is larger than that of Gutman's fixed-length test. Semi-sequential refers to the case where the testing sequence arrives in a streaming manner while the training sequences are available in a fixed-length manner. Subsequently, for the full sequential setting where the testing and both training sequences arrive sequentially, Hsu, Li and Wang~\cite{Ihwang2022sequential} proposed non-parametric tests and confirmed the benefit of sequentiality. Recently, Li and Wang~\cite{li2025} refined the results in~\cite{haghifam2019sequential,Ihwang2022sequential} by proposing a unified test for all settings and comprehensively revealing the benefit of sequentiality. The above results, especially~\cite{zhou2018binary,Ihwang2022sequential}, inspire the test design and performance analyses in this paper for discrete observed sequences. For continuous observed sequences, our analysis uses the MMD metric in~\cite{gretton2012jmlr} and is inspired by~\cite{zhu2025report}.

\subsection{Organization for Rest of the Paper}
The rest of the paper is organized as follows. In Section \ref{sec:pf}, we formulate the problem by specifying the system model and performance metrics. The main results are presented in Sections \ref{sec:known} and \ref{sec:unknown} for known and unknown number of matches, respectively. Subsequently, the proofs are given in Sections \ref{sec:proof:known} and \ref{sec:proof:unknown} for known and unknown number of matches, respectively. Finally, Section \ref{sec:conc} summarizes the paper and discusses future directions.

\section{Problem Formulation}
\label{sec:pf}
\subsection*{Notation}
Random variables and their realizations are in upper case (e.g.,  $X$) and lower case (e.g.,  $x$), respectively. All sets are denoted in calligraphic font (e.g.,  $\mathcal{X}$). We use $\bbR$, $\bbR_+$, and $\bbN$ to denote the set of real numbers, non-negative real numbers, and natural numbers, respectively. Given any positive integer $a\in\bbN$, we use $[a]$ to denote the collection of natural numbers between $1$ and $a$, use $\bbN_a$ to denote the collection of natural numbers that are greater than or equal to $a$ and use $a!$ to denote the factorial $\prod_{i\in[a]}i$. We use superscripts to denote the length of vectors, e.g., $X^n:=(X_1,\ldots,X_n)$. All logarithms are base $e$. The set of all distributions defined on the alphabet $\calX$ is denoted as $\calP(\calX)$. Notation concerning the types follow~\cite{TanBook,ZhouBook}. Specifically, given a finite set $\calX$ and a vector $x^n = (x_1,x_2,\ldots,x_n) \in\calX^n$, the {\em type} or {\em empirical distribution} is denoted as $\hatT_{x^n}(a)=\frac{1}{n}\sum_{i=1}^n \mathbbm{1}\{x_i=a\},a\in\calX$. The set of types formed with length-$n$ sequences on the alphabet $\calX$ is denoted as $\calP_{n}(\calX)$. Given a type $P\in\calP_{n}(\calX)$, the set of all sequences of length $n$ with type $P$, the type class, is denoted as $\calT^n_P$. Given any real number $\mu\in\bbR$ and any positive real number $\sigma\in\bbR_+$, we use $\calN(\mu,\sigma^2)$ to denote the Gaussian distribution with mean $\mu$ and variance $\sigma^2$ and use $\mathrm{Exp}(\sigma)$ to denote the exponential distribution with parameter $\sigma$. Given any two real numbers $(a,b)\in\bbR^2$ such that $a<b$, we use $\rmU(a,b)$ to denote the uniform distribution in the set of $[a,b]$ and use $\rmU(b)$ to denote $\rmU(0,b)$ if $b>0$. Finally, given any number $a\in(0,1)$, we use $\mathrm{Bern}(a)$ to denote the Bernoulli distribution with the probability $a$ of having one and probability of $1-a$ of having zero.

\subsection{Case of Known Number of Matches}
\subsubsection{System Model}
Fix three integers $(M_1,M_2,K)\in\bbN^3$ such that $M_1\geq M_2\geq K$, two positive real numbers $(\alpha,\beta)\in\bbR_+$ and an alphabet $\calX$. Consider any two tuples of distinct distributions $P^{M_1}:=(P_1,\ldots,P_{M_1})\in(\calP(\calX))^{M_1}$ and $Q^{M_2}:=(Q_1,\ldots,Q_{M_2})\in(\calP(\calX))^{M_2}$ such that for any $(i,j)\in[M_1]$, if $i\neq j$, $P_i\neq P_j$ and for any $(i,j)\in[M_2]$ such that $i\neq j$, $Q_j\neq Q_j$. When $\calX$ is finite, e.g., binary, each distribution is a probability mass function (pmf) while when $\calX$ is continuous, e.g., the real set, each distribution is a probability density function (pdf). The assumption of distinct generating distribution for each sequence in a database is consistent with existing studies~\cite{unnikrishnan2015asymptotically,zhou2024tit,zhou2025csm,zhou2025seq}.

Fix an integer $N\in\bbN$. Define two sequences $\xi_N:=\lceil \alpha N\rceil $ and $\chi_N:=\lceil \beta N\rceil$. Consider the first database $\bX^{\xi_N}:=\{X_1^{\xi_N},\ldots,X_{M_1}^{\xi_N}\}$ that consists of $M_1$ sequences, where for each $i\in[M_1]$, $X_i^{\xi_N}=(X_{i,1},\ldots,X_{i,\xi_N})$ is generated i.i.d. from distribution $P_i$. Analogously, the second database consists of $M_2$ sequences, $\bY^{\chi_N}:=\{Y_1^{\chi_N},\ldots,Y_{M_2}^{\chi_N}\}$, where for each $j\in[M_2]$, $Y_j^{\chi_N}=(Y_{j,1},\ldots,Y_{j,\chi_N})$ is generated i.i.d. from distribution $Q_i$. For any $(i,j)\in[M_1]\times[M_2]$, if $X_i^{\xi_N}$ and $Y_j^{\chi_N}$ are generated from the same distribution, i.e., $P_i=Q_j$, then the pair of sequences $(X_i^{\xi_N},Y_j^{\chi_N})$ is said matched. The task of sequence matching is to identify all matched pairs of sequences between the two databases when the generating distributions $(P^{M_1},Q^{M_2})$ are \emph{unknown}. 

Let us first consider the case where the number of matches $K\in\bbN$ is known. In this case, there are in total $T_K={M_1\choose K}{M_2\choose K}K!$ possibilities for all $K$ pairs of matched sequences between the two databases. One such possibility specifies a $K$-match. For ease of notation, given each $l\in[T_K]$, let $\calA_l\in([M_1]\times[M_2])^K$ collect the indices of matched pairs for one possible $K$-match, i.e., for each $(i,j)\in\calA_l$, the $i$-th sequence of the first database is mapped to the $j$-th sequence in the second database. Let $\calM_K$ collect all $T_K$ possibilities, i.e., $\calM_K=\{\calA_1,\ldots,\calA_{T_K}\}^K$. To illustrate the above definitions, we provide two examples below. 

\begin{itemize}
\item Set $M_1=3$, $M_2=2$ and $K=1$. In this case, $T_K={3 \choose 1}{2\choose 1} 1!=6$ and $\calM_K$ is given by 
\begin{align}
\begin{array}{lll}
\calA_1=\{(1,1)\} & \calA_2=\{(2,1)\} & \calA_3=\{(3,1)\} \\
\calA_4=\{(1,2)\} & \calA_5=\{(2,2)\} & \calA_6=\{(3,2)\}.
\end{array}
\label{eg:1}
\end{align}
When $l=2$, $\calA_l=\{(2,1)\}$, which means that the second sequence of the first database is matched to the first sequence of the second database.

\item Set $M_1=3$, $M_2=3$ and $K=3$. In this case, $T_K={3\choose 3}{3\choose 3}3!=6$ and $\calM_K$ is given as follows:
\begin{align}
\begin{array}{ll}
\calA_1=\{(1,1),(2,2),(3,3)\} & \calA_2=\{(1,1),(2,3),(3,2)\}\\
\calA_3=\{(1,2),(2,1),(3,3)\} & \calA_4=\{(1,2),(2,3),(3,1)\}\\
\calA_5=\{(1,3),(2,1),(3,2)\} & \calA_6=\{(1,3),(2,2),(3,1)\}.
\end{array}
\end{align}
When $l=3$, $\calA_l=\{(1,2),(2,1),(3,3)\}$, which means that the first sequence of the first database is matched to the second sequence of the second database, the second sequence of the first database is matched to the first sequence of the second database and the third sequence of the first database is matched to the third sequence of the second database.
\end{itemize}

\subsubsection{Performance Metric}
We consider two types of tests: fixed-length and sequential tests. In a fixed-length test, the sample size parameter $N$ is fixed while in a sequential test, the sample size parameter is a random number that depends on the filtration $\{(\bX^{\xi_n},\bY^{\chi_n})\}_{n\in\bbN}$. To unify the notation, a test $\Phi=(\tau,\phi)$ consists of a potentially random sample size parameter, a.k.a., the stopping time, $\tau$ and a decision $\phi:~\calX^{M_1\xi_\tau}\times\calX^{M_2\chi_\tau}\to\calM_K$. When $\tau$ equals a constant, the test is a fixed-length test~\cite{zhou2024tit}; when $\tau$ is a random number, the test is a sequential test~\cite{zhou2025seq}.

For both tests, when the number of matches is known, the performance criterion is the mismatch probability that quantifies the probability that the test identifies the matched pairs incorrectly. Fix any $\calA\in\calM_K$ and assume that the indices of matched pairs of sequences are given by $\calA$. Define the following set of distributions:
\begin{align}
\calP_\calA:=
\big\{(\tilP^{M_1},\tilQ^{M_2})\in\calP(\calX)^{M_1+M_2}:~\forall~(i,j)\in\calA,~\tilP_i=\tilQ_j~\mathrm{and}~~\forall~(\bari,\barj)\notin\calA,~\tilP_{\bari}\neq \tilQ_{\barj}\big\}\label{def:calp:cala}.
\end{align}
The set $\calP_\calA$ collects all possible tuples of generating distributions when the indices of matched pairs of sequences are given by the set $\calA\in\calM_K$. This way, under any tuple of generating distributions $(P^{M_1},Q^{M_2})\in\calP_\calA$, the mismatch probability is defined as
\begin{align}
\theta_\calA(\Phi|P^{M_1},Q^{M_2})&:=\bbP_\calA\big\{\phi(\bX^{\xi_\tau},\bY^{\chi_\tau})\neq \calA\big\}\label{def:mismatch},
\end{align}
where $\bbP_\calA$ denotes the joint generating distribution of all sequences.

Furthermore, since the stopping time can be stochastic, one would like to constraint its expected value in the worst case. A test is said to satisfy the expected stopping time universality constraint~\cite{Ihwang2022sequential} if there exists an integer $N\in\bbN$ such that the worst case expected stopping time is no greater than $N$, i.e.,
\begin{align}
\max_{\calA\in\calM_K}\sup_{(P^{M_1},Q^{M_2})\in\calP_\calA}\bbE_{\bbP_\calA}[\tau]\leq N\label{constraint:est}.
\end{align}
Note that a fixed-length test naturally satisfies this condition by setting $\tau=N$.

In this paper, we propose low-complexity exponentially consistent fixed-length and sequential tests and characterize the exponential decay rates of the mismatch probabilities of both tests.

\subsection{Case of Unknown Number of Matches}
When the number of matches is unknown, it could range from zero to $\min\{M_1,M_2\}=M_2$. Thus, the task is more complicated: one needs to estimate the number of matches and identifies all pairs of matched sequences if the estimated number is positive. Specifically, the number of possible outcomes of a test increases from $T_K$ to $T=\sum_{\hatK\in[M_2]}T_{\hatK}+1$. Recall the definition of $\calM_K$ above Eq. \eqref{eg:1}. Let $\calM:=\bigcup_{\hatK\in[M_2]}\calM_{\hatK}$. In this setting, one needs to design a test $\Phi_\rmu=(\tau_\rmu,\phi_\rmu)$ with potentially stochastic stopping time $\tau_\rmu$ and a corresponding test $\phi_\rmu$, with input of the two databases and the output being either a set in $\calM$ or the empty set $\emptyset$, the latter of which indicates that there is no matched pair of sequences.

Correspondingly, there are three error probabilities: mismatch probability, false reject probability and false alarm probability. The mismatch and false reject probabilities are defined when the number of matches is positive while the false alarm probability is defined when the number of matches is zero. Fix any set $\calA\in\calM$ that collects the indices of matched sequences. Recall the definition of $\calP_\calA$ in \eqref{def:calp:cala}. For any $(P^{M_1},Q^{M_2})\in\calP_\calA$, the mismatch and false reject probabilities are defined as follows:
\begin{align}
\bar{\theta}_\calA(\Phi_\rmu|P^{M_1},Q^{M_2})&:=\bbP_\calA\big\{\phi_\rmu(\bX^{\xi_\tau},\bY^{\chi_\tau})\notin\{\calA,\emptyset\}\big\},\label{def:mismatch:unknown}\\
\zeta_\calA(\Phi_\rmu|P^{M_1},Q^{M_2})&:=\bbP_\calA\big\{\phi_\rmu(\bX^{\xi_\tau},\bY^{\chi_\tau})=\emptyset\big\}\label{def:freject}.
\end{align}
Note that the mismatch probability $\bar{\theta}_\calA(\Phi_\rmu|P^{M_1},Q^{M_2})$ quantifies the probability that the test identifies a wrong set of matched pairs of sequences while the false reject probability $\zeta_\calA(\Phi_\rmu|P^{M_1},Q^{M_2})$ quantifies the probability that the test incorrectly claims that there is no matched pairs of sequences. The difference in the definitions of the mismatch probability in \eqref{def:mismatch} and \eqref{def:mismatch:unknown} results from the setting of known and unknown number of matches. When the number of matches is unknown, we need to remove the case of no match from the mismatch event, which corresponds to the false reject event.

Analogously to \eqref{def:calp:cala}, define the following set of distributions:
\begin{align}
\calP_\emptyset:=
\big\{(\tilP^{M_1},\tilQ^{M_2})\in\calP(\calX)^{M_1+M_2}:~\forall~(i,j)\in[M_1]\times[M_2],~\tilP_i\neq \tilQ_j\big\}\label{def:calp:emptyset}.
\end{align}
Given any $(P^{M_1},Q^{M_2})\in\calP_\emptyset$, the false alarm probability is defined as 
\begin{align}
\eta(\Phi_\rmu|P^{M_1},Q^{M_2}):=\bbP_\emptyset\Big\{\phi_\rmu(\bX^{\xi_\tau},\bY^{\chi_\tau})\neq \emptyset\Big\}\label{def:etar},
\end{align}
where $\bbP_\emptyset$ denotes the joint generating distribution of all sequences.

Since the stopping time can be stochastic, we would like its expected value to be constrained as in  \eqref{constraint:est}. Specifically, a test is said to satisfy the expected stopping time universality constraint if there exists an integer $N\in\bbN$ such that
\begin{align}
\max_{\calA\in\calM}\sup_{(P^{M_1},Q^{M_2})\in\calP_\calA}\bbE_{\bbP_\calA}[\tau_\rmu]&\leq N,\\
\sup_{(P^{M_1},Q^{M_2})\in\calP_\emptyset}\bbE_{\bbP_\emptyset}[\tau_\rmu]&\leq N.
\end{align}
In this paper, we propose low-complexity fixed-length and sequential tests and characterize the exponential decay rates of all three error probabilities. A fixed-length test naturally satisfies the expected stopping time universality constraint. However, when the number of matches is unknown, our sequential test could not satisfy the constraint. Instead, we derive a sufficient condition for our sequential tests to have bounded expected stopping time. The same sufficient condition is needed by the fixed-length test to be exponentially consistent.

\section{Main Results for Known Number of Matches}
\label{sec:known}
\subsection{Scoring Functions}
\label{sec:prelim}
\subsubsection{GJS Divergence}
Fix a finite alphabet $\calX$ and a pair of distributions $(P,Q)\in\calP(\calX)^2$ with full support. The KL divergence is defined as
\begin{align}
D(P\|Q):=\sum_{x\in\calX}P(x)\log\frac{P(x)}{Q(x)}.
\end{align}
Fix any two positive real numbers $(\alpha,\beta)\in\bbR_+^2$. Define the following linear combination of distributions $(P,Q)$:
\begin{align}
R_{\alpha,\beta}^{P,Q}:=\frac{\alpha P+\beta Q}{\alpha+\beta}\label{def:Rab},
\end{align}
and define the following linear combination of KL divergence:
\begin{align}
\mathrm{GJS}(P,Q,\alpha,\beta):=\alpha D\big(P\|R_{\alpha,\beta}^{P,Q}\big)+\beta D\big(Q\|R_{\alpha,\beta}^{P,Q}\big)\label{def:GJS}.
\end{align}
The GJS Divergence has the following variational form~\cite[Eq. (6)]{Ihwang2022sequential}:
\begin{align}
\label{gjs:variational}
\mathrm{GJS}(P,Q,\alpha,\beta):=\min_{V\in\calP(\calX)}\big(\alpha D(P||V)+\beta D(Q||V)\big).
\end{align}
Fix any two positive real numbers $(\alpha,\beta)\in\bbR_+^2$ and an integer $Nin\bbN$. Recall that $\xi_N=\lceil \alpha N\rceil$ and $\chi_N=\lceil \beta N\rceil$. Given two sequences $x^{\xi_N}=[x_1,\ldots,x_{\xi_N}]$ and $y^{\chi_N}=[y_1\ldots,y_{\chi_N}]$, the GJS scoring function is defined as 
\begin{align}
\mathrm{GJS}(x^{\xi_N},y^{\chi_N}):=\frac{1}{n}\mathrm{GJS}(\hatT_{x^{\xi_N}},\hatT_{y^{\chi_N}},\xi_N,\chi_N)\label{gjscompute}.
\end{align}
As $N\to\infty$, $\mathrm{GJS}(P,Q,x^{\xi_N},y^{\chi_N})$ converges in probability to $\mathrm{GJS}(P,Q,\alpha,\beta)$. Thus, asymptotically, it follows from the weak law of large numbers if $P=Q$, the value of $\mathrm{GJS}(P,Q,x^{\xi_N},y^{\chi_N})$ vanishes while if $P\neq Q$, the value of $\mathrm{GJS}(P,Q,x^{\xi_N},y^{\chi_N})$ is strictly positive. This way, $\mathrm{GJS}(P,Q,x^{\xi_N},y^{\chi_N})$ can be used as a scoring function to check whether two discrete i.i.d. sequences are generated from the same distribution or not, consistent with~\cite{gutman1989asymptotically,zhou2018binary,zhou2022second,zhou2024tit,zhou2025seq}.

\subsubsection{Maximum Mean Discrepancy}
Fix a positive real number $\Theta\in\bbR_+$. Let $\kappa:\calR^2\to\calR_+$ be a bounded characteristic kernel function associated with the Reproducing Kernel Hilbert Spaces~\cite{Fukumizu2007KernelMO,Sriperumbudur2008InjectiveHS} such that $\max_{(a,b)\in\bbR^2}\kappa(a,b)\leq \Theta$. Fix a continuous alphabet $\calX$. Given any two distributions $(P,Q)\in\calP(\calX)^2$, the square population MMD~\cite[Lemma 6]{gretton2012kernel} is defined as
\begin{align}
\mmd^2(P,Q)
&:=\bbE_{PP}[\kappa(X,X')]-2\bbE_{PQ}[\kappa(X,Y)]+\bbE_{QQ}[\kappa(Y,Y')]\label{def:mmd},
\end{align}
where the random variables $(X,X',Y,Y')\sim PPQQ $ are independent of each other. It follows that $\mmd^2(P,Q)=0$ if and only if $P=Q$. A usually adopted kernel function is the following Gaussian kernel function
\begin{align}
\kappa(x,y):=\exp\left(-\frac{(x-y)^2}{2\sigma^2}\right)\label{Gaussiankernel},
\end{align}
where $(x,y)\in\calR^2$ and $\sigma\in\bbR_+$ is an positive real number. In  this case, $\Theta=1$. For all numerical simulations using the MMD metric, the kernel function is chosen as the Gaussian kernel with $\sigma^2=\frac{1}{2}$.

Fix any integer $N\in\bbN$. Given two sequences $x^{\xi_N}=[x_1,\ldots,x_{\xi_N}]$ and $y^{\chi_N}=[y_1\ldots,y_{\chi_N}]$, the MMD scoring function is defined as
\begin{align}
\mmd^2(x^{\xi_N},y^{\chi_N})
\nn&:=\frac{1}{\xi_N(\xi_N-1)}\sum_{(i,j)\in[\xi_N]^2:~i\neq j}\kappa(x_i,x_j)+\frac{1}{\chi_N(\chi_N-1)}\sum_{(i,j)\in[\chi_N]^2:~i\neq j}\kappa(y_i,y_j)\\*
&\qquad-\frac{2}{\xi_N\chi_N}\sum_{(i,j)\in[\xi_n]\times [\chi_N]}\kappa(x_i,y_j)\label{MMDcompute}.
\end{align}
It was shown in~\cite[Lemma 6]{gretton2012kernel} that $\mmd^2(x^{\xi_N},y^{\chi_N})$ is an unbiased estimator for $\mmd^2(P,Q)$. Thus, as $N\to\infty$, if $P_1=P_2$, the value of $\mmd^2(x^{\xi_N},y^{\chi_N})$ vanishes; otherwise, if $P_1\neq P_2$, the value of $\mmd^2(x^{\xi_N},y^{\chi_N})$ is strictly positive. Consistent with \cite{gretton2012jmlr,zhu2025tit,zhou2025csm}, MMD can be used to construct scoring functions to check whether two continuous i.i.d. sequences are generated from the same distribution or not.

\subsection{Fixed-Length Test}
\label{sec:fl:intuition}
\subsubsection{Test Design and Asymptotic Intuition}
We first consider the fixed-length test. The existing tests, including \cite[Eq. (22)]{zhou2024tit} and \cite[Eq. (39)]{zhou2025seq} for discrete sequences and \cite[Eq. (16)]{zhou2025csm} for continuous sequences, search exhaustively over all possible sets of matched sequences to make a reliable decision. Recall from the problem setting that the total number of possibilities equals $T_K={M_1\choose K}{M_2\choose K}K!$. Thus, the computational complexity of these tests scales exponentially with respect to the numbers of sequences $(M_1,M_2)$ and the number of matches $K$. When $M_1=100$, $M_2=50$ and $K=10$, $T_K\approx 6.4526\times 10^{29}$, which is computationally intractable. To solve this problem, in this paper, we propose low-complexity tests and characterize their achievable mismatch exponents. 

\begin{algorithm}[tb]
\caption{Low complexity fixed-length test $\Phi_\rmf$}
\label{low_com:fltest}
\begin{algorithmic}[1]
\REQUIRE Parameter $N\in\bbN$, $M_1$ sequences of the first database, $M_2$ sequences of the second database, the number of matches $K$.
\ENSURE A set of indices for matched pairs of sequences.
\STATE For each $(i,j)\in[M_1]\times[M_2]$, calculate the scoring function value $f(X_i^{\xi_N},Y_j^{\chi_N})$.
\STATE Set $\hatcalA$ as the indices of sequence pairs that have smallest $K$ scoring function values.
\RETURN The set $\hatcalA$.
\end{algorithmic}
\end{algorithm}

Fix any integer $N\in\bbN$. Given any two sequences $(X^{\xi_N},Y^{\chi_N})\in\calX^{\xi_N+\chi_N}$, let $f:\calX^{\xi_N+\chi_N}\to \bbR_+$ be a scoring function, which can be the GJS scoring function in \eqref{gjscompute} or the MMD scoring function in \eqref{MMDcompute}. Our low-complexity test is summarized in Algorithm \ref{low_com:fltest}. The key idea is to calculate the values of all $M_1 M_2$ pairwise scoring functions and identify the matched pairs as those sequences having $K$ smallest scoring function values. This way, the computation complexity is $M_1M_2$ for the calculation of scoring functions and the computational complexity of sorting the scoring functions is on average $O(M_1M_2\log(M_1M_2))$ with the worst case of $O(M_1^2M_2^2)$. For the same value of $M_1=100$, $M_2=50$ and $K=10$, the number of calculations required for our test is at roughly $5000+5000^2=2.5005\times 10^7$, which is much less than $6.4526\times 10^{29}$. 

We first explain the asymptotic intuition why the fixed-length test in Algorithm \ref{low_com:fltest} works. Fix $\calA\in\calM_K$ as the set collecting indices of matched pairs of sequences. It follows from the results in Section \ref{sec:prelim} that, when $n$ is sufficiently large, for any $(i,j)\in\calA$, the scoring function value $f(X_i^{\xi_N},Y_j^{\chi_N})$  for a matched pair of sequences vanishes while for any $(\bari,\barj)\notin\calA$, the scoring function value $f(X_{\bari}^{\xi_N},Y_{\barj}^{\chi_N})$ for an unmatched pair of sequences is strictly positive. Thus, asymptotically, choosing the pairs of sequences as those having the $K$ smallest scoring function values results in the correct set of matched sequences. As we show in the next section, our test is exponentially consistent for both discrete and continuous observed sequences, regardless of the number of matches and the unknown generating distributions.

\subsubsection{Theoretical Benchmarks and Discussions}
Fix any $\calA\in\calM_K$ and tuple of distributions $(P^{M_1},Q^{M_2})\in\calP_\calA$. Given any $(i,j)\in\calA$, $(\bari,\barj)\notin\calA$ and distributions $(\Omega_1,\Omega_2,\Psi_1,\Psi_2)\in(\calP(\calX))^4$, define the linear combination of KL divergence as
\begin{align}
E_{(i,j)}^{(\bari,\barj)}(\Omega_1,\Omega_2,\Psi_1,\Psi_2)
&:=\alpha D(\Omega_1\|P_i)+\beta D(\Psi_1\|P_i)+\alpha  D(\Omega_2\|P_{\bari})+\beta D(\Psi_2\|Q_{\barj})\label{def:eij:bij}.
\end{align}
Next define the following two exponent functions
\begin{align}
E_\calA^\rmf(P^{M_1},Q^{M_2})&:=\min_{\substack{(\Omega_1,\Omega_2,\Psi_1,\Psi_2)\in(\calP(\calX))^4:\\\gjs(\Omega_1,\Psi_1,\alpha,\beta)\geq \gjs(\Omega_2,\Psi_2,\alpha,\beta)
}}\min_{\substack{(i,j)\in\calA\\(\bari,\barj)\notin\calA}}E_{(i,j)}^{\bari,\barj}(\Omega_1,\Omega_2,\Psi_1,\Psi_2)\label{def:fl:exponent},\\
E_{\calA}^{\rmf,\rmc}(P^{M_1},Q^{M_2})&:=
\frac{\min\{\alpha,\beta\}\min_{(\bari,\barj)\notin\calA}(\mmd^2(P_{\bari},Q_{\barj}))^2}{128\Theta^2}\label{def:ea2:fl}.
\end{align}
As we show below, $E_\calA^\rmf(P^{M_1},Q^{M_2})$ and $E_{\calA}^{\rmf,\rmc}(P^{M_1},Q^{M_2})$ lower bound the mismatch exponent of our fixed-length test $\Phi_\rmf$ for discrete and continuous observed sequences, respectively. Both exponents are strictly positive. 
The positiveness of $E_{\calA}^{\rmf,\rmc}(P^{M_1},Q^{M_2})$ is clear since $\mmd^2(P_{\bari},Q_{\barj})$ is strictly positive for any $(\bari,\barj)\notin\calA$. The positiveness of $E_\calA^\rmf(P^{M_1},Q^{M_2})$ requires clarification. Note that to achieve a zero value of $E_\calA^\rmf(P^{M_1},Q^{M_2})$, one needs to find $(i,j)\in\calA$, $(\bari,\barj)\notin\calA$ and $(\Omega_1,\Omega_2,\Psi_1,\Psi_2)\in(\calP(\calX))^4$ such that $\Omega_1=\Psi_1=P_i$, $\Omega_2=P_{\bari}$, $\Psi_2=Q_{\barj}$ and $\gjs(\Omega_1,\Psi_1,\alpha,\beta)\geq \gjs(\Omega_2,\Psi_2,\alpha,\beta)$. This is impossible because the above conditions cannot be satisfied simultaneously. Specifically, when $\Omega_1=\Psi_1$, the condition $\gjs(\Omega_1,\Psi_1,\alpha,\beta)\geq \gjs(\Omega_2,\Psi_2,\alpha,\beta)$ implies that $\Omega_2=\Psi_2$ while the condition $(\bari,\barj)\notin\calA$ implies that $P_{\bari}\neq Q_{\barj}$. Thus, one can not satisfy the condition $\Omega_2=P_{\bari}$ and $\Psi_2=Q_{\barj}$ simultaneously.

\begin{theorem}
\label{theorem:fl}
Fix any $\calA\in\calM_K$ and any tuple of distributions $(P^{M_1},Q^{M_2})\in\calP_\calA$.
\begin{itemize}
\item When $\calX$ is finite, using the GJS scoring function in \eqref{gjscompute}, the mismatch exponent satisfies
\begin{align}
\liminf_{N\to\infty}-\frac{1}{N}\log\theta(\Phi_\rmf|P^{M_1},Q^{M_2})\geq E_\calA^\rmf(P^{M_1},Q^{M_2}).
\end{align}
\item When $\calX$ is continuous, using the MMD scoring function in \eqref{MMDcompute}, the mismatch exponent satisfies
\begin{align}
\liminf_{N\to\infty}-\frac{1}{N}\log\theta(\Phi_\rmf|P^{M_1},Q^{M_2})\geq E_{\calA}^{\rmf,\rmc}(P^{M_1},Q^{M_2}).
\end{align}
\end{itemize}
\end{theorem}
The proof of Theorem \ref{theorem:fl} is provided in Section \ref{proof:theorem:fl}.

Theorem \ref{theorem:fl} shows that our fixed-length test $\Phi_\rmf$ is exponentially consistent for both discrete and continuous sequences and the exponent is always positive. In the proof of Theorem \ref{theorem:fl}, we bound the probability of the event where there exists a matched pair of sequences and an unmatched pair sequences and  such that the scoring function of the unmatched pair of sequences is smaller than that of the matched pair of sequences, i.e., the event $\{\exists~(i,j)\in\calA\mathrm{~and~}(\bari,\barj)\notin\calA:~f(X_i^{\xi_N},Y_j^{\chi_N})\geq f(X_{\bari}^{\xi_N},Y_{\barj}^{\chi_N})\}$. The analysis for discrete observed sequences uses the method of types~\cite{csiszar1998mt} and results in the exponent $E_\calA^\rmf(P^{M_1},Q^{M_2})$ while the analysis for continuous observed sequences uses the McDiarmid's inequality~\cite{mcdiarmid1989method} and results in the exponent $E_{\calA}^{\rmf,\rmc}(P^{M_1},Q^{M_2})$. 

Compared with the exhaustive search tests--\cite[Eq. (39)]{zhou2025seq} for discrete sequences and \cite[Eq. (16)]{zhou2025csm} for continuous sequences, the mismatch exponents of our low-complexity test are generally smaller. However, there are cases where both exponents are the same, including the case of $K=1$.
\begin{itemize}
\item For discrete observed sequences, an equivalent form of the achievable mismatch exponent~\cite[Theorem 2]{zhou2025seq} of the test in \cite[Eq. (39)]{zhou2025seq} is $F_\calA^\rmf(P^{M_1},Q^{M_2})$, where
\begin{align}
F_\calA^\rmf(P^{M_1},Q^{M_2})
&=\min_{\substack{(\Omega^{M_1},\Psi^{M_2})\in\calQ(\calA)}}E(P^{M_1},Q^{M_2},\Omega^{M_1},\Psi^{M_2}),\label{com:fl}\\
\calQ(\calA)
\nn&=\Big\{\Omega^{M_1},\Psi^{M_2})\in(\calP(\calX))^{M_1+M_2}:~\exists~\calB\in\calM_K\setminus\{\calA\},~\\
&\qquad\qquad\sum_{(i,j)\in\calA}\mathrm{GJS}(\Omega_i,\Psi_j,\alpha,\beta)\geq \sum_{(\bari,\barj)\in\calB}\mathrm{GJS}(\Omega_{\bari},\Psi_{\barj},\alpha,\beta)\Big\},\\
E(P^{M_1},Q^{M_2},\Omega^{M_1},\Psi^{M_2})
&=\Big(\sum_{i\in[M_1]}\alpha D(\Omega_i\|P_i)+\sum_{j\in[M_2]}\beta D(\Psi_j\|Q_j)\Big).
\end{align}
An equivalent form of our mismatch exponent is
\begin{align}
E_\calA^\rmf(P^{M_1},Q^{M_2})&
=\min_{\substack{(\Omega^{M_1},\Psi^{M_2})\in\calQ'(\calA)}}E(P^{M_1},Q^{M_2},\Omega^{M_1},\Psi^{M_2})\label{alt:eaf},\\
\calQ'(\calA)
\nn&=\Big\{\Omega^{M_1},\Psi^{M_2})\in(\calP(\calX))^{M_1+M_2}:~\exists~(i,j)\in\calA,~(\bari,\barj)\notin\calA,\\
&\qquad\qquad\gjs(\Omega_i,\Psi_j,\alpha,\beta)\geq \gjs(\Omega_{\bari},\Psi_{\barj},\alpha,\beta)\Big\}.
\end{align}
Comparing \eqref{com:fl} and \eqref{alt:eaf}, we find that the mismatch exponent of our low complexity test is generally smaller, i.e., $E_\calA^\rmf(P^{M_1},Q^{M_2})\leq F_\calA^\rmf(P^{M_1},Q^{M_2})$. The reason is as follows. Fix any $\Omega^{M_1},\Psi^{M_2})\in(\calP(\calX))^{M_1+M_2}$ and $\calB\in\calM_K\setminus\{\calA\}$. The condition that $\sum_{(i,j)\in\calA}\mathrm{GJS}(\Omega_i,\Psi_j,\alpha,\beta)\geq \sum_{(\bari,\barj)\in\calB}\mathrm{GJS}(\Omega_{\bari},\Psi_{\barj},\alpha,\beta)$ implies that there exists $(i,j)\in\calA\setminus\{\calB\}$ and $(\bari,\barj)\in\calB\setminus\calA$ such that $\gjs(\Omega_i,\Psi_j,\alpha,\beta)\geq \gjs(\Omega_{\bari},\Psi_{\barj},\alpha,\beta)$. Thus, $\calQ(\calA)\subseteq\calQ'(\calA)$ and $F_\calA^\rmf(P^{M_1},Q^{M_2})\geq E_\calA^\rmf(P^{M_1},Q^{M_2})$. When $K=1$, the equality $\calQ(\calA)=\calQ'(\calA)$ holds and both exponents are equal.

\item For continuous observed sequences, the achievable mismatch exponent of the test in \cite[Eq. (16)]{zhou2025csm} is given by~\cite[Theorem 1]{zhou2025csm}:
\begin{align}
F_\calA^{\rmf,\rmc}(P^{M_1},Q^{M_2})
&=\frac{\min\{\alpha,\beta\}\big(\Lambda_\calA^K(P^{M_1},Q^{M_2})\big)^2}{128\Theta^2},\\
\Lambda_\calA^K(f^{M_1},Q^{M_2})&=
\min_{\calB'\in\calM_K:~\calB'\neq\calA}
\max_{(i,j)\in\calB'\setminus\calA}\mathrm{MMD}^2(P_i,Q_j).
\end{align}
One can verify that
\begin{align}
\frac{E_{\calA}^{\rmf,\rmc}(P^{M_1},Q^{M_2})}{F_\calA^{\rmf,\rmc}(f^{M_1},g^{M_2})}
&=\frac{\min_{(\bari,\barj)\notin\calA}(\mmd^2(P_{\bari},Q_{\barj}))^2}{\min_{\calB'\in\calM_K:~\calB'\neq\calA}
\max_{(i,j)\in\calB'\setminus\calA}(\mathrm{MMD}^2(P_i,Q_j))^2}\\
&\leq \frac{\min_{(\bari,\barj)\notin\calA}(\mmd^2(P_{\bari},Q_{\barj}))^2}{\min_{\calB'\in\calM_K:~\calB'\neq\calA}\min_{(i,j)\in\calB'\setminus\calA}(\mathrm{MMD}^2(P_i,Q_j))^2}\label{com:c:fl}\\
&\leq \frac{\min_{(\bari,\barj)\notin\calA}(\mmd^2(P_{\bari},Q_{\barj}))^2}{\min_{(\bari,\barj)\notin\calA}(\mmd^2(P_{\bari},Q_{\barj}))^2}\\
&=1.
\end{align}
Thus, our achievable mismatch exponent is generally smaller. However, there are cases where both exponents are same. For example, when $K=1$, the inequality in \eqref{com:c:fl} is an equality since the set $\{(i,j)\in\calB'\setminus\calA\}$ contains only one pair, implying that the maximum and minimum operation leads to the same value and leading to $E_{\calA}^{\rmf,\rmc}(P^{M_1},Q^{M_2})=F_\calA^{\rmf,\rmc}(f^{M_1},g^{M_2})$. For large values of $K$, it is still possible that the equality holds. For example, when $M_1=M_2=K=3$, set $\alpha=\beta=1$ and $P^{M_1}=Q^{M_2}=(\mathrm{Exp}(2),\mathrm{Exp}(5),\calN(-2,1))$, and consider the Gaussian kernel function with $\sigma=\sqrt{\frac{1}{2}}$, both exponents are the same and equal approximately\footnote{The approximation here is obtained via calculating the MMD metric in \eqref{def:mmd} via its unbiased estimator in \eqref{MMDcompute} since the original integral form is hard to compute.} $4.8\times 10^{-5}$.
\end{itemize}

Compared with exhaustive search tests in \cite{zhou2025seq,zhou2025csm}, our low-complexity test has great advantage in terms of the tradeoff between the computational complexity and the mismatch probability. As discussed at the beginning of Section \ref{sec:fl:intuition}, the exhaustive search tests have exponential complexity with respect to the number of sequences $(M_1,M_2)$ and the number of matches $K$. In contrast,  the complexity of our low complexity test is polynomial in the number of sequences $(M_1,M_2)$ regardless of the number of matches $K$. Thus, for slightly large numbers of $(M_1,M_2,K)$, the exhaustive search tests are infeasible due to prohibitively high complexity while our test can still be used even for very large numbers $(M_1,M_2,K)$. In summary, although exhaustive search tests have large mismatch exponents, the overly high computational complexity makes it infeasible while our low complexity tests achieve slightly worse performance with much smaller computational complexity. As a result, our low complexity tests strike a much better tradeoff between the computational complexity and mismatch probability. 

\begin{figure}[tb]
\centering
\begin{tabular}{cc}
\includegraphics[width=.5\columnwidth]{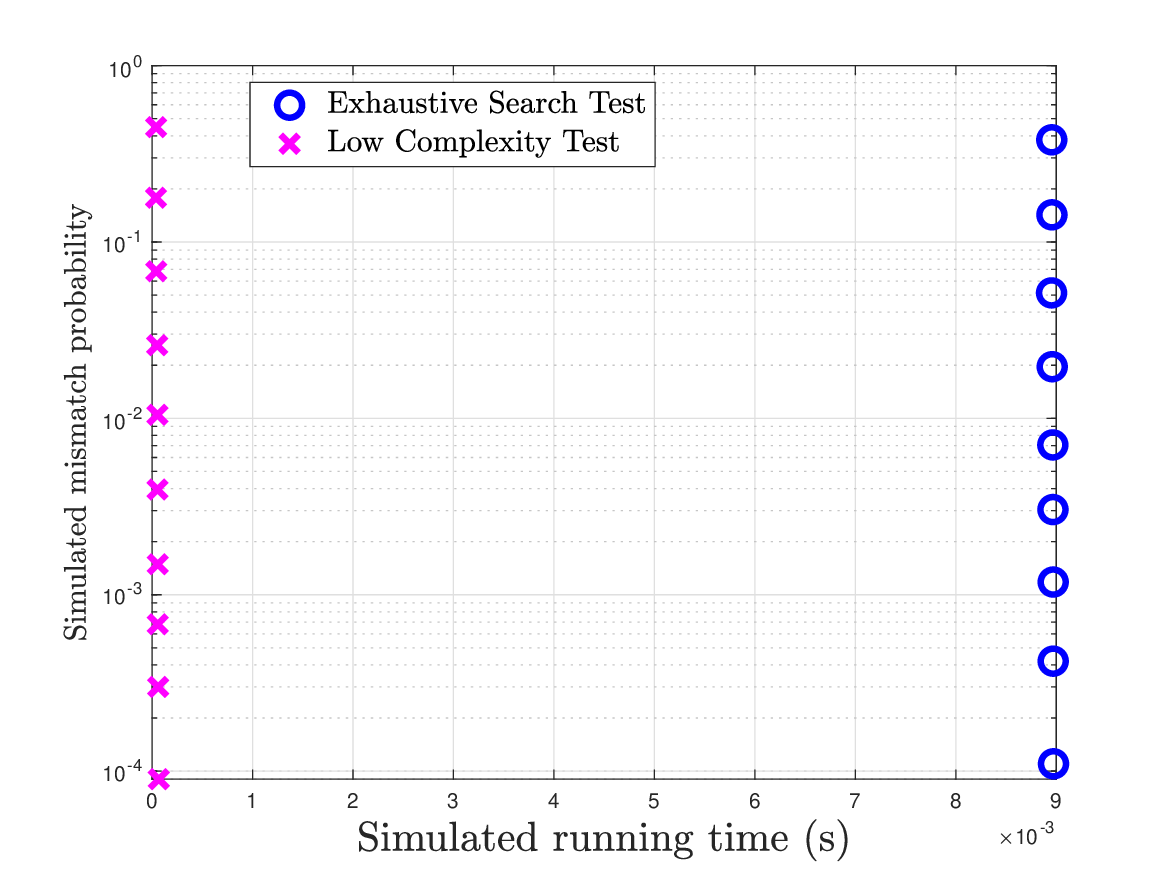}&\includegraphics[width=.5\columnwidth]{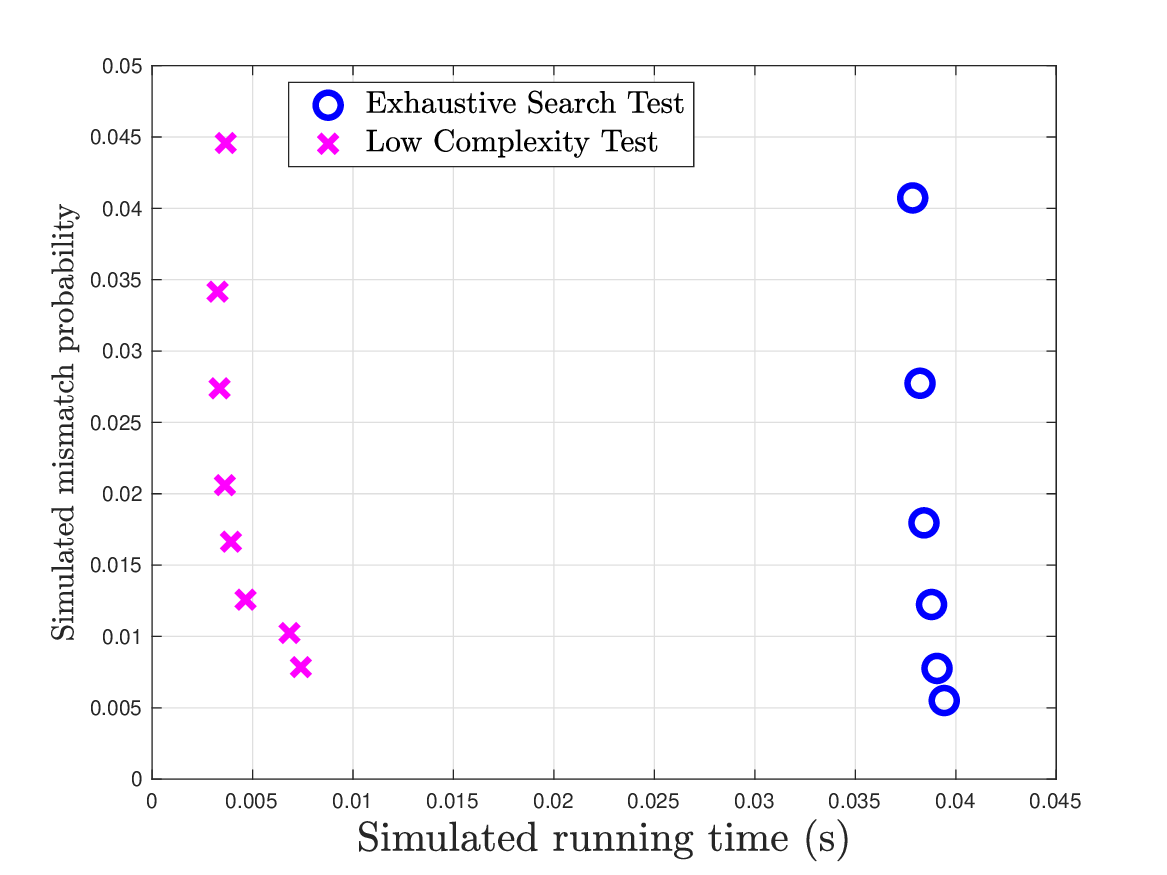}\\
{(a) Discrete observed sequences}& {(b) Continuous observed sequences}
\end{tabular}
\caption{Plot of the simulated mismatch probabilities as a function of running times for the low complexity test in Algorithm \ref{low_com:fltest} and the fixed-length tests in \cite{zhou2025csm,zhou2025seq} when $\alpha=1$ and $\beta=2$. The left figure is for discrete observed sequences using the GJS scoring function in \eqref{gjscompute} when $M_1=6$, $M_2=3$, $K=2$, $\calX=\{1,2,3\}$ and $(P^{M_1},Q^{M_2})$ satisfies that $P_1=Q_1=(0.1,0.3,0.6)$, $P_2=Q_2=(0.2,0.3,0.5)$, $P_3=(\frac{1}{3},\frac{1}{3},\frac{1}{3})$, $P_4=(0.1,0.35,0.55)$, $P_5=(0.15,0.25,0.6)$, $P_6=(0.25,0.2,0.55)$ and $Q_3=(0.3,0.2,0.5)$. The right figure is for continuous observed sequences using the MMD scoring function in \eqref{MMDcompute} when $M_1=M_2=K=3$ and $P^{M_1}=Q^{M_2}=(\mathrm{Exp}(2),\rmU(5),\calN(-2,1))$. As observed, our low complexity fixed-length test achieves a much better tradeoff between the mismatch probability and the running time.}
\label{fl_lct}
\end{figure}
To illustrate the advantage of our low-complexity test, in Fig. \ref{fl_lct}, we plot the simulated mismatch probabilities versus the average running times of our low complexity test in Algorithm \ref{low_com:fltest} and the exhaustive search fixed-length tests in \cite{zhou2025csm,zhou2025seq}. As observed, to achieve the same mismatch probability, our low complexity test requires much less time. For example, for binary observed sequences in Fig. \ref{fl_lct}(a) when $M_1=8$, $M_2=4$, $K=2$, in order to achieve a mismatch probability of $10^{-4}$, our low complexity requires $6.65\times10^{-5}$ second while the exhaustive search fixed-length test in \cite[Eq. (39)]{zhou2025seq} requires $0.009$ second, which implies that when we run the exhaustive search fixed-length test in \cite[Eq. (39)]{zhou2025seq} once, we can run our low complexity test for more than $135$ times. Analogously, for continuous observed sequences in Fig \ref{fl_lct}(b) when $M_1=M_2=K=3$, in order to achieve a mismatch probability of $0.01$, our low complexity requires $0.0074$ second while the exhaustive search fixed-length test in \cite[Eq. (16)]{zhou2025csm} requires $0.039$ second, which implies that when we run the exhaustive search fixed-length test in \cite[Eq. (16)]{zhou2025csm} once, we can run our low complexity test for more than 5 times. For continuous observed sequences, the advantage of low-complexity test is relatively insignificant. This is because the running time of calculating the MMD scoring function in \eqref{MMDcompute} dominates the running times of both our low complexity test in Algorithm \ref{low_com:fltest} and the exhaustive search fixed-length test in \cite[Eq. (16)]{zhou2025csm}, which significantly decreases the difference of the running time of the detection steps. For continuous observed sequences, the advantage of our low-complexity test lies in being feasible even for large values of $(M_1,M_2,K)$.

\subsection{Sequential Tests}

\subsubsection{Test Design}
For the sequential test, one has the freedom to determine the number of samples until a reliable decision could be made. For this purpose, we need a positive threshold $\lambda\in\bbR_+$. The procedure of our sequential test is summarized in Algorithm \ref{low_com:seqtest}. The key idea is to stop only when enough pairs of matched sequences are believed to be found and then run similar minimal scoring function test as the fixed-length test in Algorithm \ref{low_com:fltest}. Specifically, the initial sample size $n$ is set to $N-1$ and increased by one after each iteration. In each iteration, one calculates the scoring function of all $M_1 M_2$ pairs of sequences and calculate the number of the pair of sequences whose scoring function is less than the threshold $\lambda$. If the number is above the number of matches $K$, the test stops collecting more samples and makes a decision. Otherwise, the sample size parameter $n$ is increased by one and the test continues. As we shall show below, with this slight freedom to determine the sample size, the performance of the test can be improved significantly.

The asymptotic intuition of our sequential test $\Phi_\rms$ is similar to the fixed-length test $\Phi_\rmf$. When the sample size parameter $n$ is large, all matched pairs have vanishing scoring function value while any unmatched pair of sequences have strictly positive scoring function value. Thus, the test can stop as desired and makes the identifies all matched pairs successfully.

\begin{algorithm}[tb]
\caption{Low complexity sequential test $\Phi_\rms$}
\label{low_com:seqtest}
\begin{algorithmic}[1]
\REQUIRE $M_1$ sequences of the first database and $M_2$ sequences of the second database, the number of matches $K$, test design parameters $(\lambda,N)\in\bbR_+\times\bbN$.
\ENSURE Stopping time $\tau$ and a set of indices for matched pairs of sequences
\STATE Set $t=0$ and $n=N-1$.
\WHILE{$t=0$}
\STATE For each $(i,j)\in[M_1]\times[M_2]$, calculate the scoring function value $f(X_i^{\xi_n},Y_j^{\chi_n})$.
\STATE Calculate the number $r$ of scoring function values that are smaller than $\lambda$.
\IF{$r\geq K$}
\STATE Set $t=1$;
\ELSE 
\STATE Set $n=n+1$.
\STATE Obtain additional symbols from each observed sequence.
\ENDIF
\ENDWHILE
\STATE Set $\hatcalA$ as indices of sequence pairs that have smallest $K$ scoring function values.
\RETURN Stopping time $\tau=n$ and $\hatcalA$.
\end{algorithmic}
\end{algorithm}

\subsubsection{Theoretical Benchmarks and Discussions}
Given any integer $n\in\bbN$ and define
\begin{align}
g(n):=\frac{\log(n\alpha+2)+\log(n\beta+2)}{n}\label{def:g}.
\end{align}
For any positive real number $\lambda\in\bbR_+$, define the following exponent
\begin{align}
E(\lambda)
&:=\frac{\min\{\alpha,\beta\}\lambda^2}{64\Theta^2}\label{def:E:lambda}.
\end{align}
As we show below, $g(n)$ and $E(\lambda)$ are critical to bound the expected stopping time of our sequential test for discrete and continuous observed sequences, respectively.

Fix any $\calA\in\calM_K$ and $(P^{M_1},Q^{M_2})\in\calP_\calA$. Given any $\lambda\in\bbR_+$, define the exponent functions
\begin{align}
E_\calA^\rms(\lambda,P^{M_1},Q^{M_2})
&:=\min_{{\substack{(\Omega,\Psi)\in(\calP(\calX))^2:\\
\gjs(\Omega,\Psi,\alpha,\beta)\leq \lambda}}}\min_{(\bari,\barj)\notin\calA}\big(\alpha D(\Omega\|P_{\bari})+\beta D(\Psi\|Q_{\barj})\big)\label{def:ea:rms},\\
E_{\calA}^{\rms,\rmc}(\lambda,P^{M_1},Q^{M_2})
&:=\frac{\min\{\alpha,\beta\}\min_{(\bari,\barj)\notin\calA}(\mmd^2(P_{\bari},Q_{\barj})-\lambda)^2}{64\Theta^2}\label{def:e2a:rms}.
\end{align}
As we show below, $E_\calA^\rms(\lambda,P^{M_1},Q^{M_2})$ and $E_{\calA}^{\rms,\rmc}(\lambda,P^{M_1},Q^{M_2})$ lower bound the mismatch exponents of our sequential test for discrete and continuous observed sequences, respectively. Furthermore, $E_\calA^\rms(\lambda,P^{M_1},Q^{M_2})$ decrease in $\lambda$ and equals zero if $\lambda\geq \min_{(\bari,\barj)\notin\calA}\gjs(P_{\bari},Q_{\barj},\alpha,\beta)$ while $E_{\calA}^{\rms,\rmc}(\lambda,P^{M_1},Q^{M_2})$ decreases in $\lambda$ when $0\leq \lambda< \min_{(\bari,\barj)\notin\calA}\mmd^2(P_{\bari},Q_{\barj})$.

\begin{theorem}
\label{theorem:seq}
Fix any $\calA\in\calM_K$ and any tuple of distributions $(P^{M_1},Q^{M_2})\in\calP_\calA$.
\begin{itemize}
\item When $\calX$ is finite, $f(\cdot)$ is the GJS scoring function in \eqref{gjscompute}. 
\begin{itemize}
\item When $N$ satisfies $\frac{K\exp(-(N-1)(\lambda-g(N-1)))}{1-\exp(-(\lambda-g(N-1)))}\leq 1$, the expected stopping time of the sequential test satisfies $\bbE_{\bbP_\calA}[\tau]\leq N$.
\item Asymptotically, the mismatch exponent satisfies 
\begin{align}
\liminf_{N\to\infty}-\frac{1}{N}\log\theta(\Phi_\rms|P^{M_1},Q^{M_2})\geq E_\calA^\rms(\lambda,P^{M_1},Q^{M_2}).
\end{align}
\end{itemize}
\item When $\calX$ is continuous, $f(\cdot)$ is the MMD scoring function in \eqref{MMDcompute}.
\begin{itemize}
\item When $N$ satisfies $\frac{K\exp(-(N-1)E(\lambda))}{1-\exp(-E(\lambda))}\leq 1$, we have $\bbE_{\bbP_\calA}[\tau]\leq N$.
\item Asymptotically, when $\lambda<\min_{(\bari,\barj)\notin\calA}\mmd^2(P_{\bari},Q_{\barj})$, the mismatch exponent satisfies 
\begin{align}
\liminf_{N\to\infty}-\frac{1}{N}\log\theta(\Phi_\rms|P^{M_1},Q^{M_2})\geq E_{\calA}^{\rms,\rmc}(\lambda,P^{M_1},Q^{M_2}).
\end{align}
\end{itemize}
\end{itemize}
\end{theorem}
The proof of Theorem \ref{theorem:seq} is provided in Section \ref{proof:theorem:seq}.

For discrete observed sequences, the condition $\frac{K\exp(-(N-1)(\lambda-g(N-1)))}{1-\exp(-(\lambda-g(N-1)))}\leq 1$ implies that $K\exp(-(N-1)(\lambda-g(N-1)))\leq 1$, which is equivalent to $\lambda\geq \frac{\log K}{N-1}+g(N-1)$. Thus, if we set $\lambda=\frac{\log K}{N-1}+g(N-1)$ in our test, the expected stopping time is always upper bounded by $N$, satisfying the expected stopping time universality constraint in \eqref{constraint:est}. Following the proof of Theorem \ref{theorem:seq}, the mismatch exponent for discrete observed sequences is improved to 
\begin{align}
E_\calA^\rms(0,P^{M_1},Q^{M_2})
&=\min_{{\substack{(\Omega,\Psi)\in(\calP(\calX))^2:\\
\mathrm{GJS}(\Omega,\Psi,\alpha,\beta)\leq 0}}}\min_{(\bari,\barj)\notin\calA}\big(\alpha D(\Omega\|P_{\bari})+\beta D(\Psi\|Q_{\barj})\big).
\end{align}
Compared with the mismatch exponent of the fixed-length test in Theorem \ref{theorem:fl}, we reveal the benefit of sequentiality since the mismatch exponent $E_\calA^\rms(0,P^{M_1},Q^{M_2})$ of our sequential test in Algorithm \ref{low_com:seqtest} is larger than the mismatch exponent $E_\calA^\rmf(P^{M_1},Q^{M_2})$ of our fixed-length test in \ref{low_com:fltest}. This is because the GJS divergence in \eqref{def:GJS} is non-negative and thus
\begin{align}
E_\calA^\rmf(P^{M_1},Q^{M_2})
&=\min_{\substack{(\Omega_1,\Omega_2,\Psi_1,\Psi_2)\in(\calP(\calX))^4:\\\gjs(\Omega_1,\Psi_1,\alpha,\beta)\geq \gjs(\Omega_2,\Psi_2,\alpha,\beta)
}}\min_{\substack{(i,j)\in\calA\\(\bari,\barj)\notin\calA}}E_{(i,j)}^{\bari,\barj}(\Omega_1,\Omega_2,\Psi_1,\Psi_2)\\
&\leq\min_{\substack{(\Omega_1,\Omega_2,\Psi_1,\Psi_2)\in(\calP(\calX))^4:\\\gjs(\Omega_2,\Psi_2,\alpha,\beta)\leq 0}}\min_{\substack{(i,j)\in\calA\\(\bari,\barj)\notin\calA}}
E_{(i,j)}^{\bari,\barj}(\Omega_1,\Omega_2,\Psi_1,\Psi_2)\\
&=E_\calA^\rms(0,P^{M_1},Q^{M_2}).
\end{align}

Analogously, for continuous observed sequences, the condition $\frac{K\exp(-(N-1)E(\lambda))}{1-\exp(-E(\lambda))}\leq 1$ implies that $K\exp(-(N-1)E(\lambda))\leq 1$, which is equivalent to $N\geq\frac{\log K}{E(\lambda)}+1=\frac{64\Theta^2\log K}{\min\{\alpha,\beta\}\lambda^2}+1$. Thus, the mismatch exponent of the sequential test can be improved to 
\begin{align}
E_{\calA}^{\rms,\rmc}(0,P^{M_1},Q^{M_2})
&:=\frac{\min\{\alpha,\beta\}\min_{(\bari,\barj)\notin\calA}(\mmd^2(P_{\bari},Q_{\barj}))^2}{64\Theta^2}.
\end{align}

\begin{figure}[tb]
\centering
\begin{tabular}{cc}
\includegraphics[width=.5\columnwidth]{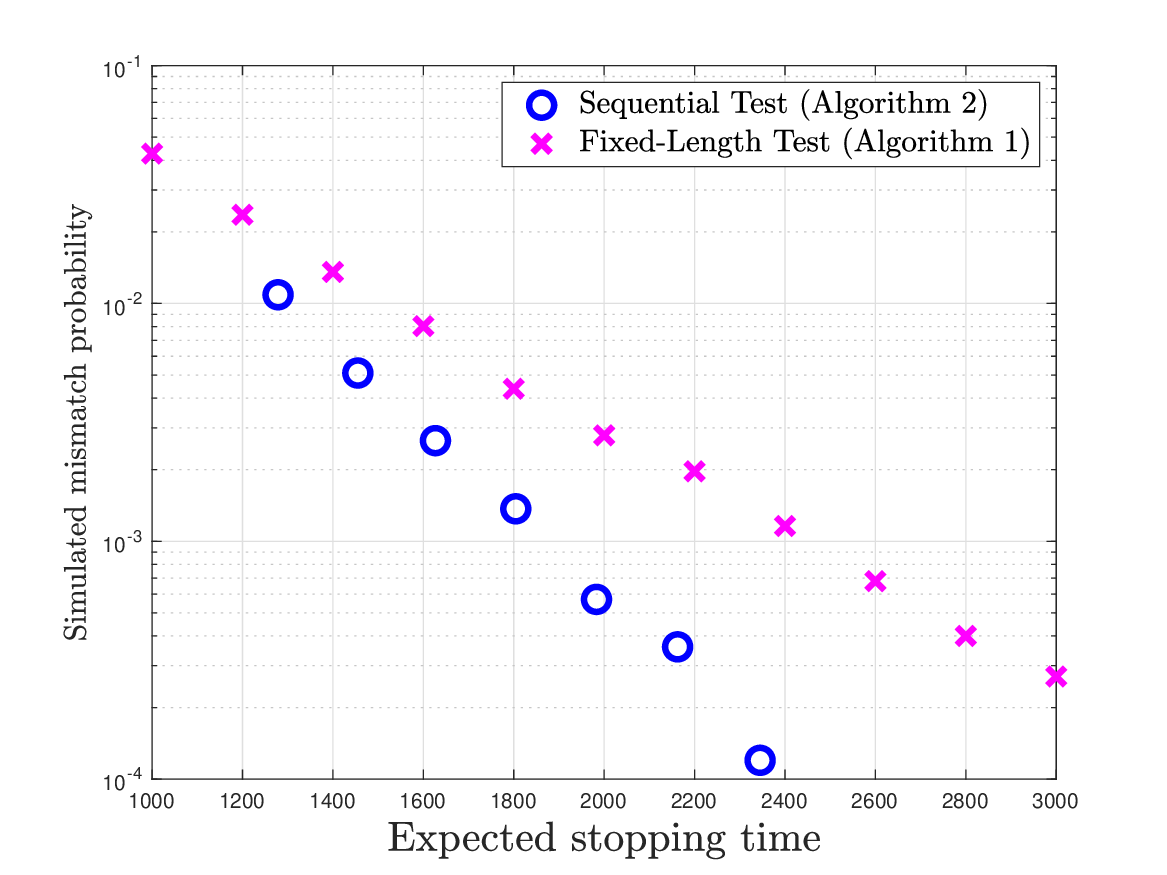}&\includegraphics[width=.5\columnwidth]{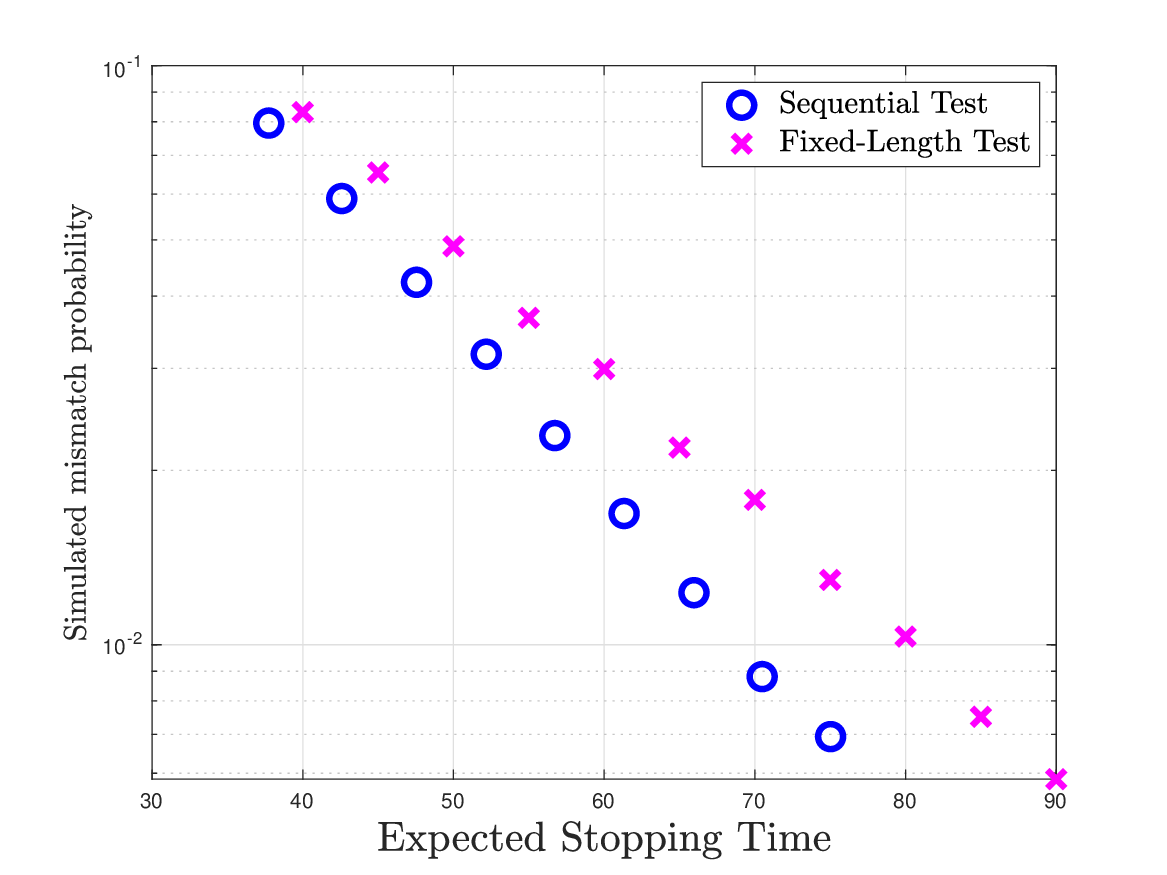}\\
{(a) Discrete Observed Sequences} & {(b) Continuous Observed Sequences} 
\end{tabular}
\caption{Plot of the simulated mismatch probabilities as a function of expected stopping time for the low complexity fixed-length test in Algorithm \ref{low_com:fltest} and the low-complexity sequential test in Algorithm \ref{low_com:seqtest} with $\lambda=0.01$ when $\alpha=1$, $\beta=2$. The left figure is for $M_1=3$, $M_2=K=2$, $P^{M_1}=\mathrm{Bern}(0.1,0.2,0.15)$ and $Q^{M_2}=\mathrm{Bern}(0.1,0.2)$ while the right figure is for $M_1=M_2=3$, $K=2$, $P^{M_1}=(\mathrm{Exp}(2),\rmU(10),\calN(-5,1))$ and $Q^{M_2}=(\mathrm{Exp}(2),\rmU(10),\calN(5,4))$. As observed, with the same expected stopping time, our sequential test achieves better performance than the fixed-length test.
}
\label{bseq_d}
\end{figure}

To illustrate the benefit of sequentiality, in Fig. \ref{bseq_d}, we plot the simulated mismatch probabilities as a function of the expected stopping time for the low complexity fixed-length test in Algorithm \ref{low_com:fltest} and the low-complexity sequential test in Algorithm \ref{low_com:seqtest} with $\lambda=0.01$. As observed, the sequential test achieves smaller mismatch probability under the same expected stopping time, validating the benefit of sequentiality in the non-asymptotic case.

\section{Main Results for Unknown Number of Matches}
\label{sec:unknown}
\subsection{Fixed-Length Tests}
When the number of matches is unknown, it can be any value from zero to $\min\{M_1,M_2\}$, which equals to $M_2$ since we assume $M_1\geq M_2$. In this case, we need to estimate the number of matches and identify all matched pairs of sequences. Our low-complexity fixed-length test for this case is summarized in Algorithm \ref{low_com:fltest:unknown}. The key idea is to identify all pairs of sequences that have small enough scoring function values as the matched pairs and claim there is no matched pair of sequences if the scoring functions of all $M_1M_2$ pairs of sequences are large enough. Specifically, a positive threshold $\lambda\in\bbR_+$ is introduced. Similarly to the case of known number of matches as in Algorithm \ref{low_com:fltest}, we first calculate all $M_1M_2$ pairwise scoring function values. Subsequently, we collect the indices of potential matched pairs as the set $\hatcalA$, which contain indices of pairs of sequences that have scoring function value no greater than $\lambda$. If $\hatcalA$ is the empty set, the decision of no match is made; otherwise, the matched pairs of sequences are claimed to be those having indices in $\hatcalA$. 

The asymptotic intuition of the test in Algorithm \ref{low_com:fltest:unknown} is analogously to that of Algorithm \ref{low_com:fltest}. It follows from the weak law of large numbers that the scoring function value of any matched pairs of sequences vanishes while the scoring function value of any unmatched pairs of sequences is strictly positive. Therefore, when $\lambda$ is strictly positive, we can always make the correct decision, either find there is no matched pair or identify all matched pairs of sequences. In what follows, we explicitly lower bound the exponential decay rates of three error probabilities of the test: mismatch, false reject and false alarm.

\begin{algorithm}[tb]
\caption{Low complexity fixed-length test $\Phi^\rmu_\rmf$ with unknown number of matches}
\label{low_com:fltest:unknown}
\begin{algorithmic}[1]
\REQUIRE Parameter $N\in\bbN$, $M_1$ sequences of the first database, $M_2$ sequences of the second database, test design parameter $\lambda\in\bbR_+$.
\ENSURE A set of indices for matched pairs of sequences.
\STATE For each $(i,j)\in[M_1]\times[M_2]$, calculate the scoring function value $f(X_i^{\xi_N},Y_j^{\chi_N})$.
\STATE Set $\hatcalA$ as indices of all pairs of sequences that have scoring function value no greater than $\lambda$.
\IF{$\hatcalA=\emptyset$}
\RETURN The empty set $\emptyset$.
\ELSE
\RETURN The set $\hatcalA$.
\ENDIF
\end{algorithmic}
\end{algorithm}

Fix any positive real number $\lambda\in\bbR_+$. Given any $(P^{M_1},Q^{M_2})\in\calP_\emptyset$, define the exponent function 
\begin{align}
E_{\rm{fa}}(\lambda,P^{M_1},Q^{M_2})
&:=\min_{(i,j)\in[M_1]\times[M_2]}\min_{\substack{(\Omega,\Psi)\in(\calP(\calX))^2:\\\gjs(\Omega,\Psi,\alpha,\beta)\leq \lambda}}\Big(\alpha D(\Omega\|P_i)+\beta D(\Psi\|Q_j)\Big)\label{def:e:fa},\\
E_{\rm{fa}}^\rmc(\lambda,P^{M_1},Q^{M_2})
&:=
\min_{(i,j)\in[M_1]\times[M_2]}\frac{\min\{\alpha,\beta\}(\mmd^2(P_i,Q_j)-\lambda)^2}{64\Theta^2}\label{def:e2:fa}.
\end{align}
As we show below, $E_{\rm{fa}}(\lambda,P^{M_1},Q^{M_2})$ and $E_{\rm{fa}}^\rmc(\lambda,P^{M_1},Q^{M_2})$ lower bound the false alarm exponents for discrete and continuous observed sequences respectively. The exponent function $E_{\rm{fa}}(\lambda,P^{M_1},Q^{M_2})$ decreases in $\lambda$, is strictly positive if $0\leq \lambda<\min_{(i,j)\in[M_1]\times[M_2]}\gjs(P_i,Q_j,\alpha,\beta)$ and achieves the maximum value when $\lambda=0$. Furthermore, when $\lambda<\min_{(i,j)\in[M_1]\times[M_2]}\mmd^2(P_i,Q_j)$, $E_{\rm{fa}}^\rmc(\lambda,P^{M_1},Q^{M_2})$ decreases in $\lambda$ and is strictly positive.

Recall the definitions of the exponent functions $E(\cdot)$, $E_\calA^\rms(\cdot)$, and $E_{\calA}^{\rms,\rmc}(\cdot)$ in \eqref{def:E:lambda}, \eqref{def:ea:rms} and \eqref{def:e2a:rms}, respectively.
\begin{theorem}
\label{theorem:fl:u}
Our fixed-length test in Algorithm \ref{low_com:fltest:unknown} is exponentially consistent under mild conditions. Fix any $\lambda\in\bbR_+$.
\begin{itemize}
\item For discrete observed sequences, $f(\cdot)$ is the GJS scoring function in \eqref{gjscompute}. 
\begin{itemize}
\item For any $(P^{M_1},Q^{M_2})\in\calP_\emptyset$,  the false alarm exponent satisfies
\begin{align}
\liminf_{N\to\infty}-\frac{1}{N}\log \eta(\Phi^\rmu_\rmf|P^{M_1},Q^{M_2})
&\geq E_{\rm{fa}}(\lambda,P^{M_1},Q^{M_2}).
\end{align}
\item Fix any $\calA\in\calM$ and tuple of distributions $(P^{M_1},Q^{M_2})\in\calP_\calA$. The mismatch and false reject exponents satisfy
\begin{align}
\liminf_{N\to\infty}-\frac{1}{N}\log \bar{\theta}(\Phi^\rmu_\rmf|P^{M_1},Q^{M_2})
&\geq \min\big\{\lambda,E_\calA^\rms(\lambda,P^{M_1},Q^{M_2})\big\},\\
\liminf_{N\to\infty}-\frac{1}{N}\log \eta(\Phi^\rmu_\rmf|P^{M_1},Q^{M_2})
&\geq \lambda.
\end{align} 
\end{itemize}
\item For continuous observed sequences, $f(\cdot)$ is the MMD scoring function in \eqref{MMDcompute}. 
\begin{itemize}
\item For any $(P^{M_1},Q^{M_2})\in\calP_\emptyset$,  the false alarm exponent satisfies
\begin{align}
\liminf_{N\to\infty}-\frac{1}{N}\log \eta(\Phi^\rmu_\rmf|P^{M_1},Q^{M_2})
&\geq E_{\rm{fa}}^\rmc(\lambda,P^{M_1},Q^{M_2}).
\end{align}
\item Fix any $\calA\in\calM$ and tuple of distributions $(P^{M_1},Q^{M_2})\in\calP_\calA$. The mismatch and false reject exponents satisfy
\begin{align}
\liminf_{N\to\infty}-\frac{1}{N}\log \bar{\theta}(\Phi^\rmu_\rmf|P^{M_1},Q^{M_2})
&\geq \min\big\{E(\lambda),E_{\calA}^{\rms,\rmc}(\lambda,P^{M_1},Q^{M_2})\big\},\\
\liminf_{N\to\infty}-\frac{1}{N}\log \eta(\Phi^\rmu_\rmf|P^{M_1},Q^{M_2})
&\geq E(\lambda).
\end{align} 
\end{itemize}
\end{itemize}
\end{theorem}
The proof of Theorem \ref{theorem:fl:u} is provided in Section \ref{proof:fl:u}.

We first explain why each exponent function appears. For our fixed-length test in Algorithm \ref{low_com:fltest:unknown}, a false alarm event occurs if there exists a pair of sequences that has scoring function value no greater than $\lambda$ when the unknown generating distributions satisfy $(P^{M_1},Q^{M_2})\in\calP_\emptyset$. The exponent functions $E_{\rm{fa}}(\lambda,P^{M_1},Q^{M_2})$ and $E_{\rm{fa}}^\rmc(\lambda,P^{M_1},Q^{M_2})$ lower bound the exponential decay rates of the probabilities of such event for discrete and continuous observed sequences, respectively. Furthermore, when there are matched pairs of sequences, a mismatch event occurs if i) a matched pair of sequences has scoring function value greater than $\lambda$ or ii) an unmatched pair of sequence has scoring function value no greater than $\lambda$. The exponential decay rates of these two error events are lower bounded by $(\lambda,E_\calA^\rms(\lambda,P^{M_1},Q^{M_2}))$ for discrete observed sequences and by $(E(\lambda),E_{\calA}^{\rms,\rmc}(\lambda,P^{M_1},Q^{M_2}))$ for continuous observed sequences. Finally, a false reject event occurs if all matched pairs of sequences have scoring function values greater than $\lambda$ and the exponent functions $(\lambda,E(\lambda))$ lower bound the exponential decay rates of the probabilities of the false reject event for discrete and continuous observed sequences, respectively.

\begin{figure}[tb]
\centering
\begin{tabular}{cc}
\includegraphics[width=.5\columnwidth]{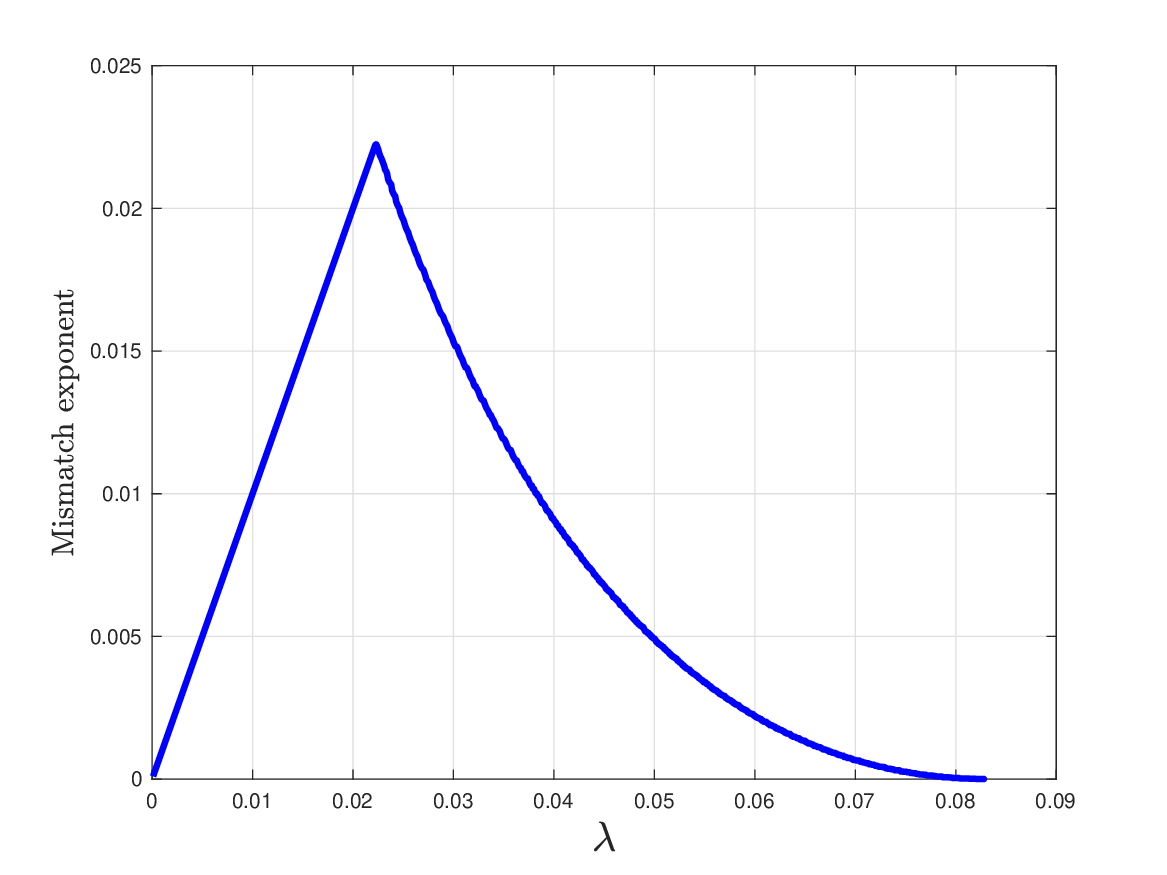} &\includegraphics[width=.5\columnwidth]{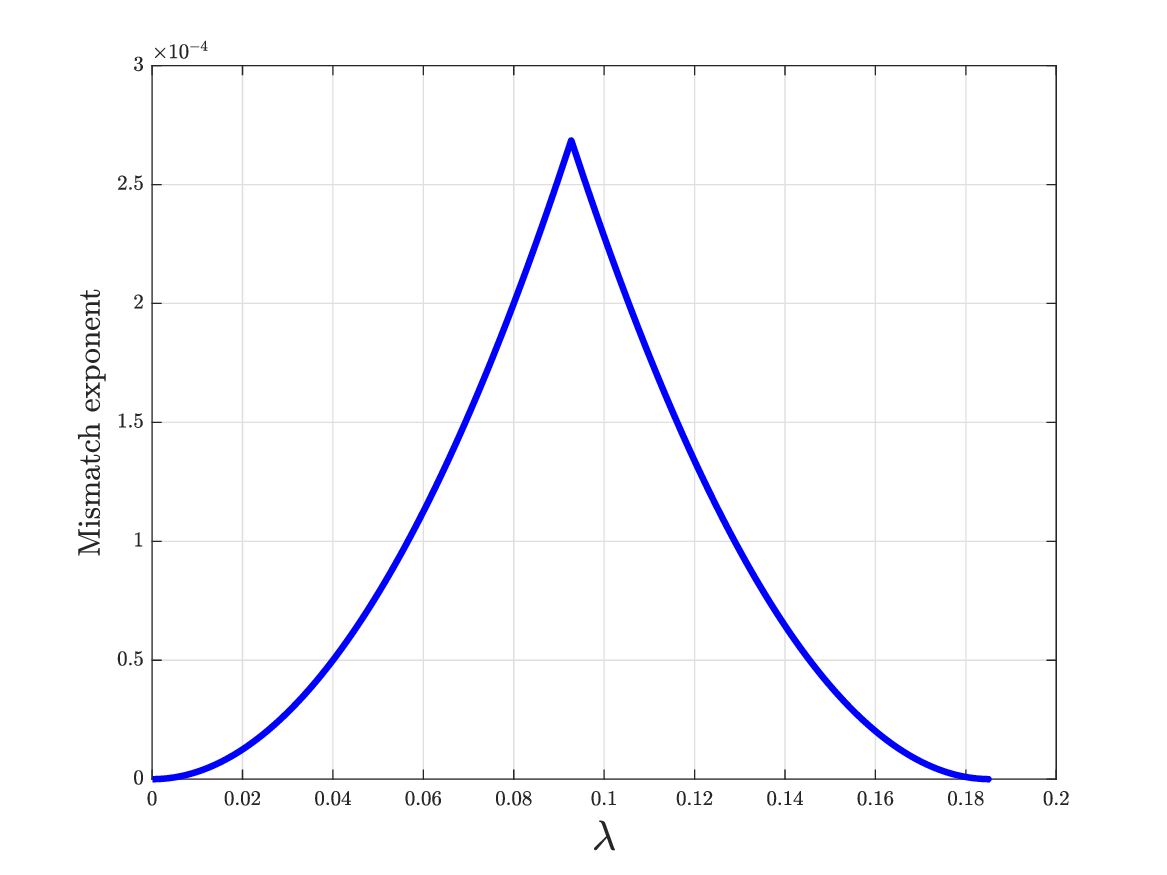}  \\
{(a) Discrete observed sequences}& {(b) Continuous observed sequences} 
\end{tabular}
\caption{Plot of the achievable mismatch exponent in Theorem \ref{theorem:seq} as a function of our sequential test in Algorithm \ref{low_com:seqtest} with parameter $\lambda$. The left figure is for discrete observed sequences with $\alpha=1$, $\beta=2$, $M_1=3$, $M_2=K=2$, $P^{M_1}=\mathrm{Bern}(0.1,0.3,0.6)$ and $Q^{M_2}=\mathrm{Bern}(0.1,0.3)$ while the right figure is for continuous observed sequences with $\alpha=\beta=5$, $M_1=M_2=K=3$, $P^{M_1}=Q^{M_2}=(\mathrm{Exp}(1),\rmU(10,30),\calN(-10,25))$. As observed, the mismatch exponent first increases in $\lambda$ and then decreases in $\lambda$.}
\label{mis_exponent}
\end{figure}

The test design parameter $\lambda$ trades off the exponential decay rates of the three error probabilities: false alarm, mismatch and false reject. The false alarm exponent decreases in $\lambda$ while the false reject exponent increases in $\lambda$. The mismatch exponent is a bit complicated as a minimum of two terms: one term increases in $\lambda$ while the other term decreases in $\lambda$. As we shall, for sequential test, such tradeoff is resolved due to the freedom to choose the stopping time. To clarify, in Fig. \ref{mis_exponent}, we plot the mismatch exponent as a function of $\lambda$. As observed, the mismatch exponent first increases in $\lambda$ and then decreases in $\lambda$. This is because, when $\lambda$ is small, the exponent function $E_{\calA}^{\rms}(\lambda,P^{M_1},Q^{M_2})$ for discrete observed sequences and $E_{\calA}^{\rms,\rmc}(\lambda,P^{M_1},Q^{M_2})$ is larger than $\lambda$ and thus the mismatch exponent increases in $\lambda$. But as $\lambda$ increases further, $E_{\calA}^{\rms,\rmc}(\lambda,P^{M_1},Q^{M_2})$ is smaller than $\lambda$ and the mismatch exponent is given by $E_{\calA}^{\rms,\rmc}(\lambda,P^{M_1},Q^{M_2})$, which decreases in $\lambda$.

In contrast to Theorem \ref{theorem:fl}, when the number of matches is unknown, our fixed-length test is no longer exponentially consistent for any tuple of generating distributions. Instead, Theorem \ref{theorem:fl:u} provides a sufficient condition for our tests to have strictly positive exponents. Specifically, when there is no match, for any 
$(P^{M_1},Q^{M_2})\in\calP_\emptyset$, the false alarm exponent of our fixed-length test is strictly positive when $\lambda<\min_{(i,j)\in[M_1]\times[M_2]}\gjs(P_i,Q_j,\alpha,\beta)$ for discrete sequences and $\lambda<\min_{(i,j)\in[M_1]\times[M_2]}\mmd^2(P_i,Q_j)$ for continuous sequences. When there exists matched pairs whose indices are given a set $\calA\in\calM$, for any $(P^{M_1},Q^{M_2})\in\calP_\calA$, both the misclassification and false reject exponents are positive when $0<\lambda<\min_{(\bari,\barj)\notin\calA}\mathrm{GJS}(P_{\bari},Q_{\barj},\alpha,\beta)$ for discrete sequences and when $0<\lambda<\min_{(\bari,\barj)\notin\calA}\mmd^2(P_{\bari},Q_{\barj})$ for continuous sequences.

\begin{figure}[tb]
\centering
\begin{tabular}{cc}
\includegraphics[width=.5\columnwidth]{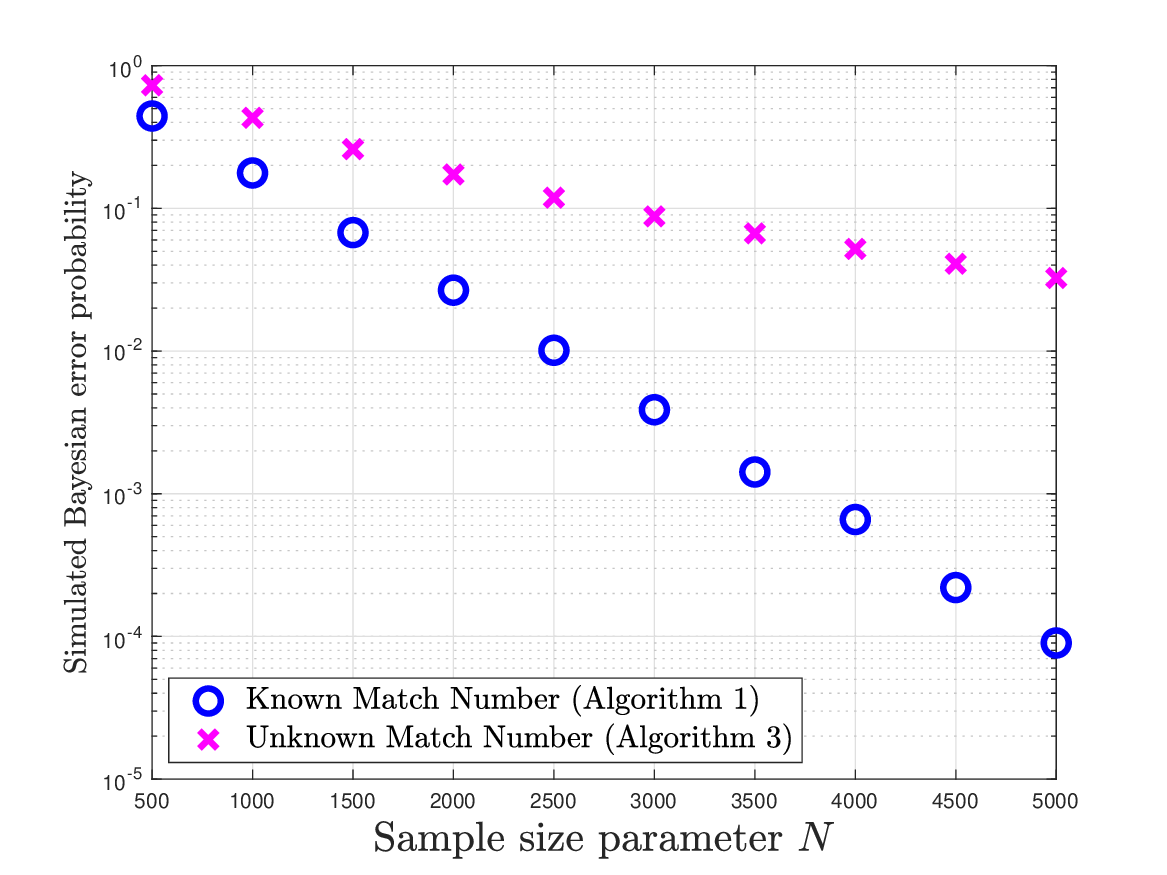}&\includegraphics[width=.5\columnwidth]{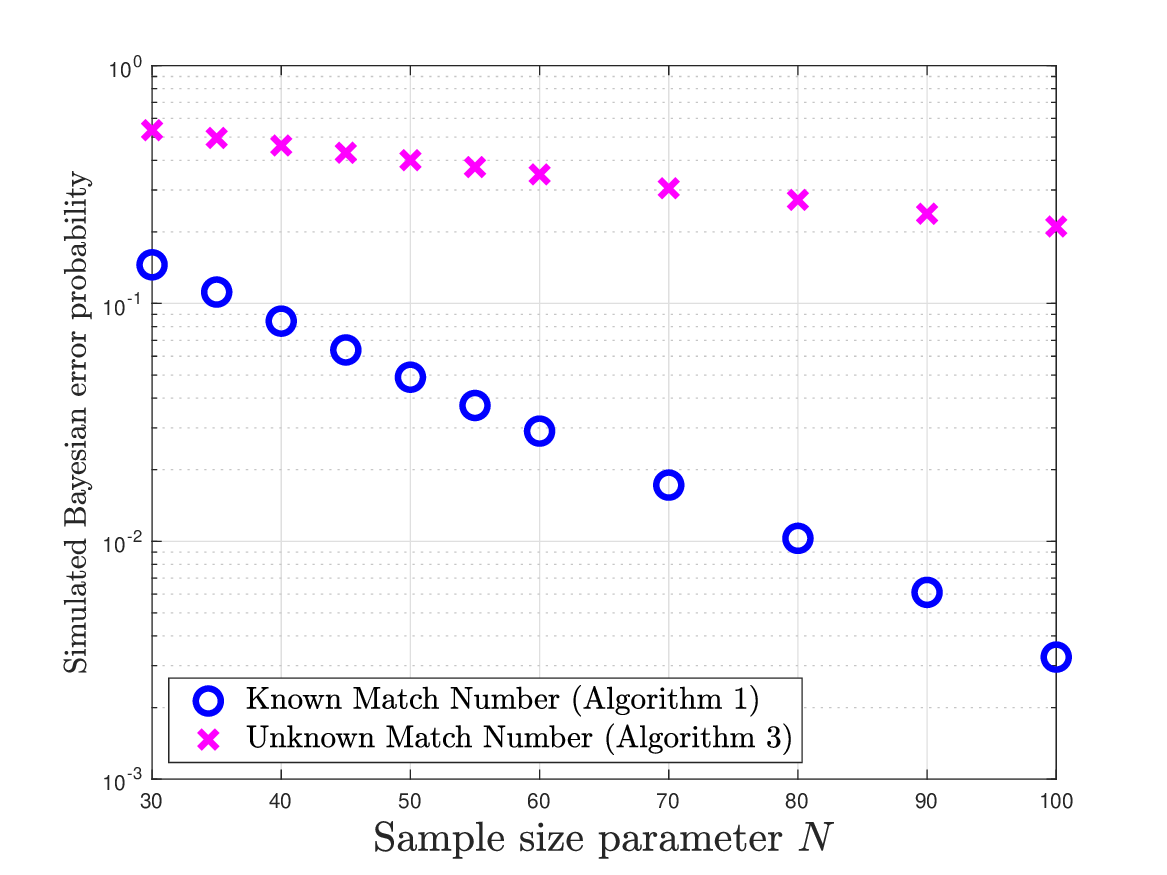}\\
{(a) Discrete observed sequences}& {(b) Continuous observed sequences}
\end{tabular}
\caption{Plot of the simulated mismatch probability of the fixed-length test in Algorithm \ref{low_com:fltest} and the Bayesian error probability of the fixed-length test in Algorithm \ref{low_com:fltest:unknown} when there are matched pairs of sequences, $\alpha=1$ and $\beta=2$. The left figure is for discrete observed sequences using the GJS scoring function in \eqref{gjscompute} when $\lambda=0.002$, $M_1=6$, $M_2=3$, $K=2$, $\calX=\{1,2,3\}$ and $(P^{M_1},Q^{M_2})$ satisfies that $P_1=Q_1=(0.1,0.3,0.6)$, $P_2=Q_2=(0.2,0.3,0.5)$, $P_3=(\frac{1}{3},\frac{1}{3},\frac{1}{3})$, $P_4=(0.1,0.35,0.55)$, $P_5=(0.15,0.25,0.6)$, $P_6=(0.25,0.2,0.55)$ and $Q_3=(0.3,0.2,0.5)$. The right figure is for continuous observed sequences using the MMD scoring function in \eqref{MMDcompute} when $\lambda=0.008$, $M_1=M_2=K=3$ and $P^{M_1}=Q^{M_2}=(\mathrm{Exp}(2),\rmU(5),\calN(-2,1))$. As observed, there is a penalty of not knowing the number of matches: the test in Algorithm \ref{low_com:fltest} that knows the number of matches achieves smaller Bayesian error probability than the test in Algorithm \ref{low_com:fltest:unknown} that does not know the number of matches.}
\label{fl_penalty}
\end{figure}

Comparing Theorems \ref{theorem:fl} and \ref{theorem:fl:u}, we reveal the penalty on the test performance when the number of matches is unknown. For fair comparison, when the number of matches is unknown and positive, we need to consider the Bayesian exponent, which is the smaller one of the mismatch and the false reject exponent. For continuous sequences, the Bayesian exponent is given by $\min\big\{E(\lambda),E_{\calA}^{\rms,\rmc}(\lambda,P^{M_1},Q^{M_2})\big\}$. The maximum Bayesian exponent is achieved by $\lambda$ such that $E(\lambda)=E_{\calA}^{\rms,\rmc}(\lambda,P^{M_1},Q^{M_2})$. Using the definitions of $E(\cdot)$ in in \eqref{def:E:lambda} and $E_{\calA}^{\rms,\rmc}(\cdot)$ in \eqref{def:e2a:rms}, we conclude that the maximum Bayesian exponent is achieved when $\lambda=\frac{\min_{(i,j)\notin\calA\mmd^2(P_{\bari},Q_{\barj})}}{2}$, leading to the Bayesian exponent of $\frac{\min\{\alpha,\beta\}\min_{(\bari,\barj)\notin\calA}(\mmd^2(P_{\bari},Q_{\barj}))^2}{256\Theta^2}$, which is exactly half of the mismatch exponent in Theorem \ref{theorem:fl} when the number of matches is known. Similarly, for discrete observed sequences, the maximum Bayesian exponent is achieved by $\lambda$ such that $\lambda=E_\calA^\rms(\lambda,P^{M_1},Q^{M_2})$. In contrast to the case of continuous observed sequences, the theoretical comparison of the Bayesian exponent with the mismatch exponent in Theorem \ref{theorem:fl} for discrete observed sequences is less obvious due to the complicated expressions of the exponent functions. Instead, we provide a numerical example. Specifically, in Fig. \ref{fl_penalty}, we plot the simulated mismatch probability of the test in Algorithm \ref{low_com:fltest} and the Bayesian error probability of the test in Algorithm \ref{low_com:fltest:unknown} when there are matches, for both discrete and continuous observed sequences. As observed, in both cases, there is a penalty of not knowing the number of matches: the test in Algorithm \ref{low_com:fltest} that knows the number of matches achieves smaller Bayesian error probability than the test in Algorithm \ref{low_com:fltest:unknown} that does not know the number of matches.

\subsection{Sequential Tests}
\begin{algorithm}[tb]
\caption{Low complexity sequential test $\Phi^\rmu_\rms$ with unknown number of matches}
\label{low_com:seqtest:unknown}
\begin{algorithmic}[1]
\REQUIRE $M_1$ sequences of the first database and $M_2$ sequences of the second database, test design parameters $(\lambda_1,\lambda_2,N)\in\bbR_+^2\times\bbN$ such that $\lambda_1\leq \lambda_2$.
\ENSURE Stopping time $\tau$ and a set of indices for matched pairs of sequences.
\STATE Set $t=0$ and $n=N-1$.
\WHILE{$t=0$}
\STATE For each $(i,j)\in[M_1]\times[M_2]$, calculate the scoring function value $f(X_i^{\xi_n},Y_j^{\chi_n})$.
\STATE Set $\hatcalA$ as indices of all pairs of sequences that have scoring function value no greater than $\lambda_1$.
\STATE Set $\calB$ as indices of all pairs of sequences that have scoring function value greater than $\lambda_2$.
\IF{$|\hatcalA|+|\calB|=M_1M_2$}
\STATE Set $t=1$;
\ELSE 
\STATE Set $n=n+1$.
\STATE Obtain additional symbols from each observed sequence
\ENDIF
\ENDWHILE
\IF{$\hatcalA=\emptyset$}
\RETURN Stopping time $\tau_\rmu=n$ and the empty set $\emptyset$.
\ELSE
\RETURN Stopping time $\tau_\rmu=n$ and the set $\hatcalA$.
\ENDIF
\end{algorithmic}
\end{algorithm}

When the number of matches is unknown, the construction of a sequential test is quite complicated since one needs to determine a stopping time to make a reliable decision of either no match or identifying all matched pairs of sequences. To solve the problem, we adopt the idea of clustering. Specifically, set the initial stopping time as $n=N-1$, where $N\in\bbN$ is a design parameter. The value of $n$ is increased by one in each iteration until the stopping criterion is matched. Within each iteration, we first calculate the values of all $M_1M_2$ pairwise scoring functions. Subsequently, using two positive thresholds $(\lambda_1,\lambda_2)$ such that $\lambda_1\leq \lambda_2$, we use a set $\hatcalA$ to collect indices of all pairs of sequences that having scoring function value no greater than $\lambda_1$ and use another set $\calB$ to collect indices of all pairs of sequences that having scoring function value greater than $\lambda_2$. Intuitively, $\hatcalA$ collects the indices of matched pairs while $\calB$ collects indices of unmatched pairs. When $|\hatcalA|+|\calB|$ so that all $M_1M_2$ pairs of sequences are either classified as matched or unmatched, the test stops to make a decision. Otherwise, the test takes additional samples. When the test stops, if $\hatcalA=\emptyset$, the decision of no match is made; otherwise, the set $\hatcalA$ is output as the indices of all matched pairs.

As we shall show, with the slight freedom to determine the stopping time, the sequential test achieves significantly better performance. Specifically, this freedom helps identify matched and unmatched pairs of sequences with high probability, as reflected by steps 2-10 that determine the stopping time. The asymptotic intuition of the sequential test is similar to the fixed-length test and thus omitted. Below, we explicitly bound the exponential decay rates of all three error probabilities of the sequential test.

Recall the definitions of $g(\cdot)$ in \eqref{def:g}, $E_\calA^\rms(\cdot)$ in \eqref{def:ea:rms} and $E_{\calA}^{\rms,\rmc}(\cdot)$ in \eqref{def:e2a:rms}, $E_{\rm{fa}}(\cdot)$ in \eqref{def:e:fa}, and $E_{\rm{fa}}^\rmc(\cdot)$ in \eqref{def:e2:fa}.
\begin{theorem}
\label{theorem:seq:u}
The sequential test in Algorithm \ref{low_com:seqtest:unknown} has bounded expected stopping time and is exponentially consistent under mild conditions. Fix any positive real numbers $(\lambda_1,\lambda_2)\in\bbR_+^2$ such that $\lambda_1\leq \lambda_2$.
\begin{itemize}
\item For discrete observed sequences, $f(\cdot)$ is the GJS scoring function in \eqref{gjscompute}. 
\begin{itemize}
\item For any $(P^{M_1},Q^{M_2})\in\calP_\emptyset$, when $N$ satisfies $\frac{M_1M_2\exp(-(N-1)E_{\rm{fa}}(\lambda_2,P^{M_1},Q^{M_2})-g(N-1))}{1-\exp(-(E_{\rm{fa}}(\lambda_2,P^{M_1},Q^{M_2})-g(N-1)))}\le 1$, the expected stopping time is upper bounded by $N$. Furthermore, the false alarm exponent satisfies
\begin{align}
\liminf_{N\to\infty}-\frac{1}{N}\log\eta(\Phi_\rms^\rmu|P^{M_1},Q^{M_2})&\geq E_{\rm{fa}}(\lambda_1,P^{M_1},Q^{M_2}).
\end{align}

\item Fix any $\calA\in\calM$ and tuple of distributions $(P^{M_1},Q^{M_2})\in\calP_\calA$. When $N$ satisfies $\frac{|\calA|\exp(-(N-1)(\lambda_1-g(N-1)))}{1-\exp(-(\lambda_1-g(N-1)))}+\frac{(M_1M_2-|\calA|)\exp(-(N-1)(E_\calA^\rms(\lambda_2,P^{M_1},Q^{M_2})-g(N-1)))}{1-\exp(-(E_\calA^\rms(\lambda_2,P^{M_1},Q^{M_2})-g(N-1)))}\leq 1$, the expected stopping time is upper bounded by $N$. Furthermore, the mismatch and false reject exponents satisfy
\begin{align}
\liminf_{N\to\infty}-\frac{1}{N}\log\bar{\theta}(\Phi_\rms^\rmu|P^{M_1},Q^{M_2})&\geq \min\{E_\calA^\rms(\lambda_1,P^{M_1},Q^{M_2}),\lambda_2\},\\
\liminf_{N\to\infty}-\frac{1}{N}\log\zeta(\Phi_\rms^\rmu|P^{M_1},Q^{M_2})&\geq \lambda_2.
\end{align}
\end{itemize}
\item For continuous observed sequences, $f(\cdot)$ is the MMD scoring function in \eqref{MMDcompute}. 
\begin{itemize}
\item For any $(P^{M_1},Q^{M_2})\in\calP_\emptyset$, when $N$ satisfies $\frac{M_1M_2\exp(-(N-1)E_{\rm{fa}}^\rmc(\lambda_2,P^{M_1},Q^{M_2})}{1-\exp(-E_{\rm{fa}}^\rmc(\lambda_2,P^{M_1},Q^{M_2}))}\le 1$, the expected stopping time is upper bounded by $N$. Furthermore, the false alarm exponent satisfies
\begin{align}
\liminf_{N\to\infty}-\frac{1}{N}\log\eta(\Phi_\rms^\rmu|P^{M_1},Q^{M_2})&\geq E_{\rm{fa}}^\rmc(\lambda_1,P^{M_1},Q^{M_2}).
\end{align}
\item Fix any $\calA\in\calM$ and tuple of distributions $(P^{M_1},Q^{M_2})\in\calP_\calA$. When $N$ is chosen such that
$\frac{|\calA|\exp(-(N-1)E(\lambda_1)}{1-\exp(-E(\lambda_1))}+\frac{(M_1M_2-|\calA|)\exp(-(N-1)E_{\calA}^{\rms,\rmc}(\lambda_2,P^{M_1},Q^{M_2}))}{1-\exp(-E_{\calA}^{\rms,\rmc}(\lambda_2,P^{M_1},Q^{M_2}))}\leq 1$, the expected stopping time is upper bounded by $N$. Furthermore, when $\lambda_1<\min_{(\bari,\barj)\notin\calA}\mmd^2(P_{\bari},Q_{\barj})$, the mismatch and false reject exponents satisfy
\begin{align}
\liminf_{N\to\infty}-\frac{1}{N}\log\bar{\theta}(\Phi_\rms^\rmu|P^{M_1},Q^{M_2})&\geq \min\{E_{\calA}^{\rms,\rmc}(\lambda_1,P^{M_1},Q^{M_2}),E(\lambda_2)\},\\
\liminf_{N\to\infty}-\frac{1}{N}\log\zeta(\Phi_\rms^\rmu|P^{M_1},Q^{M_2})&\geq E(\lambda_2).
\end{align}
\end{itemize}
\end{itemize}
\end{theorem}
The proof of Theorem \ref{theorem:seq:u} is provided in Section \ref{proof:seq:u}.

We first provide a proof sketch and explain why each exponent function appears. First consider the case where there is no matched pair of sequences. Our test in Algorithm \ref{low_com:seqtest:unknown} cannot stop if there exists a pair of sequences whose scoring function value lies between $\lambda_1$ and $\lambda_2$ while a false alarm event occurs if there exists a pair of sequences whose scoring function value is no greater than $\lambda_1$. For any integer $n$, the probability that the stopping time is greater than $n$ is upper bounded by the probability that there exists a pair of sequences whose scoring function value is no greater than $\lambda_2$, whose exponential decay rate is lower bounded by $E_{\rm{fa}}(\lambda_2,P^{M_1},Q^{M_2})$ for discrete observed sequences and by $E_{\rm{fa}}^\rmc(\lambda_2,P^{M_1},Q^{M_2})$ for continuous observed sequences, respectively. Thus, using the sum of geometric series, the expected stopping time is bounded as desired under mild conditions on $\lambda_2$ listed above. The exponential decay rate of the probability of the false alarm event is lower bounded by $E_{\rm{fa}}(\lambda_1,P^{M_1},Q^{M_2})$ for discrete observed sequences and by $E_{\rm{fa}}^\rmc(\lambda_1,P^{M_1},Q^{M_2})$ for continuous observed sequences, respectively. Next consider the case where there exists matched pairs of sequences. For any integer $n$, the probability that the stopping time is greater than $n$ is upper bounded by the probability that there exists a matched pair of sequences whose scoring function value is greater than $\lambda_1$ or there exists an unmatched pair of sequences whose scoring function value is no greater than $\lambda_2$. A lower bound on the exponential decay rates of these two events are $(\lambda_1,E_{\calA}^{\rms}(\lambda_2,P^{M_1},Q^{M_2}))$ for discrete observed sequences and $(E(\lambda_1),E_{\calA}^{\rms,\rmc}(\lambda_2,P^{M_1},Q^{M_2}))$ for continuous observed sequences, leading to the constraint of $(\lambda_1,\lambda_2)$ to ensure bounded expected stopping time. Furthermore, a mismatch event occurs if there exists a matched pair of sequences whose scoring function value is greater than $\lambda_2$ or there exists an unmatched pair of sequences whose scoring function value is no greater than $\lambda_1$, leading to exponents $(E_\calA^\rms(\lambda_1,P^{M_1},Q^{M_2}),\lambda_2)$ for discrete observed sequences and exponents $(E_{\calA}^{\rms,\rmc}(\lambda_1,P^{M_1},Q^{M_2}),E(\lambda_2))$ for continuous observed sequences. Finally, a false reject event occurs if all matched pairs of sequences have scoring function greater than $\lambda_2$, leading to an exponential decay rate of $\lambda_2$ for discrete observed sequences and $E(\lambda_2)$ for continuous observed sequences.

In contrast to Theorem \ref{theorem:seq}, when the number of matches is unknown, our sequential tests does not satisfy the expected stopping time universality constraint. Instead, Theorem \ref{theorem:seq:u} provides a sufficient condition for our sequential test to have bounded expected stopping time, which is also the condition for our sequential test to have strictly positive exponents. Specifically, when there is no match, asymptotically as $N\to\infty$, for any $(P^{M_1},Q^{M_2})\in\calP_\emptyset$, the expected stopping time is bounded when $0<\lambda_2<\min_{(i,j)\in[M_1]\times[M_2]}\mathrm{GJS}(P_i,Q_j,\alpha,\beta)$ for discrete observed sequences and when $0<\lambda_2<\min_{(i,j)\in[M_1]\times[M_2]}\mmd^2(P_i,Q_j)$
for continuous observed sequences. In this case, $\lambda_1$ is unconstrained and can be chosen as small as possible, leading to maximal false alarm exponents $(E_{\rm{fa}}(0,P^{M_1},Q^{M_2}),E_{\rm{fa}}^\rmc(0,P^{M_1},Q^{M_2}))$ for discrete and continuous observed sequences, respectively. Next consider the case with matched pairs of sequences whose indices are denoted by a set $\calA\in\calM$. Fix any $(P^{M_1},Q^{M_2})\in\calP_\calA$. Asymptotically as $N\to\infty$, if $\lambda_1>0$ and $\lambda_2<\min_{(\bari,\barj)\notin\calA}\mathrm{GJS}(P_{\bari},Q_{\barj},\alpha,\beta)$, the expected stopping time is bounded for discrete sequences. By properly choosing parameters $(\lambda_1,\lambda_2)$ on the boundary of the above conditions, the maximal achievable mismatch and false reject exponents are $\min\{E_\calA^\rms(0,P^{M_1},Q^{M_2}),\min_{(\bari,\barj)\notin\calA}\mathrm{GJS}(P_{\bari},Q_{\barj},\alpha,\beta)\}$ and $\min_{(\bari,\barj)\notin\calA}\mathrm{GJS}(P_{\bari},Q_{\barj},\alpha,\beta)$, respectively. Analogously, for sufficiently large $N$, if $\lambda_1>0$ and $\lambda_2<\min_{(\bari,\barj)\notin\calA}\mmd^2(P_{\bari},Q_{\barj})$, the expected stopping time is bounded for continuous observed sequences and the maximal achievable mismatch and false reject exponents are $\min\{E_\calA^{\rms\rmc}(0,P^{M_1},Q^{M_2}),E(\min_{(\bari,\barj)\notin\calA}\mmd^2(P_{\bari},Q_{\barj}))\}$ and $E(\min_{(\bari,\barj)\notin\calA}\mmd^2(P_{\bari},Q_{\barj}))$, respectively.

\begin{figure}[tb]
\centering
\begin{tabular}{cc}
\includegraphics[width=.5\columnwidth]{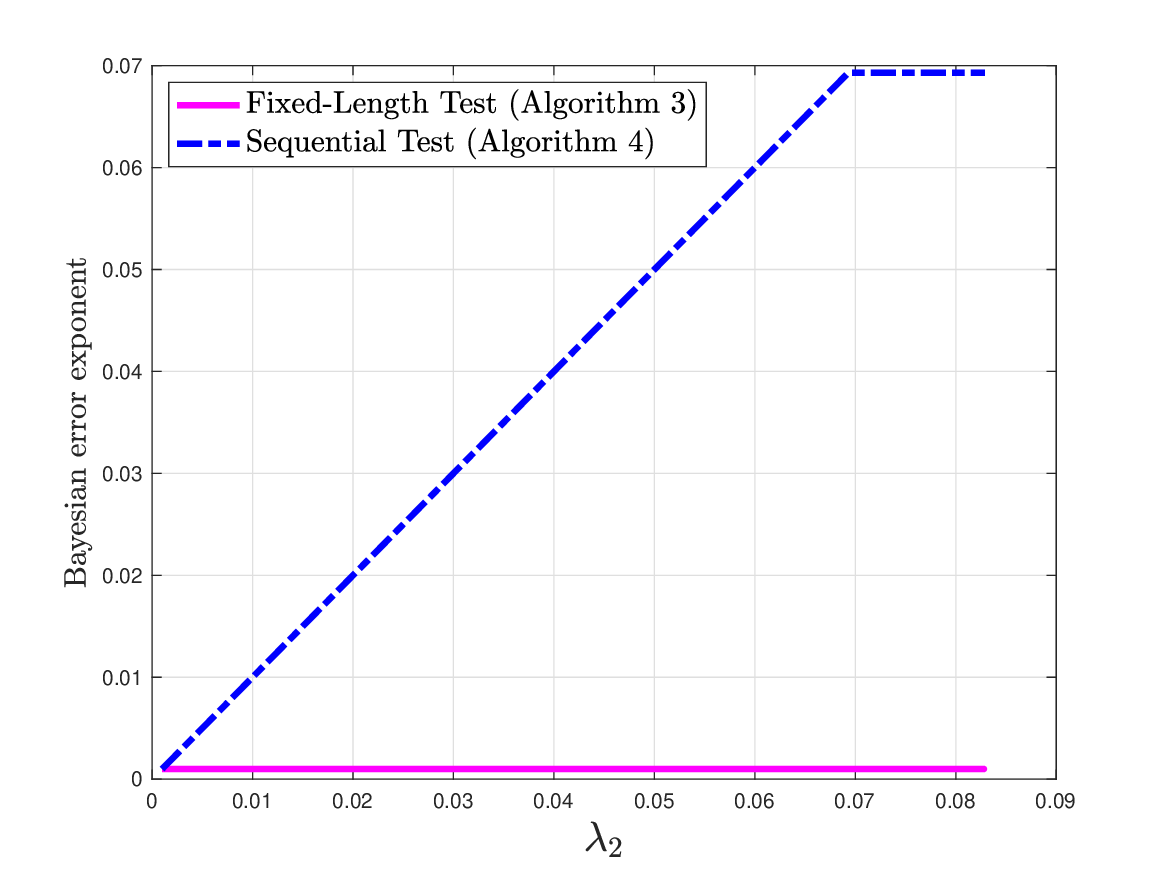} &\includegraphics[width=.5\columnwidth]{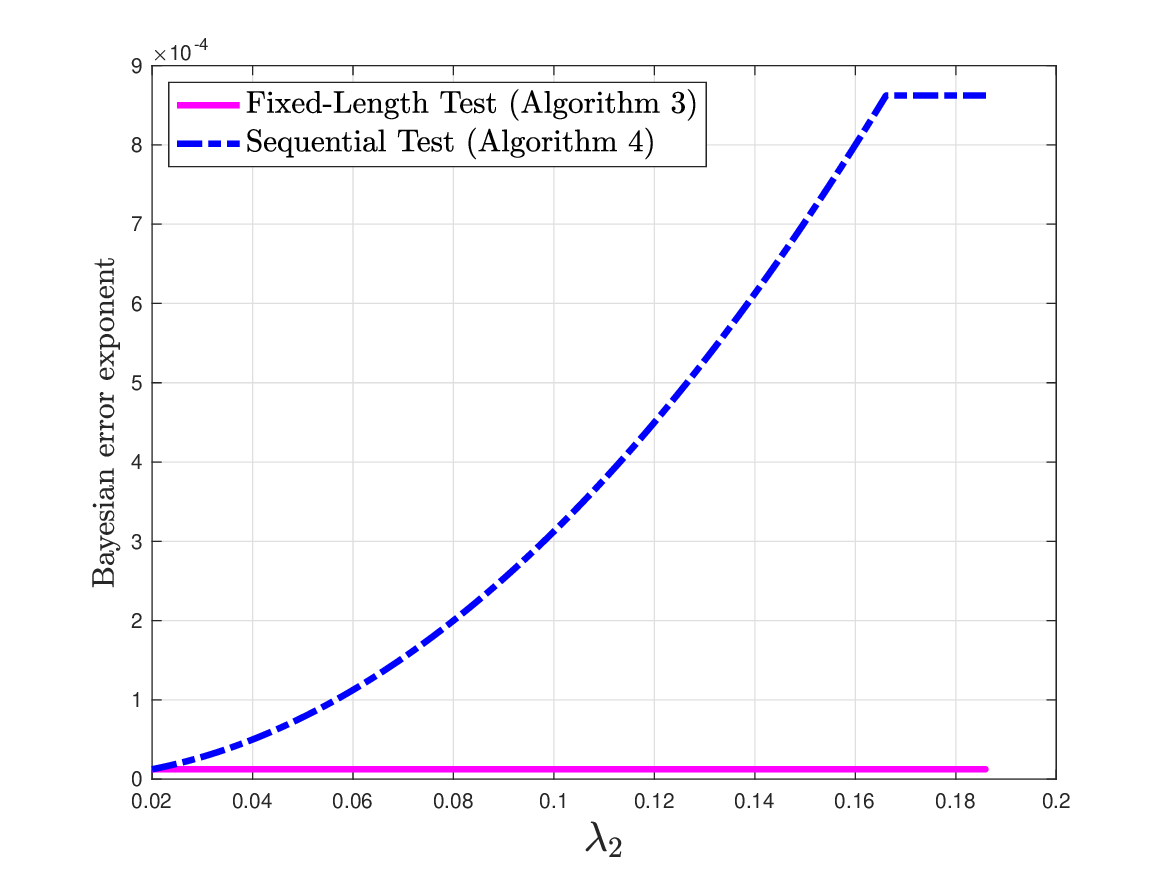} \\
{(a) Discrete observed sequences}& {(b) Continuous observed sequences} 
\end{tabular}
\caption{Plot of the achievable Bayesian error exponents of the fixed-length test in Algorithm \ref{low_com:fltest:unknown} and the sequential test in Algorithm \ref{low_com:seqtest:unknown} as a function of the test design parameter $\lambda_2$. The left figure is for discrete observed sequences with $\lambda_1=0.001$
$\alpha=1$, $\beta=2$, $M_1=3$, $M_2=K=2$, $P^{M_1}=\mathrm{Bern}(0.1,0.3,0.6)$ and $Q^{M_2}=\mathrm{Bern}(0.1,0.3)$ while the right figure is for continuous observed sequences with $\lambda_1=0.02$, $\alpha=\beta=5$, $M_1=M_2=K=3$, $P^{M_1}=Q^{M_2}=(\mathrm{Exp}(1),\rmU(10,30),\calN(-10,25))$.}
\label{bene_uk}
\end{figure}

\begin{figure}[tb]
\centering
\begin{tabular}{cc}
\includegraphics[width=.5\columnwidth]{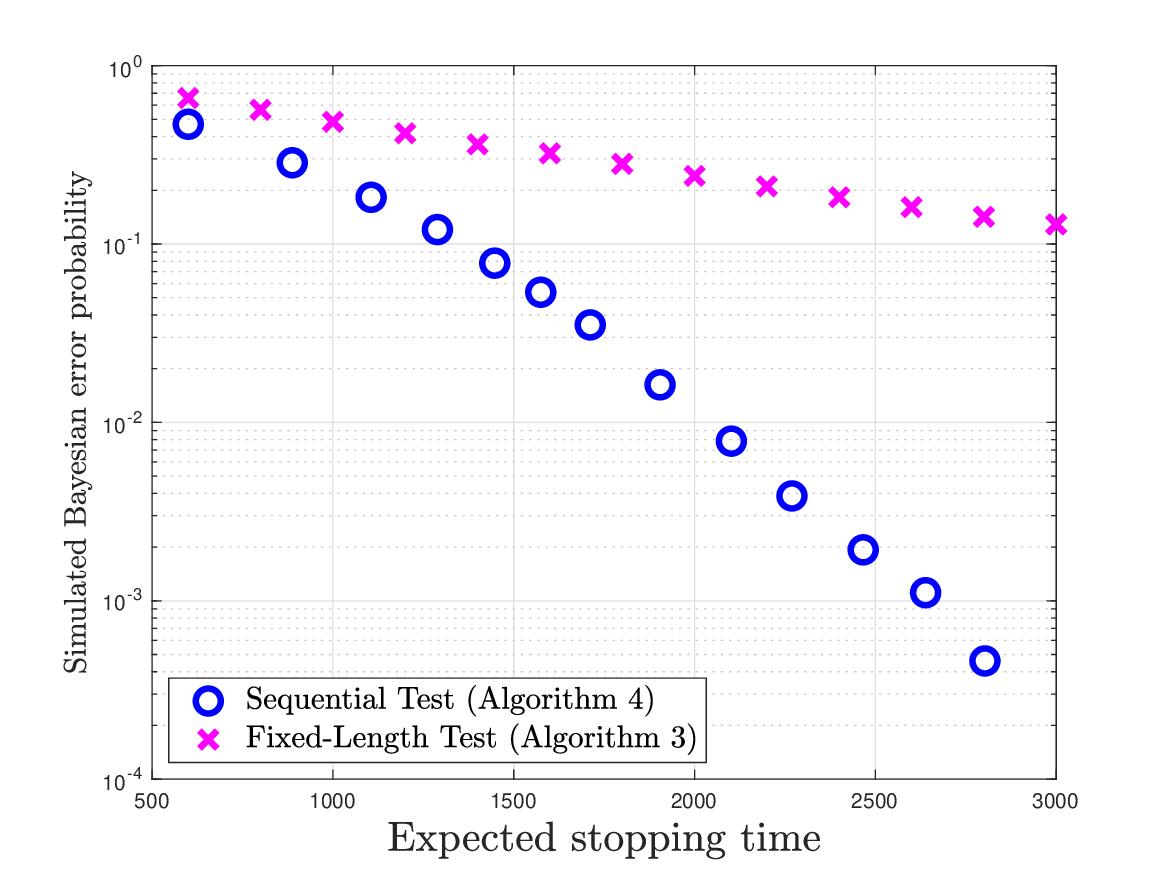}&\includegraphics[width=.5\columnwidth]{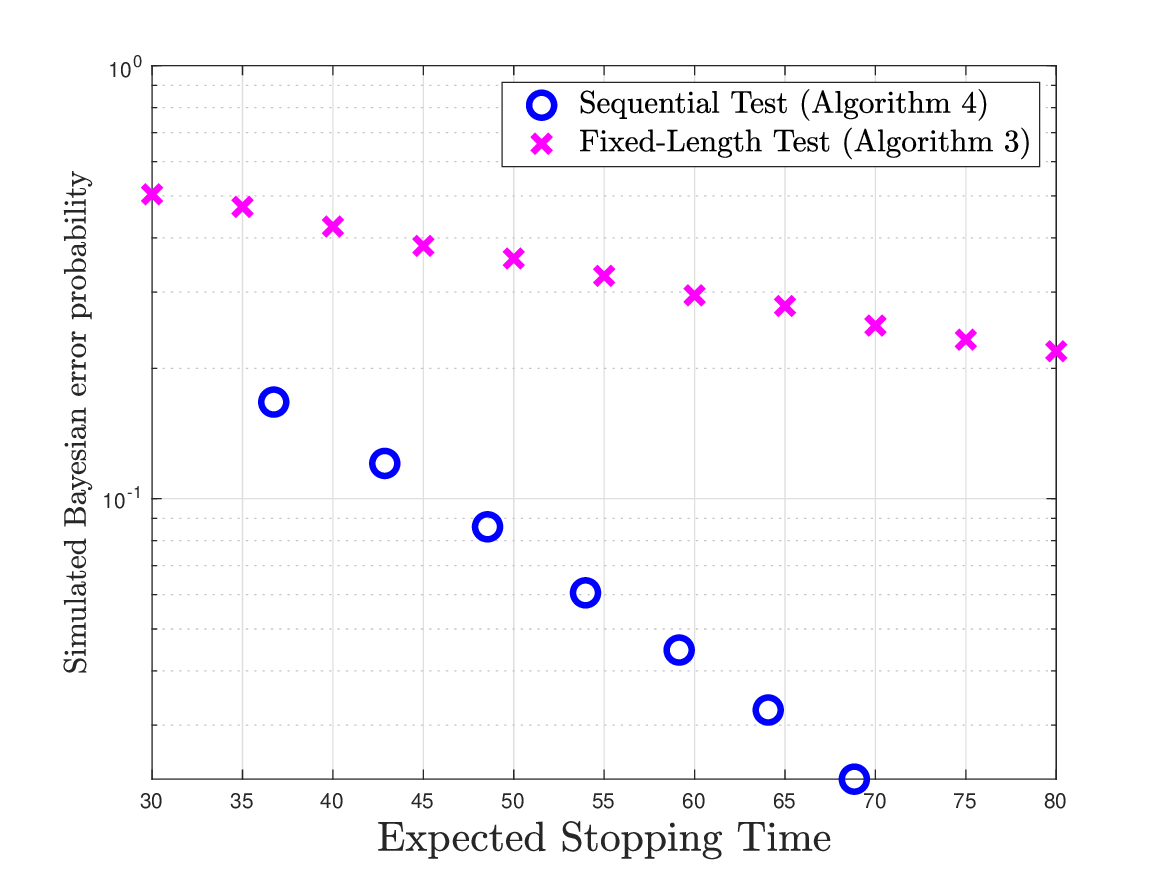}\\
{(a) Discrete Observed Sequences} & {(b) Continuous Observed Sequences} 
\end{tabular}
\caption{Plot of the simulated Bayesian error probabilities as a function of expected stopping time for the low complexity fixed-length test in Algorithm \ref{low_com:fltest:unknown} and the low-complexity sequential test in Algorithm \ref{low_com:seqtest:unknown} with $\lambda=\lambda_1=0.01$ when $\alpha=1$, $\beta=2$. The left figure is for $\lambda_2=0.03$, $M_1=3$, $M_2=K=2$, $P^{M_1}=\mathrm{Bern}(0.1,0.2,0.15)$ and $Q^{M_2}=\mathrm{Bern}(0.1,0.2)$ while the right figure is for $\lambda_2=0.05$, $M_1=M_2=3$, $K=2$, $P^{M_1}=(\mathrm{Exp}(2),\rmU(10),\calN(-5,1))$ and $Q^{M_2}=(\mathrm{Exp}(2),\rmU(10),\calN(5,4))$.
}
\label{bseq_d_uk}
\end{figure}

Comparing Theorems \ref{theorem:fl:u} and \ref{theorem:seq:u}, we reveal the benefit of sequentiality. Set $\lambda$ in Theorem \ref{theorem:fl:u} as $\lambda_1$ in Theorem \ref{theorem:seq:u}. The false alarm exponents of the fixed-length and sequential tests are the same. However, the mismatch and false reject exponents of the sequential test in Algorithm \ref{low_com:seqtest:unknown} are larger than those of the fixed-length test in Algorithm \ref{low_com:fltest:unknown}. This is because $\lambda_2\geq \lambda_1$. If one digs deeper, the benefit of sequentiality results from the freedom of the sequential test to determine the stopping time and thus avoid unreliable decisions. In Fig. \ref{bene_uk}, we plot the Bayesian exponents in Theorems \ref{theorem:fl:u} and \ref{theorem:seq:u} for both discrete and continuous observed sequences and show that the benefit of sequentiality can be strict. Furthermore, in Fig. \ref{bseq_d_uk}, we plot the simulated Bayesian error probabilities as a function of expected stopping time for the low complexity fixed-length test in Algorithm \ref{low_com:fltest:unknown} and the low-complexity sequential test in Algorithm \ref{low_com:seqtest:unknown} with $\lambda=0.01$. As observed, the sequential test achieves smaller Bayesian error probability under the same expected stopping time, validating the superiority of the sequential test even in the non-asymptotic case with finite expected stopping times.

\begin{figure}[tb]
\centering
\begin{tabular}{cc}
\includegraphics[width=.5\columnwidth]{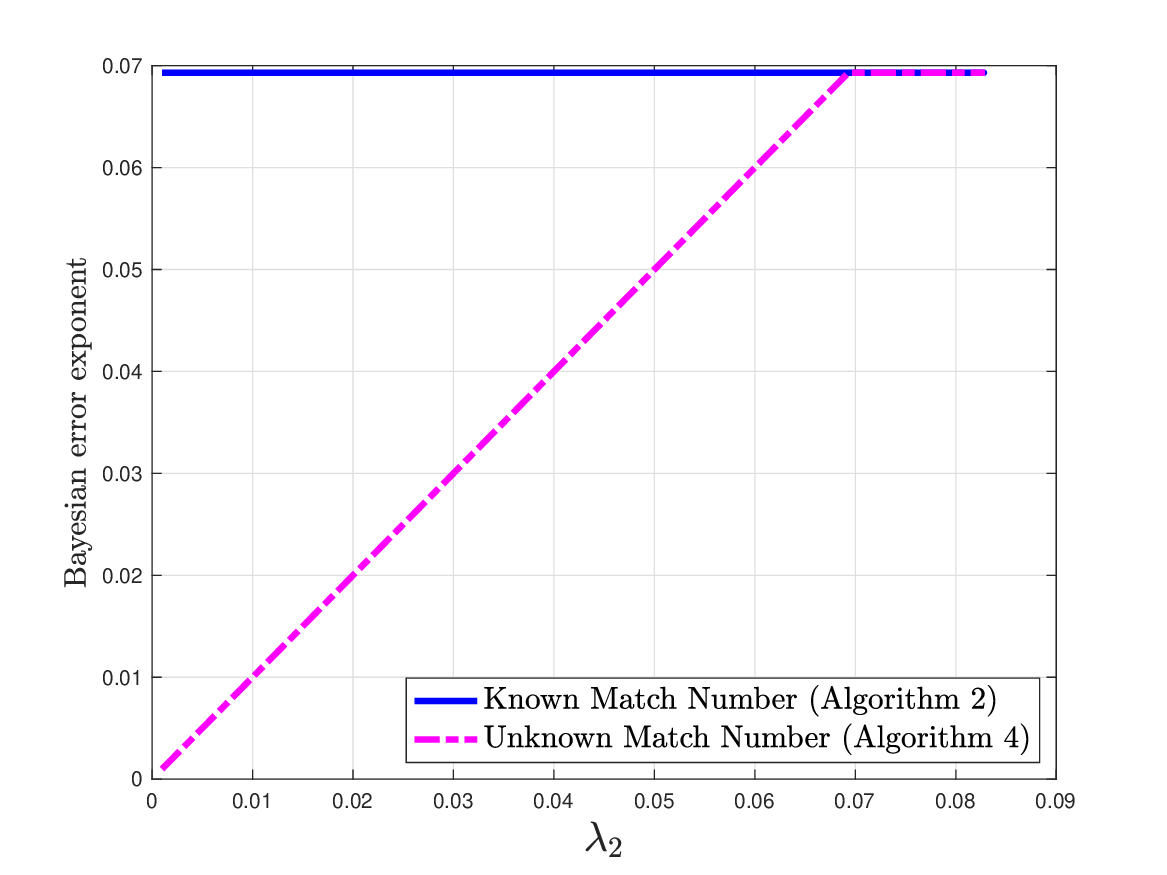} &\includegraphics[width=.5\columnwidth]{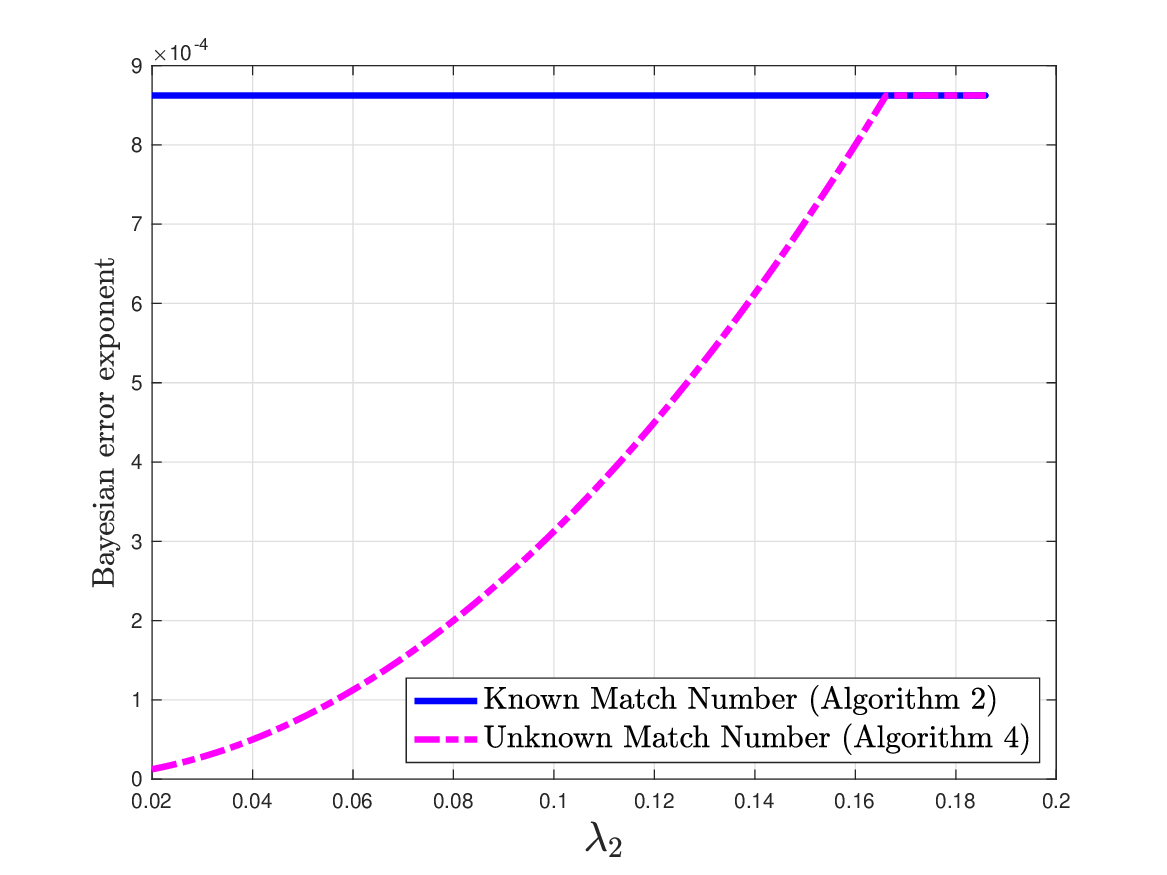}  \\
{(a) Discrete observed sequences}& {(b) Continuous observed sequences} 
\end{tabular}
\caption{Plot of the achievable Bayesian error exponent of the sequential test in Algorithm \ref{low_com:seqtest} and the Bayesian exponent of the sequential test in Algorithm \ref{low_com:seqtest:unknown} as a function of the test design parameter $\lambda_2$. The left figure is for discrete observed sequences with $\lambda=\lambda_1=0.001$
$\alpha=1$, $\beta=2$, $M_1=3$, $M_2=K=2$, $P^{M_1}=\mathrm{Bern}(0.1,0.3,0.6)$ and $Q^{M_2}=\mathrm{Bern}(0.1,0.3)$ while the right figure is for continuous observed sequences with $\lambda=\lambda_1=0.02$, $\alpha=\beta=5$, $M_1=M_2=K=3$, $P^{M_1}=Q^{M_2}=(\mathrm{Exp}(1),\rmU(10,30),\calN(-10,25))$. As observed, when the number of matches is unknown, the achievable Bayesian exponent of the sequential test in Algorithm \ref{low_com:seqtest:unknown} can be strictly smaller than the sequential test in Algorithm \ref{low_com:seqtest} that knows the number of matches.}
\label{penal_uk}
\end{figure}

\begin{figure}[tb]
\centering
\begin{tabular}{cc}
\includegraphics[width=.5\columnwidth]{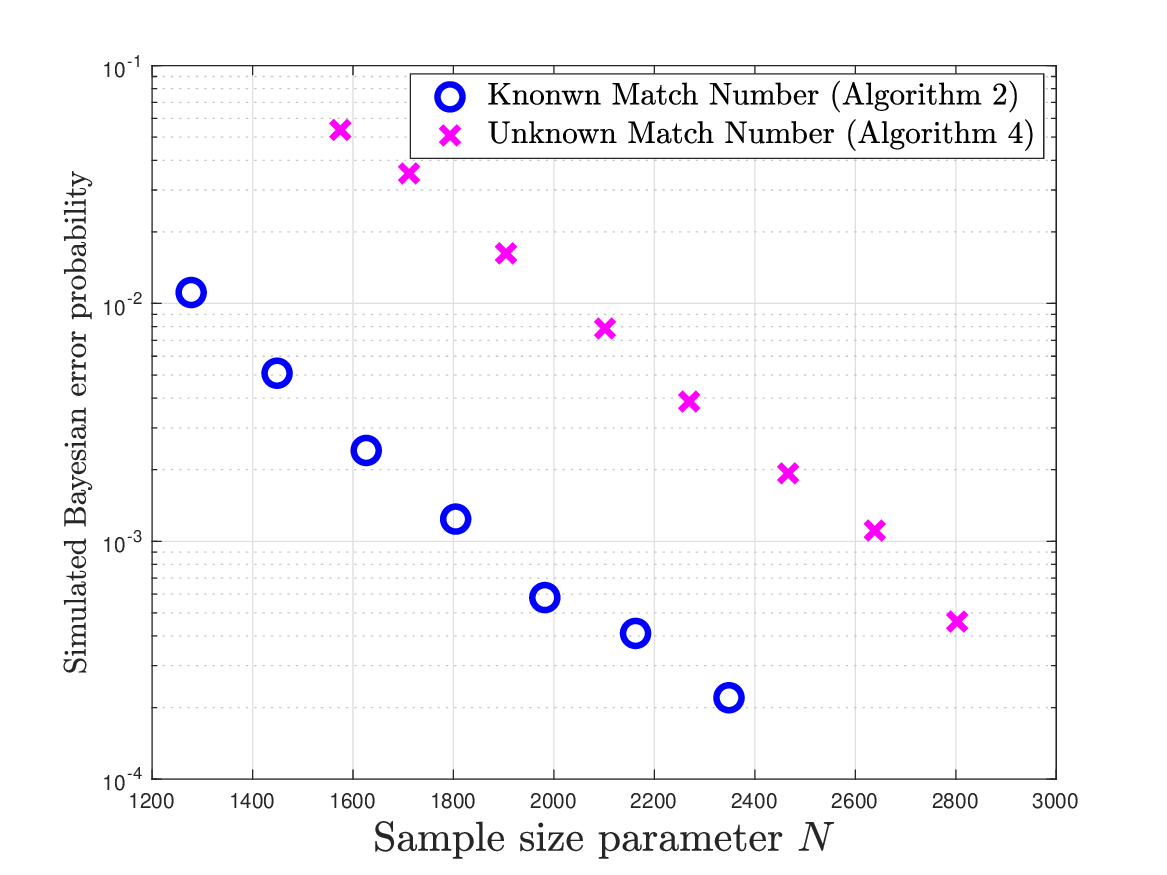}&\includegraphics[width=.5\columnwidth]{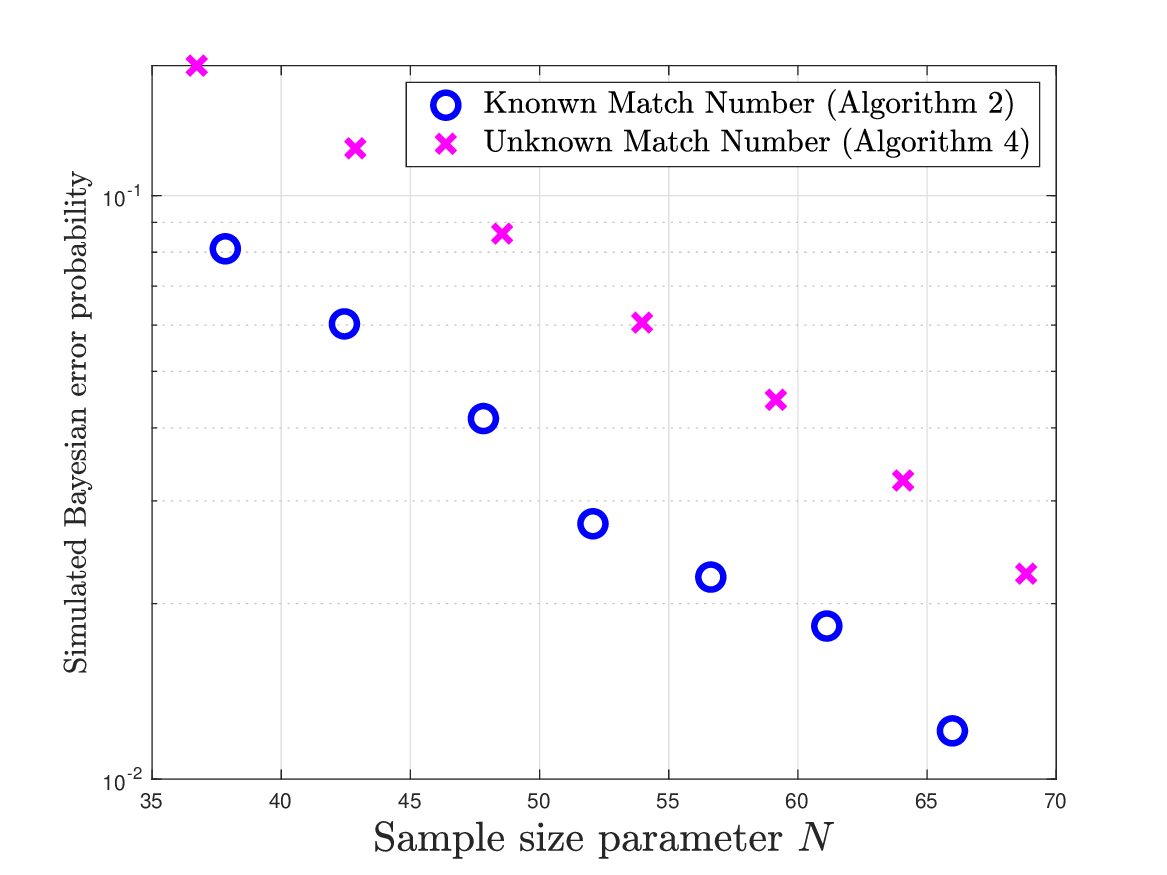}\\
{(a) Discrete Observed Sequences} & {(b) Continuous Observed Sequences} 
\end{tabular}
\caption{Plot of the simulated Bayesian error probabilities as a function of expected stopping time for the low complexity sequential test with known number of matches in Algorithm \ref{low_com:seqtest} and the low-complexity sequential test with unknown number of matches in Algorithm \ref{low_com:seqtest:unknown} with $\lambda=\lambda_1=0.01$ when $\alpha=1$, $\beta=2$. The left figure is for $\lambda_2=0.03$, $M_1=3$, $M_2=K=2$, $P^{M_1}=\mathrm{Bern}(0.1,0.2,0.15)$ and $Q^{M_2}=\mathrm{Bern}(0.1,0.2)$ while the right figure is for $\lambda_2=0.05$, $M_1=M_2=3$, $K=2$, $P^{M_1}=(\mathrm{Exp}(2),\rmU(10),\calN(-5,1))$ and $Q^{M_2}=(\mathrm{Exp}(2),\rmU(10),\calN(5,4))$. As observed, there is a penalty of not knowing the number of matches: the sequential test in Algorithm \ref{low_com:seqtest:unknown} that does not known the number of matches has larger Bayesian error probability than the sequential test in Algorithm \ref{low_com:seqtest} that knows the number of matches.
}
\label{penalty_seq}
\end{figure}

Comparing Theorems \ref{theorem:seq} and \ref{theorem:seq:u}, we find the penalty of not knowing the number of matches when there are matched pair of sequences. For discrete sequences, the achievable Bayesian exponent of the sequential test in Algorithm \ref{low_com:seqtest:unknown} is $\min\{E_\calA^\rms(\lambda_1,P^{M_1},Q^{M_2}),\lambda_2\}$, which is less than the mismatch exponent $E_\calA^\rms(\lambda_1,P^{M_1},Q^{M_2})$ of the sequential test in Algorithm \ref{low_com:seqtest} with knowledge of the number of matches. The same is also true for continuous observed sequences. In Fig. \ref{penal_uk}, we show that the penalty can be strict via a numerical example by showing that when the number of matches is unknown, the achievable Bayesian exponent of the sequential test in Algorithm \ref{low_com:seqtest:unknown} can be strictly smaller than the sequential test in Algorithm \ref{low_com:seqtest} that knows the number of matches. Furthermore, in Fig. \ref{penalty_seq}, we plot the simulated Bayesian error probabilities of the sequential tests in Algorithms \ref{low_com:seqtest} and \ref{low_com:seqtest:unknown} to illustrate the penalty of not knowing the number of matches. As observed, the sequential test in Algorithm \ref{low_com:seqtest:unknown} that does not known the number of matches has larger Bayesian error probability than the sequential test in Algorithm \ref{low_com:seqtest} that knows the number of matches, demonstrating the penalty of not knowing the nubmer of matches.

\section{Proofs for Known Number of Matches}
\label{sec:proof:known}
\subsection{Proof of Theorem \ref{theorem:fl} (Fixed-Length Test)}
\label{proof:theorem:fl}
For the low-complexity test $\Phi_\rmf$ in Algorithm \ref{low_com:fltest}, a mismatch error event occurs if there exists a matched pair of sequences that has a larger scoring function than that of an unmatched pair of sequences. Fix any $\calA\in\calM_K$ and $(P^{M_1},Q^{M_2})\in\calP_\calA$. The mismatch probability satisfies
\begin{align}
\theta_\calA(\Phi_\rmf|P^{M_1},Q^{M_2})
&=\bbP_\calA\Big\{\exists~(i,j)\in\calA\mathrm{~and~}(\bari,\barj)\notin\calA:~f(X_i^{\xi_N},Y_j^{\chi_N})\geq f(X_{\bari}^{\xi_N},Y_{\barj}^{\chi_N})\Big\}\\
&\leq \sum_{(i,j)\in\calA}\sum_{(\bari,\barj)\notin\calA}\bbP_\calA\Big\{f(X_i^{\xi_N},Y_j^{\chi_N})\geq f(X_{\bari}^{\xi_N},Y_{\barj}^{\chi_N})\Big\}\label{fl:ach1}.
\end{align}
Fix any $(i,j)\in\calA$ and $(\bari,\barj)\notin\calA$. Since $(P^{M_1},Q^{M_2})\in\calP_\calA$, it follows that $P_i=Q_j$. 

When $\calX$ is finite, $f(\cdot)$ is the GJS scoring function in \eqref{gjscompute}. Using the method of types~\cite{csiszar1998mt}, it follows that each probability term in \eqref{fl:ach1} satisfies
\begin{align}
\nn&\bbP_\calA\Big\{f(X_i^{\xi_N},Y_j^{\chi_N})\geq f(X_{\bari}^{\xi_N},Y_{\barj}^{\chi_N})\Big\}\\*
&=\bbP_\calA\Big\{\gjs(X_i^{\xi_N},Y_j^{\chi_N})\geq \gjs(X_{\bari}^{\xi_N},Y_{\barj}^{\chi_N})\Big\}\\
&=\sum_{\substack{(\Omega_1,\Omega_2)\in(\calP^{\xi_N}(\calX))^2\\
(\Psi_1,\Psi_2)\in(\calP^{\chi_N}(\calX))^2}}
\sum_{\substack{
(x_i^{\xi_N},x_{\bari}^{\xi_N})\in\calT_{\xi_N}(\Omega_1)\times\calT_{\xi_N}(\Omega_2)\\
(y_j^{\chi_N},y_{\barj}^{\chi_N})\in\calT_{\chi_N}(\Psi_1)\times\calT_{\chi_N}(\Psi_2):\\~\gjs(x_i^{\xi_N},y_j^{\chi_N})\geq \gjs(x_{\bari}^{\xi_N},y_{\barj}^{\chi_N})}}
P_i^{\xi_N}(x_i^{\xi_N})P_i^{\chi_N}(y_j^{\chi_N})P_{\bari}^{\xi_N}(x_{\bari}^{\xi_N})Q_{\barj}^{\chi_N}(y_{\barj}^{\chi_N})\\
&=\sum_{\substack{(\Omega_1,\Omega_2,\Psi_1,\Psi_2)\in(\calP^{\xi_N}(\calX))^2\times (\calP^{\chi_N}(\calX))^2:\\\gjs(\Omega_1,\Psi_1,\xi_N,\chi_N)\geq \gjs(\Omega_2,\Psi_2,\xi_N,\chi_N)
}}P_i^{\xi_N}(\calT_{\xi_N}(Q_1))P_i^{\chi_N}(\calT_{\chi_N}(Q_2))P_{\bari}^{\xi_N}(\calT_{\xi_N}(\barQ_1))P_{\barj}^{\chi_N}(\calT_{\chi_N}(\barQ_2))
\label{use:defgjs}\\
&\leq \sum_{\substack{(\Omega_1,\Omega_2,\Psi_1,\Psi_2)\in(\calP^{\xi_N}(\calX))^2\times (\calP^{\chi_N}(\calX))^2:\\\gjs(\Omega_1,\Psi_1,\xi_N,\chi_N)\geq \gjs(\Omega_2,\Psi_2,\xi_N,\chi_N)
}}\exp\bigg(-\xi_N D(\Omega_1\|P_i)-\xi_N D(\Omega_2\|P_{\bari})-\chi_ND(\Psi_1\|P_i)-\chi_ND(\Psi_2\|Q_{\barj})\bigg)\label{probtypeclass}\\
\nn&\leq (\xi_N+1)^{2|\calX|}(\chi_N+1)^{2|\calX|}\\*
&\quad\times\exp\bigg(-\min_{\substack{(\Omega_1,\Omega_2,\Psi_1,\Psi_2)\in(\calP^{\xi_N}(\calX))^2\times (\calP^{\chi_N}(\calX))^2:\\\gjs(\Omega_1,\Psi_1,\xi_N,\chi_N)\geq \gjs(\Omega_2,\Psi_2,\xi_N,\chi_N)
}}\Big(\xi_N D(\Omega_1\|P_i)+\xi_N D(\Omega_2\|P_{\bari})+\chi_ND(\Psi_1\|P_i)+\chi_ND(\Psi_2\|Q_{\barj})\Big)\bigg)\label{numtypes},
\end{align}
where \eqref{use:defgjs} follows from the definition of GJS scoring function in \eqref{gjscompute}, \eqref{probtypeclass} follows from the upper bound for the probability of a type class~\cite[Theorem 11.1.4]{cover2012elements}, and \eqref{numtypes} follows from the upper bound on the number of types~\cite[Theorem 11.1.1]{cover2012elements}

Recall that $\xi_N=\lceil\alpha N\rceil$ and $\chi_N=\lceil\beta N\rceil$. Using the definition of the exponent function $E_\calA^\rmf(P^{M_1},Q^{M_2})$ in \eqref{def:fl:exponent}, it follows from \eqref{fl:ach1} and \eqref{numtypes} that the mismatch exponent satisfies
\begin{align}
\liminf_{N\to\infty}-\frac{1}{N}\log \theta_\calA(\Phi_\rmf|P^{M_1},Q^{M_2})
&\geq E_\calA^\rmf(P^{M_1},Q^{M_2}).
\end{align}

When $\calX$ is continuous,  similarly to \cite[Eq. (108), (121) and (124)]{zhou2025csm} with $(i^*,j^*)$ replaced by $(\bari,\barj)$, it follows from McDiarmid's inequality~\cite{mcdiarmid1989method} that each probability term in \eqref{fl:ach1} satisfies
\begin{align}
\bbP_\calA\Big\{\mmd^2(X_i^{\xi_N},Y_j^{\chi_N})\geq \mmd^2(X_{\bari}^{\xi_N},Y_{\barj}^{\chi_N})\Big\}
\leq \exp\bigg(-\frac{N\min\{\alpha,\beta\}(\mmd^2(P_{\bari},Q_{\barj}))^2}{128\Theta^2}\bigg).
\end{align}
Thus, the mismatch exponent satisfies
\begin{align}
\liminf_{N\to\infty}-\frac{1}{N}\log \theta_\calA(\Phi_\rmf|P^{M_1},Q^{M_2})
&\geq \min_{(\bari,\barj)\notin\calA}\frac{\min\{\alpha,\beta\}(\mmd^2(P_{\bari},Q_{\barj}))^2}{128\Theta^2}\\
&=E_{\calA}^{\rmf,\rmc}(P^{M_1},Q^{M_2}),\label{usee2afl}
\end{align}
where \eqref{usee2afl} follows from the definition of $E_{\calA}^{\rmf,\rmc}(\cdot)$ in \eqref{def:ea2:fl}.

The proof of Theorem \ref{theorem:fl} is now completed.

\subsection{Proof of Theorem \ref{theorem:seq} (Sequential Test)}
\label{proof:theorem:seq}
The analysis is two fold: the expected stopping time and the mismatch exponent. Fix any $\calA\in\calM_K$ and $(P^{M_1},Q^{M_2})\in\calP_\calA$. The expected stopping time of the sequential test in Algorithm \ref{low_com:seqtest} satisfies
\begin{align}
\bbE_{\bbP_\calA}[\tau]
&=N-1+\sum_{n\in\bbN_{N-1}}\bbP_{\calA}\{\tau>n\}\label{expected:st:step1}.
\end{align}
Each probability term in \eqref{expected:st:step1}, which quantifies the probability that the random stopping time $\tau$ is greater than $n$, satisfies
\begin{align}
\bbP_\calA\{\tau>n\}
&\leq \bbP_\calA\big\{\exists~(i,j)\in\calA:~f(X_i^{\xi_n},Y_j^{\chi_n})>\lambda\big\}\\
&\leq \sum_{(i,j)\in\calA}\bbP_\calA\big\{f(X_i^{\xi_n},Y_j^{\chi_n})>\lambda\big\}\label{seq:step1}.
\end{align}

Furthermore, a mismatch error event occurs if there exists an unmatched pair of sequences whose scoring function has value no greater than $\lambda$. Thus, the mismatch probability satisfies
\begin{align}
\theta(\Phi_\rms|P^{M_1},Q^{M_2})
&=\bbP_\calA\big\{\exists~(\bari,\barj)\notin\calA:~f(X_{\bari}^{\xi_\tau},Y_{\barj}^{\chi_\tau})\leq \lambda\big\}\\
&=\sum_{n\in\bbN_{N-1}}\bbP_\calA\{\tau=n\}\bbP_\calA\big\{\exists~(\bari,\barj)\notin\calA:~f(X_{\bari}^{\xi_n},Y_{\barj}^{\chi_n})\leq \lambda\big\}\\
&\leq \sum_{n\in\bbN_{N-1}}\bbP_\calA\big\{\exists~(\bari,\barj)\notin\calA:~f(X_{\bari}^{\xi_n},Y_{\barj}^{\chi_n})\leq \lambda\big\}\\
&\leq \sum_{n\in\bbN_{N-1}}\sum_{(\bari,\barj)\notin\calA}\bbP_\calA\big\{f(X_{\bari}^{\xi_n},Y_{\barj}^{\chi_n})\leq \lambda\big\}
\label{mismatch:seq:step1}.
\end{align}

In what follows, we upper bound the expected stopping time in \eqref{expected:st:step1} and the mismatch probability in \eqref{mismatch:seq:step1} for discrete and continuous observed sequences, respectively. 

\subsubsection{Discrete Observed Sequences}
Fix any $(i,j)\in\calA$. We first bound each probability term in \eqref{seq:step1}. When the observed sequences are discrete, $f(\cdot)$ is the GJS scoring function in \eqref{gjscompute}.  Analogously to \eqref{numtypes}, it follows from the method of types~\cite{csiszar1998mt} that
\begin{align}
\nn&\bbP_\calA\big\{f(X_i^{\xi_n},Y_j^{\chi_n})>\lambda\big\}\\*
&\leq \bbP_\calA\big\{\gjs(X_i^{\xi_n},Y_j^{\chi_n})\geq \lambda\big\}\\
&=\sum_{(\Omega,\Psi)\in\calP^{\xi_n}(\calX)\times\calP^{\chi_n}(\calX)}\sum_{\substack{(x_i^{\xi_n},y_j^{\chi_n})\in\calT_{\xi_n}(\Omega)\times\calT_{\chi_n}(\Psi):\\\gjs(x_i^{\xi_n},y_j^{\chi_n})\geq \lambda}}P_i^{\xi_n}(x_i^{\xi_n})P_i^{\chi_n}(y_j^{\chi_n})\\
&=\sum_{\substack{(\Omega,\Psi)\in\calP^{\xi_n}(\calX)\times\calP^{\chi_n}(\calX):\\
\gjs(\Omega,\Psi,\xi_n,\chi_n)\geq n\lambda}}P_i^{\xi_n}(\calT_{\xi_n}(\Omega))P_i^{\chi_n}(\calT_{\chi_n}(\Psi))\label{usegjs:seq}\\
&\leq \sum_{\substack{(\Omega,\Psi)\in\calP^{\xi_n}(\calX)\times\calP^{\chi_n}(\calX):\\
\gjs(\Omega,\Psi,\xi_n,\chi_n)\geq n\lambda}}\exp\Big(-n\big(\alpha D(\Omega\|P_i)+\beta D(\Psi\|P_i)\big)\Big)\label{usedefs:seq}\\
&\leq \sum_{\substack{(\Omega,\Psi)\in\calP(\calX)\times\calP^{\chi_n}(\calX):\\
\gjs(\Omega,\Psi,\xi_n,\chi_n)\geq n\lambda}}\exp\Big(-n\min_{\substack{(\Omega,\Psi)\in(\calP(\calX))^2}}\big(\alpha D(\Omega\|P_i)+\beta D(\Psi\|P_i)\big)\Big)\label{enlargeprob}\\
&\leq \sum_{\substack{(\Omega,\Psi)\in\calP(\calX)\times\calP^{\chi_n}(\calX):\\
\gjs(\Omega,\Psi,\alpha,\beta)\geq \lambda}}\exp(-n\lambda)\label{usegjsalt:seq}\\
&\leq (\xi_n+1)^{|\calX|}(\chi_n+1)^{|\calX|}\exp(-n\lambda)\label{numtype:seq}\\
&\leq \exp(-n(\lambda-g(n)))\label{useg:seq},
\end{align}
where \eqref{usegjs:seq} uses the definition of GJS scoring function in \eqref{gjscompute}, \eqref{usedefs:seq} follows from the upper bound on the probability of a type class in \cite[Theorem 11.1.4]{cover2012elements} and the fact that $\xi_n\ge\alpha n$, $\chi_n\geq \beta n$, \eqref{enlargeprob} follows since $\exp(-a)$ decreases in its parameter $a\in\bbR_+$, \eqref{usegjsalt:seq} follows from the alternative form of GJS divergence in \eqref{gjs:variational} which implies that $\min_{\substack{(\Omega,\Psi)\in(\calP(\calX))^2}}\big(\alpha D(\Omega\|P_i)+\beta D(\Psi\|P_i)\big)=\gjs(\Omega,\Psi,\alpha,\beta)$, \eqref{numtype:seq} follows from the upper bound on the number of types in \cite[Theorem 11.1.1]{cover2012elements} and \eqref{useg:seq} follows from the definition of $g(n)$ in \eqref{def:g}.

Combining \eqref{seq:step1} and \eqref{useg:seq} and using the fact that $\calA\in\calM_K$, it follows that 
\begin{align}
\bbP_\calA\{\tau>n\}
&\leq \sum_{(i,j)\in\calA}\exp(-n(\lambda-g(n)))\\
&\leq K\exp(-n(\lambda-g(n)))\label{expected:st:step2}.
\end{align}
Thus, it follows from \eqref{expected:st:step1} and \eqref{expected:st:step2} that the stopping time satisfies
\begin{align}
\bbE_{\bbP_\calA}[\tau]
&=N-1+\sum_{n\in\bbN_{N-1}}\bbP_{\calA}\{\tau>n\}\\
&\leq N-1+\sum_{n\in\bbN_{N-1}}K\exp(-n(\lambda-g(n)))\\
&\leq N-1+\sum_{n\in\bbN_{N-1}}K\exp(-n(\lambda-g(N-1)))\label{decreaseg:seq}\\
&=N-1+\frac{K\exp(-(N-1)(\lambda-g(N-1)))}{1-\exp(-(\lambda-g(N-1)))}\label{sumgeo:seq},
\end{align}
where \eqref{decreaseg:seq} follows since $g(n)$ decreases in its parameter $n\in\bbN$, and \eqref{sumgeo:seq} follows from the sum of geometric series. Thus, when $N$ is sufficiently large such that $\frac{K\exp(-(N-1)(\lambda-g(N-1)))}{1-\exp(-(\lambda-g(N-1)))}\leq 1$, the expected stopping time satisfies $\bbE_{\bbP_\calA}[\tau]\leq N$.

We next analyze the mismatch probability. It follows from \eqref{mismatch:seq:step1} that the mismatch probability satisfies
\begin{align}
\theta(\Phi_\rms|P^{M_1},Q^{M_2})
&\leq \sum_{n\in\bbN_{N-1}}\sum_{(\bari,\barj)\notin\calA}\bbP_\calA\{\gjs(X_{\bari}^{\xi_n},Y_{\barj}^{\chi^n})\leq \lambda\}\label{mismatch:seq:step2}.
\end{align}
Fix $n\in\bbN_{N-1}$ and $(\bari,\barj)\notin\calA$. Analogously to steps leading to \eqref{usedefs:seq}, using the method of types, it follows that
\begin{align}
\bbP_\calA\big\{\gjs(X_{\bari}^{\xi_n},Y_{\barj}^{\chi_n})\leq \lambda\big\}
&\leq \sum_{\substack{(\Omega,\Psi)\in\calP^{\xi_n}(\calX)\times\calP^{\chi_n}(\calX):\\
\gjs(\Omega,\Psi,\xi_n,\chi_n)\leq n\lambda}}\exp\Big(-n\big(\alpha D(\Omega\|P_{\bari})+\beta D(\Psi\|Q_{\barj})\big)\Big)\\
&\leq  \sum_{\substack{(\Omega,\Psi)\in\calP^{\xi_n}(\calX)\times\calP^{\chi_n}(\calX):\\
\gjs(\Omega,\Psi,\xi_n,\chi_n)\leq n\lambda}}\exp\Big(-n\min_{{\substack{(\Omega,\Psi)\in(\calP(\calX))^2:\\
\gjs(\Omega,\Psi,\xi_n,\chi_n)\leq n\lambda}}}\big(\alpha D(\Omega\|P_{\bari})+\beta D(\Psi\|Q_{\barj})\big)\Big)\\
&\leq (\xi_n+1)^{|\calX|}(\chi_n+1)^{|\calX|}\exp\Big(-n\min_{{\substack{(\Omega,\Psi)\in(\calP(\calX))^2:\\
\gjs(\Omega,\Psi,\alpha,\beta)\leq \lambda}}}\big(\alpha D(\Omega\|P_{\bari})+\beta D(\Psi\|Q_{\barj})\big)\Big)\label{mismatch:seq:step3},
\end{align}
where \eqref{mismatch:seq:step3} follows from the definition of the GJS scoring function in \eqref{gjscompute}, the fact that $\xi_n=\lceil \alpha n\rceil\geq \alpha n$,  $\chi_n=\lceil \beta n\rceil\geq \beta n$, which implies that $\frac{1}{n}\gjs(\Omega,\Psi,\xi_n,\chi_n)\geq \gjs(\Omega,\Psi,\alpha,\beta)$.

Combining \eqref{mismatch:seq:step2} and \eqref{mismatch:seq:step3} and using the definitions of $g(\cdot)$ in \eqref{def:g} and $E_\calA^\rms(\cdot)$ in \eqref{def:ea:rms}, we have
\begin{align}
\theta(\Phi_\rms|P^{M_1},Q^{M_2})
&\leq \sum_{n\in\bbN_{N-1}}\sum_{(\bari,\barj)\notin\calA}\exp\big(-n(E_\calA^\rms(\lambda,P^{M_1},Q^{M_2})-g(n))\big)\\
&\leq \sum_{n\in\bbN_{N-1}}(M_1M_2-K)\exp\big(-n(E_\calA^\rms(\lambda,P^{M_1},Q^{M_2})-g(N-1))\big)\label{mismatch:seq:step4}\\
&\leq (M_1M_2-K)\frac{\exp(-(N-1)(E_\calA^\rms(\lambda,P^{M_1},Q^{M_2})-g(N-1)))}{1-\exp(-(E_\calA^\rms(\lambda,P^{M_1},Q^{M_2})-g(N-1))}\label{mismatch:seq:step5},
\end{align}
where \eqref{mismatch:seq:step4} follows since $g(n)$ decreases in $n$ and there are $M_1M_2-K$ unmatched pairs of sequences, and \eqref{mismatch:seq:step5} follows from the sum of geometric series. Thus, the mismatch exponent for discrete observed sequences satisfies
\begin{align}
\liminf_{N\to\infty}-\frac{1}{N}\log\theta(\Phi_\rms|P^{M_1},Q^{M_2})
&\geq E_\calA^\rms(\lambda,P^{M_1},Q^{M_2}).
\end{align}

\subsubsection{Continuous Observed Sequences}
For continuous observed sequences, $f(\cdot)$ is the MMD scoring function  in \eqref{MMDcompute}.  It follows from the results in  \eqref{expected:st:step1} and \eqref{seq:step1} that the expected stopping time satisfies
\begin{align}
\bbE_{\bbP_\calA}[\tau]
&\leq N-1+\sum_{n\in\bbN_{N-1}}\bbP_{\calA}\{\mmd^2(X_i^{\xi_n},Y_j^{\chi_n})>\lambda\}\\
&\leq N-1+\sum_{n\in\bbN_{N-1}}K\exp(-nE(\lambda))\label{usemcdiar:seq:1}\\
&\leq N-1+\frac{K\exp(-(N-1)E(\lambda))}{1-\exp(-E(\lambda))}\label{est:con:sumofg},
\end{align}
where \eqref{usemcdiar:seq:1} follows from the result in \cite[Eq. (144)-(146)]{zhou2025csm} and the definition of $E(\cdot)$ in \eqref{def:E:lambda}, and \eqref{est:con:sumofg} follows from the sum of geometric series. Thus, when $N$ is sufficiently large such that $\frac{K\exp(-(N-1)E(\lambda))}{1-\exp(-E(\lambda))}\leq 1$, the expected stopping time satisfies $\bbE_{\bbP_\calA}[\tau]\leq N$.

We next bound the mismatch probability. It follows from \eqref{mismatch:seq:step1} that the mismatch probability satisfies
\begin{align}
\theta(\Phi_\rms|P^{M_1},Q^{M_2})
&\leq \sum_{n\in\bbN_{N-1}}\sum_{(\bari,\barj)\notin\calA}\bbP_\calA\{\mmd^2(X_{\bari}^{\xi_n},Y_{\barj}^{\chi^n})\leq \lambda\}\label{mmd:seq:step1}.
\end{align}
Recall the definition of $E_{\calA}^{\rms,\rmc}(\cdot)$ in \eqref{def:e2a:rms} and the fact that $\xi_n=\lceil \alpha n\rceil\geq \alpha n$ and $\chi_n=\lceil \beta n\rceil\geq \beta n$. For any $(\bari,\barj)\notin\calA$, it follows from \cite[Eq. (158)-(159)]{zhou2025csm} that when $\lambda<\mmd^2(P_{\bari},Q_{\barj})$,
\begin{align}
\bbP_\calA\{\mmd^2(X_{\bari}^{\xi_n},Y_{\barj}^{\chi^n})\leq \lambda\}
&\leq \exp\Bigg(-\frac{n\min\{\alpha,\beta\}(\mmd^2(P_{\bari},Q_{\barj})-\lambda)^2}{64\Theta^2}\Bigg)\label{usemcdiarmid}\\
&\leq \exp(-nE_{\calA}^{\rms,\rmc}(\lambda,P^{M_1},Q^{M_2}))\label{useea2s:seq}.
\end{align}
Combining \eqref{mmd:seq:step1} and \eqref{useea2s:seq} and using the sum of geometric series, it follows that when $\lambda<\min_{(\bari,\barj)\notin\calA}\mmd^2(P_{\bari},Q_{\barj})$,
\begin{align}
\theta(\Phi_\rms|P^{M_1},Q^{M_2})
&\leq \sum_{n\in\bbN_{N-1}}(M_1M_2-K)\exp(-nE_{\calA}^{\rms,\rmc}(\lambda,P^{M_1},Q^{M_2}))\\
&\leq (M_1M_2-K)\frac{\exp(-(N-1))E_{\calA}^{\rms,\rmc}(\lambda,P^{M_1},Q^{M_2})}{1-\exp(E_{\calA}^{\rms,\rmc}(\lambda,P^{M_1},Q^{M_2}))}.
\end{align}
Thus, when $\lambda<\min_{(\bari,\barj)\notin\calA}\mmd^2(P_{\bari},Q_{\barj})$, the mismatch exponent for continuous observed sequences satisfies
\begin{align}
\liminf_{N\to\infty}-\frac{1}{N}\log\theta(\Phi_\rms|P^{M_1},Q^{M_2})
&\geq E_{\calA}^{\rms,\rmc}(\lambda,P^{M_1},Q^{M_2}).
\end{align}

\section{Proof for Unknown Number of Matches}
\label{sec:proof:unknown}
\subsection{Proof of Theorem \ref{theorem:fl:u} (Fixed-Length Test)}
\label{proof:fl:u}
\subsubsection{Case with No Match}
First consider the case with no matched pair of sequences. In this case, an error occurs if there exists a pair of sequences whose scoring function value is no greater than $\lambda$. Thus, given any $(P^{M_1},Q^{M_2})\in\calP_\emptyset$, the false alarm probability satisfies
\begin{align}
\eta(\Phi^\rmu_\rmf|P^{M_1},Q^{M_2})
&=\bbP_\emptyset\big\{\exists~(i,j)\in[M_1]\times[M_2]:~f(X_i^{\xi_N},Y_j^{\chi_N})\leq \lambda\big\}\\
&\leq \sum_{(i,j)\in[M_1]\times[M_2]}\bbP_\emptyset\big\{f(X_i^{\xi_N},Y_j^{\chi_N})\leq \lambda\big\}\label{fa:u:1}.
\end{align}
Fix any $(i,j)\in[M_1]\times[M_2]$. For discrete observed sequences, $f(\cdot)$ is the GJS scoring function  in \eqref{gjscompute}. Similarly to \eqref{mismatch:seq:step3}, it follows that
\begin{align}
\bbP_\emptyset\big\{\gjs(X_i^{\xi_N},Y_j^{\chi_N})\leq \lambda\big\}
&\leq (\xi_N+1)^{|\calX|}(\chi_N+1)^{|\calX|}\exp\bigg(-N\min_{\substack{(\Omega,\Psi)\in(\calP(\calX))^2:\\\gjs(\Omega,\Psi,\alpha,\beta)\leq \lambda}}\Big(\alpha D(\Omega\|P_i)+\beta D(\Psi\|Q_j)\Big)\bigg)\label{fa:u:2}.
\end{align}
Using the definition of $E_{\rm{fa}}(\cdot)$ in \eqref{def:e:fa} and combining \eqref{fa:u:1}, \eqref{fa:u:2}, the false alarm exponent for discrete observed sequences satisfies
\begin{align}
\liminf_{N\to\infty}-\frac{1}{N}\log \eta(\Phi^\rmu_\rmf|P^{M_1},Q^{M_2})
&\geq \min_{(i,j)\in[M_1]\times[M_2]}\min_{\substack{(\Omega,\Psi)\in(\calP(\calX))^2:\\\gjs(\Omega,\Psi,\alpha,\beta)\leq \lambda}}\Big(\alpha D(\Omega\|P_i)+\beta D(\Psi\|Q_j)\Big)\\
&=E_{\rm{fa}}(\lambda,P^{M_1},Q^{M_2})\label{fa:u:3}.
\end{align}

For continuous sequences, $f(\cdot)$ is the MMD scoring function in \eqref{MMDcompute}. Similarly to \eqref{usemcdiarmid}, using the McDiarmid's inequality~\cite{mcdiarmid1989method}, when $\lambda<\mmd^2(P_i,Q_j)$, we have
\begin{align}
\bbP_\emptyset\big\{\mmd^2(X_i^{\xi_N},Y_j^{\chi_N})\leq \lambda\big\}\leq\exp\Bigg(-\frac{N\min\{\alpha,\beta\}(\mmd^2(P_i,Q_j)-\lambda)^2}{64\Theta^2}\Bigg)\label{c:fa:2},
\end{align}
Thus, using the definition of $E_{\rm{fa}}^\rmc(\cdot)$ in \eqref{def:e2:fa} and combining \eqref{fa:u:1} and \eqref{c:fa:2}, when $\lambda<\min_{(i,j)\in[M_1]\times[M_2]}\mmd^2(P_i,Q_j)$, the false alarm exponent for continuous observed sequences satisfies
\begin{align}
\liminf_{N\to\infty}-\frac{1}{N}\log \eta(\Phi^\rmu_\rmf|P^{M_1},Q^{M_2})
&\geq \min_{(i,j)\in[M_1]\times[M_2]}\frac{\min\{\alpha,\beta\}(\mmd^2(P_i,Q_j)-\lambda)^2}{64\Theta^2}\\
&=E_{\rm{fa}}^\rmc(\lambda,P^{M_1},Q^{M_2})\label{c:fa:3}.
\end{align}

\subsubsection{Case with Matched Pairs}
We next consider the case with matched pairs of sequences. Fix any $\calA\in\calM$ and tuple of distributions $(P^{M_1},Q^{M_2})\in\calP_\calA$. When the matched pairs are indexed by $\calA$, for the test in Algorithm \ref{low_com:fltest:unknown}, a mismatch error event occurs if there exists a matched pair of sequences whose scoring function value is greater than $\lambda$, i.e., the event $\calE_1:=\{\exists~(i,j)\in\calA:~f(X_i^{\xi_N},Y_j^{\chi_N})>\lambda\}$, or there exists an unmatched pair of sequences whose scoring function value is no greater than $\lambda$, i.e., the event $\calE_2:=\{\exists~(\bari,\barj)\notin\calA:~f(X_{\bari}^{\xi_N},Y_{\barj}^{\chi_N})\leq \lambda\}$. Thus, the mismatch probability satisfies
\begin{align}
\bar{\theta}(\Phi^\rmu_\rmf|P^{M_1},Q^{M_2})
&=\bbP_\calA\{\calE_1\mathrm{~or~}\calE_2\}\\
&=\bbP_\calA\{\calE_1\}+\bbP_\calA\{\calE_2\}\label{mis:u:1}.
\end{align}
Furthermore, a false reject event occurs if all $M_1M_2$ pairs of sequences have scoring function values greater than $\lambda$. Thus, the false reject probability satisfies
\begin{align}
\zeta(\Phi^\rmu_\rmf|P^{M_1},Q^{M_2})  
&\leq \bbP_\calA\{\forall~(i,j)\in[M_1]\times[M_2],~f(X_i^{\xi_N},Y_j^{\chi_N})\geq \lambda\}\\
&\leq \bbP_\calA\{\forall~(i,j)\in\calA,~f(X_i^{\xi_N},Y_j^{\chi_N})\geq \lambda\}\\
&\leq \min_{(i,j)\in\calA}\bbP_\calA\{f(X_i^{\xi_N},Y_j^{\chi_N})\geq \lambda\}\label{fr:u:1}.
\end{align}

We first bound the mismatch probability for discrete observed sequences. In this case, $f(\cdot)$ is the GJS scoring function in \eqref{gjscompute}. Using the definition of $g(\cdot)$ in \eqref{def:g} and the result in \eqref{useg:seq}, we have
\begin{align}
\bbP_\calA\{\calE_1\}
&\leq \sum_{(i,j)\in\calA}\bbP_\calA\{\gjs(X_i^{\xi_N},Y_j^{\chi_N})>\lambda\}\\
&\leq \sum_{(i,j)\in\calA}\exp(-N(\lambda-g(N)))\\
&\leq |\calA|\exp(-N(\lambda-g(N)))\label{mis:u:2}.
\end{align}
Furthermore, it follows from \eqref{mismatch:seq:step3} that for any $(\bari,\barj)\notin\calA$, 
\begin{align}
\bbP_\calA\{\gjs(X_{\bari}^{\xi_N},Y_{\barj}^{\chi_N})\leq \lambda\}
&\leq(\xi_N+1)^{|\calX|}(\chi_N+1)^{|\calX|}\exp\Big(-N\min_{{\substack{(\Omega,\Psi)\in(\calP(\calX))^2:\\
\gjs(\Omega,\Psi,\alpha,\beta)\leq \lambda}}}\big(\alpha D(\Omega\|P_{\bari})+\beta D(\Psi\|Q_{\barj})\big)\Big).
\end{align}
Thus, using the definitions of $E_\calA^\rms(\cdot)$ in \eqref{def:ea:rms} and $\calE_2$, we have
\begin{align}
\bbP_\calA\{\calE_2\}
&\leq (M_1M_2-|\calA|)\exp(-NE_\calA^\rms(\lambda,P^{M_1},Q^{M_2}))\label{mis:u:3}.
\end{align}
Combining \eqref{mis:u:1}, \eqref{mis:u:2} and \eqref{mis:u:3}, the mismatch exponent for discrete observed sequences satisfies
\begin{align}
\liminf_{N\to\infty}-\frac{1}{N}\log \bar{\theta}(\Phi^\rmu_\rmf|P^{M_1},Q^{M_2})
&\geq \min\big\{\lambda,E_\calA^\rms(\lambda,P^{M_1},Q^{M_2})\big\}.
\end{align}

Next we bound the false reject probability for discrete observed sequences. It follows from \eqref{useg:seq} that for any $(i,j)\in\calA$,
\begin{align}
\bbP_\calA\{f(X_i^{\xi_N},Y_j^{\chi_N})\geq \lambda\}
&\leq \exp(-N(\lambda-g(N)))\label{fr:u:2}.
\end{align}
Thus, combining \eqref{fr:u:1} and \eqref{fr:u:2}, the false reject exponent for discrete observed sequences satisfies
\begin{align}
\liminf_{N\to\infty}-\frac{1}{N}\log \eta(\Phi^\rmu_\rmf|P^{M_1},Q^{M_2})
&\geq \lambda.
\end{align}

We now bound the mismatch probability for continuous observed sequences. In this case, $f(\cdot)$ is the MMD scoring function in \eqref{MMDcompute}. It follows from the definition of $E(\cdot)$ in \eqref{def:E:lambda} and the result in \cite[Eq. (144)-(146)]{zhou2025csm} that
\begin{align}
\bbP_\calA\big\{\calE_1\big\}
&\leq \sum_{(i,j)\in\calA}\bbP_\calA\big\{\mmd^2(X_i^{\xi_N},Y_j^{\chi_N})>\lambda\big\}\\
&\leq \sum_{(i,j)\in\calA}\exp(-NE(\lambda))\\
&\leq |\calA|\exp(-NE(\lambda))\label{c:mis:u:2}.
\end{align}
Recall the definition of $E_{\calA}^{\rms,\rmc}(\cdot)$ in \eqref{def:e2a:rms}. For any $(\bari,\barj)\notin\calA$, when $\lambda<\mmd^2(P_{\bari},Q_{\barj})$, it follows from \eqref{useea2s:seq} that
\begin{align}
\bbP_\calA\{\mmd^2(X_{\bari}^{\xi_N},Y_{\barj}^{\chi_N})\leq \lambda\}
&\leq \exp(-NE_{\calA}^{\rms,\rmc}(\lambda,P^{M_1},Q^{M_2})).
\end{align}
Thus, when $\lambda<\min_{(\bari,\barj)\notin\calA}\mmd^2(P_{\bari},Q_{\barj})$,
\begin{align}
\bbP_\calA\big\{\calE_2\big\}
&\leq (M_1M_2-|\calA|)\exp(-NE_{\calA}^{\rms,\rmc}(\lambda,P^{M_1},Q^{M_2}))\label{c:mis:u:3}.
\end{align}
Combining \eqref{mis:u:1}, \eqref{c:mis:u:2} and \eqref{c:mis:u:3}, when $\lambda<\min_{(\bari,\barj)\notin\calA}\mmd^2(P_{\bari},Q_{\barj})$, the mismatch exponent for continuous observed sequences satisfies
\begin{align}
\liminf_{N\to\infty}-\frac{1}{N}\log \bar{\theta}(\Phi^\rmu_\rmf|P^{M_1},Q^{M_2})
&\geq \min\big\{E(\lambda),E_{\calA}^{\rms,\rmc}(\lambda,P^{M_1},Q^{M_2})\big\}.
\end{align}

Finally, we bound the false reject probability for continuous observed sequences. Using the definition of $E(\cdot)$ in \eqref{def:E:lambda}, it follows from \cite[Eq. (144)-(146)]{zhou2025csm} that for any $(i,j)\in\calA$,
\begin{align}
\bbP_\calA\{\mmd^2(X_i^{\xi_N},Y_j^{\chi_N})\geq \lambda\}
&\leq \exp(-NE(\lambda))\label{fr:u:3}.
\end{align}
Thus, combining \eqref{fr:u:1} and \eqref{fr:u:3}, the false reject exponent for continuous observed sequences satisfies
\begin{align}
\liminf_{N\to\infty}-\frac{1}{N}\log\zeta(\Phi^\rmu_\rmf|P^{M_1},Q^{M_2})
&\geq E(\lambda).
\end{align}

\subsection{Proof of Theorem \ref{theorem:seq:u} (Sequential Test)}
\label{proof:seq:u}
\subsubsection{Case with No Match}
First consider the case with no match. Fix any $(P^{M_1},Q^{M_2})\in\calP_\emptyset$. In this case, the expected stopping time satisfies
\begin{align}
\bbE_{\bbP_\emptyset}[\tau_\rmu]
&=N-1+\sum_{n\in\bbN_{N-1}}\bbP_\emptyset\{\tau_\rmu>n\}\label{st:u:1}.
\end{align}
Fix any $n\in\bbN$. It follows from the sequential test design in Algorithm \ref{low_com:seqtest:unknown} that 
\begin{align}
\bbP_\emptyset\{\tau_\rmu>n\}
&\leq \bbP_\emptyset\big\{\exists~(i,j)\in[M_1]\times[M_2],~\lambda_1<f(X_i^{\xi_n},Y_j^{\chi_n})\leq \lambda_2\big\}\\
&\leq \sum_{(i,j)\in[M_1]\times[M_2]}\bbP_\emptyset\big\{\lambda_1<f(X_i^{\xi_n},Y_j^{\chi_n})\leq \lambda_2\big\}\\
&\leq \sum_{(i,j)\in[M_1]\times[M_2]}\bbP_\emptyset\big\{f(X_i^{\xi_n},Y_j^{\chi_n})\leq \lambda_2\big\}\label{st:u:2}.
\end{align}
Furthermore, the false alarm probability satisfies
\begin{align}
\eta(\Phi_\rms^\rmu|P^{M_1},Q^{M_2})
&\leq \bbP_\emptyset\big\{\exists~(i,j)\in[M_1]\times[M_2],~f(X_i^{\xi_{\tau_\rmu}},Y_j^{\chi_{\tau_\rmu}})\leq \lambda_1\big\}\\
&=\sum_{n\in\bbN_{N-1}}\bbP_\emptyset\{\tau_\rmu=n\}\bbP_\emptyset\big\{\exists~(i,j)\in[M_1]\times[M_2],~f(X_i^{\xi_n},Y_j^{\chi_n})\leq \lambda_1\big\}\\
&\leq \sum_{n\in\bbN_{N-1}}\bbP_\emptyset\big\{\exists~(i,j)\in[M_1]\times[M_2],~f(X_i^{\xi_n},Y_j^{\chi_n})\leq \lambda_1\big\}\\
&\leq \sum_{n\in\bbN_{N-1}}\sum_{(i,j)\in[M_1]\times[M_2]}\bbP_\emptyset\big\{f(X_i^{\xi_n},Y_j^{\chi_n})\leq \lambda_1\big\}\label{fa:seq:u:1}.
\end{align}

Fix any $n\in\bbN$ and $(i,j)\in[M_1]\times[M_2]$. For discrete sequences, $f(\cdot)$ is the GJS scoring function in \eqref{gjscompute}.  Using the definitions of $g(\cdot)$ in \eqref{def:g} and $E_{\rm{fa}}(\cdot)$ in \eqref{def:e:fa}, and the result in \eqref{fa:u:2}, each probability term in \eqref{st:u:1} satisfies
\begin{align}
\bbP_\emptyset\big\{\gjs(X_i^{\xi_n},Y_j^{\chi_n})\leq \lambda_2\big\}
&\leq\exp\big(-n(E_{\rm{fa}}(\lambda_2,P^{M_1},Q^{M_2})-g(n))\big)\label{st:u:3}.
\end{align}
Combining \eqref{st:u:1}, \eqref{st:u:2} and \eqref{st:u:3} and using the sum of geometric series, the expected stopping time of the test satisfies
\begin{align}
\bbE_{\bbP_\emptyset}[\tau_\rmu]
&=N-1+M_1M_2\sum_{n\in\bbN_{N-1}}\exp\big(-n(E_{\rm{fa}}(\lambda_2,P^{M_1},Q^{M_2})-g(n))\big)\\
&\le N-1+M_1M_2\sum_{n\in\bbN_{N-1}}\exp\big(-n(E_{\rm{fa}}(\lambda_2,P^{M_1},Q^{M_2})-g(N-1))\big)\label{st:u:4}\\
&\le N-1+\frac{M_1M_2\exp(-(N-1)(E_{\rm{fa}}(\lambda_2,P^{M_1},Q^{M_2})-g(N-1)))}{1-\exp(-(E_{\rm{fa}}(\lambda_2,P^{M_1},Q^{M_2})-g(N-1)))}\label{st:u:5},
\end{align}
where \eqref{st:u:4} follows since $g(n)$ decreases in $n$, and \eqref{st:u:5} follows from the sum of geometric series. Hence, when $N$ is sufficiently large such that $\frac{M_1M_2\exp(-(N-1)E_{\rm{fa}}(\lambda_2,P^{M_1},Q^{M_2})-g(N-1))}{1-\exp(-(E_{\rm{fa}}(\lambda_2,P^{M_1},Q^{M_2})-g(N-1)))}\le 1$, the expected stopping time satisfies $\bbE_{\bbP_\emptyset}[\tau_\rmu]\le N$.

Analogously to \eqref{st:u:3} and \eqref{st:u:5}, it follows from \eqref{fa:seq:u:1} that the false alarm probability satisfies
\begin{align}
\eta(\Phi_\rms^\rmu|P^{M_1},Q^{M_2})
&\leq \sum_{n\in\bbN_{N-1}}\sum_{(i,j)\in[M_1]\times[M_2]}\bbP_\emptyset\big\{\gjs(X_i^{\xi_n},Y_j^{\chi_n})\leq \lambda_1\big\}\\
&\leq \sum_{n\in\bbN_{N-1}}\sum_{(i,j)\in[M_1]\times[M_2]}\exp\big(-n(E_{\rm{fa}}(\lambda_1,P^{M_1},Q^{M_2})-g(n))\big)\\
&\leq \frac{M_1M_2\exp(-(N-1)(E_{\rm{fa}}(\lambda_1,P^{M_1},Q^{M_2})-g(N-1)))}{1-\exp(-(E_{\rm{fa}}(\lambda_1,P^{M_1},Q^{M_2})-g(N-1)))}\label{fa:seq:u:2}.
\end{align}
Thus, the false alarm exponent for discrete observed sequences satisfies
\begin{align}
\liminf_{N\to\infty}-\frac{1}{N}\log\eta(\Phi_\rms^\rmu|P^{M_1},Q^{M_2})&\geq E_{\rm{fa}}(\lambda_1,P^{M_1},Q^{M_2}).
\end{align}

For continuous observed sequences, $f(\cdot)$ is the MMD scoring function in \eqref{MMDcompute}. Using the definition of $E_{\rm{fa}}^\rmc(\cdot)$ in \eqref{def:e2:fa} and the result in \eqref{c:fa:2}, for any 
$n\in\bbN$ and positive real number $\lambda<\min_{(i,j)\in[M_1]\times[M_2]}\mmd^2(P_i,Q_j)$, it follows that
\begin{align}
\bbP_\emptyset\big\{\mmd^2(X_i^{\xi_n},Y_j^{\chi_n})\leq \lambda\big\}\leq\exp(-nE_{\rm{fa}}^\rmc(\lambda,P^{M_1},Q^{M_2}))\label{st:seq:u:6}.
\end{align}
Thus, it follows from \eqref{st:u:1}, \eqref{st:u:2} and \eqref{st:seq:u:6} that when $\lambda_2<\min_{(i,j)\in[M_1]\times[M_2]}\mmd^2(P_i,Q_j)$, the expected stopping time satisfies
\begin{align}
\bbE_{\bbP_\emptyset}[\tau_\rmu]
&=N-1+\sum_{n\in\bbN_{N-1}}\sum_{(i,j)\in[M_1]\times[M_2]}\bbP_\emptyset\big\{\mmd^2(X_i^{\xi_n},Y_j^{\chi_n})\leq \lambda_2\big\}\\
&\leq N-1+\sum_{n\in\bbN_{N-1}}M_1M_2\exp(-nE_{\rm{fa}}^\rmc(\lambda_2,P^{M_1},Q^{M_2}))\\
&\le N-1+\frac{M_1M_2\exp(-(N-1)E_{\rm{fa}}^\rmc(\lambda_2,P^{M_1},Q^{M_2})}{1-\exp(-E_{\rm{fa}}^\rmc(\lambda_2,P^{M_1},Q^{M_2}))},
\end{align}
which is less than $N$ when $\frac{M_1M_2\exp(-(N-1)E_{\rm{fa}}^\rmc(\lambda_2,P^{M_1},Q^{M_2})}{1-\exp(-E_{\rm{fa}}^\rmc(\lambda_2,P^{M_1},Q^{M_2}))}\le 1$.

Furthermore, in a similar manner, it follows from \eqref{fa:seq:u:1} that when $\lambda_1<\min_{(i,j)\in[M_1]\times[M_2]}\mmd^2(P_i,Q_j)$, the false alarm probability satisfies
\begin{align}
\eta(\Phi_\rms^\rmu|P^{M_1},Q^{M_2})
&\leq \sum_{n\in\bbN_{N-1}}\sum_{(i,j)\in[M_1]\times[M_2]}\bbP_\emptyset\big\{\mmd^2(X_i^{\xi_n},Y_j^{\chi_n})\leq \lambda_1\big\}\\
&\leq \sum_{n\in\bbN_{N-1}}\sum_{(i,j)\in[M_1]\times[M_2]}\exp(-nE_{\rm{fa}}^\rmc(\lambda_1,P^{M_1},Q^{M_2}))\\
&\leq \frac{M_1M_2\exp(-(N-1)E_{\rm{fa}}^\rmc(\lambda_1,P^{M_1},Q^{M_2})}{1-\exp(-E_{\rm{fa}}^\rmc(\lambda_1,P^{M_1},Q^{M_2}))}\label{fa:seq:u:3}.
\end{align}
Thus, when $\lambda_1<\min_{(i,j)\in[M_1]\times[M_2]}\mmd^2(P_i,Q_j)$, the false alarm exponent for continuous observed sequences satisfies
\begin{align}
\liminf_{N\to\infty}-\frac{1}{N}\log\eta(\Phi_\rms^\rmu|P^{M_1},Q^{M_2})&\geq E_{\rm{fa}}^\rmc(\lambda_1,P^{M_1},Q^{M_2}).
\end{align}

\subsubsection{General Expressions for the Case with Matched Pairs}
We next consider the case with matched pairs of sequences. Fix any $\calA\in\calM$ and tuple of distributions $(P^{M_1},Q^{M_2})\in\calP_\calA$.  In this case, the expected stopping time of the test in Algorithm \ref{low_com:seqtest:unknown} satisfies
\begin{align}
\bbE_{\bbP_\calA}[\tau_\rmu]
&=N-1+\sum_{n\in\bbN_{N-1}}\bbP_\calA\{\tau_\rmu>n\}\label{st:u:6},
\end{align}
where
\begin{align}
\bbP_\calA\{\tau_\rmu>n\}
&=\bbP_\calA\{\exists~(i,j)\in[M_1]\times[M_2],~\lambda_1<f(X_i^{\xi_n},Y_j^{\chi_n})\leq \lambda_2\}\\
&\leq \sum_{(i,j)\in[M_1]\times[M_2]}\bbP_\calA\{\lambda_1<f(X_i^{\xi_n},Y_j^{\chi_n})\leq \lambda_2\}\\
&\leq \sum_{(i,j)\in\calA}\bbP_\calA\{f(X_i^{\xi_n},Y_j^{\chi_n})>\lambda_1\}+\sum_{(\bari,\barj)\notin\calA}\bbP_\calA\{f(X_{\bari}^{\xi_n},Y_{\barj}^{\chi_n})\leq \lambda_2\}\label{st:u:7}.
\end{align}

When the matched pairs are indexed by $\calA$, a mismatch error event occurs if there exists a matched pair of sequences whose scoring function value is greater than $\lambda_2$, i.e., the event $\calF_1:=\{\exists~(i,j)\in\calA:~f(X_i^{\xi_{\tau_\rmu}},Y_j^{\chi_{\tau_\rmu}})>\lambda_2\}$, or there exists an unmatched pair of sequences whose scoring function value is no greater than $\lambda_1$, i.e., the event $\calF_2:=\{\exists~(\bari,\barj)\notin\calA:~f(X_{\bari}^{\xi_{\tau_\rmu}},Y_{\barj}^{\chi_{\tau_\rmu}})\leq \lambda_1\}$. Thus, the mismatch probability satisfies
\begin{align}
\bar{\theta}(\Phi_\rms^\rmu|P^{M_1},Q^{M_2})
&=\bbP_\calA\{\calF_1\}+\bbP_\calA\{\calF_2\}\label{mis:u:seq:1}.
\end{align}
Furthermore,
\begin{align}
\bbP_\calA\{\calF_1\}
&=\sum_{n\in\bbN_{N-1}}\bbP_{\calA}\{\tau_\rmu=n\}\bbP_\calA\{\exists~(i,j)\in\calA:~f(X_i^{\xi_n},Y_j^{\chi_n})>\lambda_2\}\\
&\leq \sum_{n\in\bbN_{N-1}}\sum_{(i,j)\in\calA}\bbP_\calA\{f(X_i^{\xi_n},Y_j^{\chi_n})>\lambda_2\}\label{mis:u:seq:2}.
\end{align}
Similarly,
\begin{align}
\bbP_\calA\{\calF_2\}
&\leq \sum_{n\in\bbN_{N-1}}\sum_{(\bari,\barj)\notin\calA}\bbP_\calA\{f(X_{\bari}^{\xi_n},Y_{\barj}^{\chi_n})\leq \lambda_1\}\label{mis:u:seq:3}.
\end{align}

A false reject event occurs if all pairs of sequences have scoring function values greater than $\lambda_2$. Thus, the false alarm probability satisfies
\begin{align}
\zeta(\Phi_\rms^\rmu|P^{M_1},Q^{M_2})
&\leq \bbP_\calA\big\{\forall~(i,j)\in[M_1]\times[M_2],~f(X_i^{\xi_{\tau_\rmu}},Y_j^{\chi_{\tau_\rmu}})>\lambda_2\big\}\\
&\leq \bbP_\calA\big\{\forall~(i,j)\in\calA,~f(X_i^{\xi_{\tau_\rmu}},Y_j^{\chi_{\tau_\rmu}})>\lambda_2\big\}\\
&\leq \min_{(i,j)\in\calA}\bbP_\calA\big\{f(X_i^{\xi_{\tau_\rmu}},Y_j^{\chi_{\tau_\rmu}})>\lambda_2\big\}\\
&\leq \min_{(i,j)\in\calA}\sum_{n\in\bbN_{N-1}}\bbP_\calA\{f(X_i^{\xi_n},Y_j^{\chi_n})>\lambda_2\}\label{fr:u:seq:1}.
\end{align}

\subsubsection{Case with Matched Pairs for Discrete Observed Sequences}
For discrete observed sequences, $f(\cdot)$ is the GJS scoring function in \eqref{gjscompute}. We first bound the expected stopping time in \eqref{st:u:6}. Recall the definition of $g(\cdot)$ in \eqref{def:g}. Fix any $n\in\bbN$. It follows from \eqref{useg:seq} with $\lambda$ replaced by $\lambda_1$ that
\begin{align}
\sum_{(i,j)\in\calA}\bbP_\calA\{\gjs(X_i^{\xi_n},Y_j^{\chi_n})>\lambda_1\}
&\leq \sum_{(i,j)\in\calA}\exp(-n(\lambda_1-g(n)))\\
&\leq |\calA|\exp(-n(\lambda_1-g(n)))\label{st:u:8}.
\end{align}
Furthermore, it follows from the definition of $E_\calA^\rms(\cdot)$ in \eqref{def:ea:rms} and the result in \eqref{mismatch:seq:step3} with $\lambda$ replaced with $\lambda_2$ that
\begin{align}
\sum_{(\bari,\barj)\notin\calA}\bbP_\calA\{\gjs(X_{\bari}^{\xi_n},Y_{\barj}^{\chi_n})\leq \lambda_2\}
&\leq \sum_{(\bari,\barj)\notin\calA}\exp\big(-n(E_\calA^\rms(\lambda_2,P^{M_1},Q^{M_2})-g(n))\big)\\
&=(M_1M_2-|\calA|)\exp\big(-n(E_\calA^\rms(\lambda_2,P^{M_1},Q^{M_2})-g(n))\big)\label{st:u:9}.
\end{align}
Combining \eqref{st:u:6}, \eqref{st:u:7}, \eqref{st:u:8} and \eqref{st:u:9} and using the sum of geometric series, the expected stopping time satisfies 
\begin{align}
\bbE_{\bbP_\calA}[\tau_\rmu]
&\leq N-1+\sum_{n\in\bbN_{N-1}}\big(|\calA|\exp(-n(\lambda_1-g(n)))+(M_1M_2-|\calA|)\exp\big(-n(E_\calA^\rms(\lambda_1,P^{M_1},Q^{M_2})-g(n))\big)\big)\\
\nn&\leq N-1+\frac{|\calA|\exp(-(N-1)(\lambda_1-g(N-1)))}{1-\exp(-(\lambda_1-g(N-1)))}\\*
&\qquad+\frac{(M_1M_2-|\calA|\exp(-(N-1)(E_\calA^\rms(\lambda_2,P^{M_1},Q^{M_2})-g(N-1))))}{1-\exp(-(E_\calA^\rms(\lambda_2,P^{M_1},Q^{M_2})-g(N-1)))}.
\end{align}
Thus, when $N$ is sufficiently large such that $\frac{|\calA|\exp(-(N-1)(\lambda_1-g(N-1)))}{1-\exp(-(\lambda_1-g(N-1)))}+\frac{(M_1M_2-|\calA|)\exp(-(N-1)(E_\calA^\rms(\lambda_2,P^{M_1},Q^{M_2})-g(N-1)))}{1-\exp(-(E_\calA^\rms(\lambda_2,P^{M_1},Q^{M_2})-g(N-1)))}\leq 1$, the expected stopping time satisfies $\bbE_{\bbP_\calA}[\tau_\rmu]\leq N$.

We next bound the mismatch probability in \eqref{mis:u:seq:1}.  It follows from \eqref{useg:seq} with $\lambda$ replaced by $\lambda_2$, the result in \eqref{mis:u:seq:2} and the sum of geometric series that
\begin{align}
\bbP_\calA\{\calF_1\}
&\leq \sum_{n\in\bbN_{N-1}}\sum_{(i,j)\in\calA}\bbP_\calA\{\gjs(X_i^{\xi_n},Y_j^{\chi_n})>\lambda_2\}\\
&\leq \sum_{n\in\bbN_{N-1}}\sum_{(i,j)\in\calA}\exp(-n(\lambda_2-g(n)))\\
&\leq |\calA| \sum_{n\in\bbN_{N-1}}\exp(-n(\lambda_2-g(N-1)))\\
&\leq \frac{|\calA|\exp(-(N-1)(\lambda_2-g(N-1)))}{1-\exp(-(\lambda_2-g(N-1)))}\label{mis:u:seq:4}.
\end{align}
Analogously, it follows from the definition of $E_\calA^\rms(\cdot)$ in \eqref{def:ea:rms}, the result in \eqref{mismatch:seq:step3} with $\lambda$ replaced with $\lambda_1$ and the result in \eqref{mis:u:seq:3} that
\begin{align}
\bbP_\calA\{\calF_2\}
&\leq \sum_{n\in\bbN_{N-1}}\sum_{(\bari,\barj)\notin\calA}\bbP_\calA\{\gjs(X_{\bari}^{\xi_n},Y_{\barj}^{\chi_n})\leq \lambda_1\}\\
&\leq \sum_{n\in\bbN_{N-1}}\sum_{(\bari,\barj)\notin\calA}\exp\big(-n(E_\calA^\rms(\lambda_1,P^{M_1},Q^{M_2})-g(n))\big)\\
&\leq (M_1M_2-|\calA|)\sum_{n\in\bbN_{N-1}}\sum_{(\bari,\barj)\notin\calA}\exp\big(-n(E_\calA^\rms(\lambda_1,P^{M_1},Q^{M_2})-g(N-1))\big)\\
&\le \frac{(M_1M_2-|\calA|)\exp(-(N-1)(E_\calA^\rms(\lambda_1,P^{M_1},Q^{M_2})-g(N-1)))}{1-\exp(-(E_\calA^\rms(\lambda_1,P^{M_1},Q^{M_2})-g(N-1)))}\label{mis:u:seq:5}.
\end{align}
Thus, combining \eqref{mis:u:seq:1}, \eqref{mis:u:seq:2}, \eqref{mis:u:seq:3}, \eqref{mis:u:seq:4} and \eqref{mis:u:seq:5}, the mismatch exponent for discrete observed sequences satisfies
\begin{align}
\liminf_{N\to\infty}-\frac{1}{N}\log\bar{\theta}(\Phi_\rms^\rmu|P^{M_1},Q^{M_2})&\geq \min\{E_\calA^\rms(\lambda_1,P^{M_1},Q^{M_2}),\lambda_2\}.
\end{align}

Finally, we bound the false reject probability. Analogously to \eqref{mis:u:seq:4}, it follows from \eqref{useg:seq} with $\lambda$ replaced by $\lambda_2$, the result in \eqref{fr:u:seq:1} and the sum of geometric series that the false reject probability satisfies
\begin{align}
\zeta(\Phi_\rms^\rmu|P^{M_1},Q^{M_2})
&\leq\min_{(i,j)\in\calA}\sum_{n\in\bbN_{N-1}}\bbP_\calA\{\gjs(X_i^{\xi_n},Y_j^{\chi_n})>\lambda_2\}\\
&\leq \sum_{n\in\bbN_{N-1}}\exp(-n(\lambda_2-g(n)))\\
&\leq \frac{\exp(-(N-1)(\lambda_2-g(N-1)))}{1-\exp(-(\lambda_2-g(N-1)))}.
\end{align}
Thus, the false reject exponent for discrete observed sequences satisfies
\begin{align}
\liminf_{N\to\infty}-\frac{1}{N}\log\zeta(\Phi_\rms^\rmu|P^{M_1},Q^{M_2})&\geq \lambda_2.
\end{align}

\subsubsection{Case with Matched Pairs for Continuous Observed Sequences}
For continuous observed sequences, $f(\cdot)$ is the MMD scoring function in \eqref{MMDcompute}. We first bound the expected stopping time in \eqref{st:u:6}. It follows from the definition of $E(\cdot)$ in \eqref{def:E:lambda} and the result in \cite[Eq. (144)-(146)]{zhou2025csm} that
\begin{align}
\sum_{(i,j)\in\calA}\bbP_\calA\{\mmd^2(X_i^{\xi_n},Y_j^{\chi_n})>\lambda_1\}
&\leq \sum_{(i,j)\in\calA}\exp(-nE(\lambda_1))\\
&\leq |\calA|\exp(-nE(\lambda_1))\label{st:u:10}.
\end{align}
Furthermore, it follows from the result in \eqref{useea2s:seq} and the definition of $E_{\calA}^{\rms,\rmc}(\cdot)$ in \eqref{useea2s:seq} that when $\lambda_2<\min_{(\bari,\barj)\notin\calA}\mmd^2(P_{\bari},Q_{\barj})$
\begin{align}
\sum_{(\bari,\barj)\notin\calA}\bbP_\calA\{\mmd^2(X_{\bari}^{\xi_n},Y_{\barj}^{\chi_n})\leq \lambda_2\}
&\leq \sum_{(\bari,\barj)\notin\calA}\exp(-nE_{\calA}^{\rms,\rmc}(\lambda_2,P^{M_1},Q^{M_2}))\\
&\leq (M_1M_2-|\calA|)\exp(-nE_{\calA}^{\rms,\rmc}(\lambda_2,P^{M_1},Q^{M_2}))\label{st:u:11}.
\end{align}
Combining \eqref{st:u:6}, \eqref{st:u:7}, \eqref{st:u:10} and \eqref{st:u:11} and using the sum of geometric series, when $\lambda_2<\min_{(\bari,\barj)\notin\calA}\mmd^2(P_{\bari},Q_{\barj})$, the expected stopping time satisfies
\begin{align}
\bbE_{\bbP_\calA}[\tau_\rmu]
&\leq N-1+\frac{|\calA|\exp(-(N-1)E(\lambda_1)}{1-\exp(-E(\lambda_1))}+\frac{(M_1M_2-|\calA|)\exp(-(N-1)E_{\calA}^{\rms,\rmc}(\lambda_2,P^{M_1},Q^{M_2}))}{1-\exp(-E_{\calA}^{\rms,\rmc}(\lambda_2,P^{M_1},Q^{M_2}))},
\end{align}
which is less than $N$ if $\frac{|\calA|\exp(-(N-1)E(\lambda_1)}{1-\exp(-E(\lambda_1))}+\frac{(M_1M_2-|\calA|)\exp(-(N-1)E_{\calA}^{\rms,\rmc}(\lambda_2,P^{M_1},Q^{M_2}))}{1-\exp(-E_{\calA}^{\rms,\rmc}(\lambda_2,P^{M_1},Q^{M_2}))}\leq 1$.

We next bound the mismatch probability. Using \cite[Eq. (144)-(146)]{zhou2025csm}, analogously to \eqref{est:con:sumofg}, it follows from the definition of $E(\cdot)$ in \eqref{def:E:lambda} and the result in \eqref{mis:u:seq:2} that
\begin{align}
\bbP_\calA\{\calF_1\}
&\leq \sum_{n\in\bbN_{N-1}}\sum_{(i,j)\in\calA}\bbP_\calA\{\mmd^2(X_i^{\xi_n},Y_j^{\chi_n})>\lambda_2\}\\
&\leq \sum_{n\in\bbN_{N-1}}\sum_{(i,j)\in\calA}\exp(-nE(\lambda_2))\\
&\leq \frac{|\calA|\exp(-(N-1)E(\lambda_2))}{1-\exp(-E(\lambda_2))}\label{mis:u:seq:6}.
\end{align}
Furthermore, it follows from \eqref{useea2s:seq} and \eqref{mis:u:seq:3} that when $\lambda_1<\min_{(\bari,\barj)\notin\calA}\mmd^2(P_{\bari},Q_{\barj})$
\begin{align}
\bbP_\calA\{\calF_2\}
&\leq \sum_{n\in\bbN_{N-1}}\sum_{(\bari,\barj)\notin\calA}\bbP_\calA\{\mmd^2(X_{\bari}^{\xi_n},Y_{\barj}^{\chi_n})\leq \lambda_1\}\\
&\leq \sum_{n\in\bbN_{N-1}}\sum_{(\bari,\barj)\notin\calA}\exp(-nE_{\calA}^{\rms,\rmc}(\lambda_1,P^{M_1},Q^{M_2}))\\
&\leq \frac{(M_1M_2-|\calA|)\exp(-(N-1)E_{\calA}^{\rms,\rmc}(\lambda_1,P^{M_1},Q^{M_2}))}{1-\exp(-E_{\calA}^{\rms,\rmc}(\lambda_1,P^{M_1},Q^{M_2}))}\label{mis:u:seq:7}.
\end{align}
Thus, combining \eqref{mis:u:seq:1}, \eqref{mis:u:seq:2}, \eqref{mis:u:seq:3}, \eqref{mis:u:seq:6} and \eqref{mis:u:seq:7}, when $\lambda_1<\min_{(\bari,\barj)\notin\calA}\mmd^2(P_{\bari},Q_{\barj})$, the mismatch exponent for continuous observed sequences satisfies
\begin{align}
\liminf_{N\to\infty}-\frac{1}{N}\log\bar{\theta}(\Phi_\rms^\rmu|P^{M_1},Q^{M_2})&\geq \min\{E_{\calA}^{\rms,\rmc}(\lambda_1,P^{M_1},Q^{M_2}),E(\lambda_2)\}.
\end{align}

Finally, we bound the false reject probability for continuous observed sequences. Analogously to \eqref{mis:u:seq:6}, it follows from \cite[Eq. (144)-(146)]{zhou2025csm} with $\lambda$ replaced by $\lambda_2$, the result in \eqref{fr:u:seq:1} and the sum of geometric series that
\begin{align}
\zeta(\Phi_\rms^\rmu|P^{M_1},Q^{M_2})
&\leq\min_{(i,j)\in\calA}\sum_{n\in\bbN_{N-1}}\bbP_\calA\{\mmd^2(X_i^{\xi_n},Y_j^{\chi_n})>\lambda_2\}\\
&\leq \sum_{n\in\bbN_{N-1}}\exp(-nE(\lambda_2))\\
&\leq \frac{\exp(-(N-1)E(\lambda_2))}{1-\exp(-E(\lambda_2))}.
\end{align}
Thus, the false reject exponent for continuous observed sequences satisfies
\begin{align}
\liminf_{N\to\infty}-\frac{1}{N}\log\zeta(\Phi_\rms^\rmu|P^{M_1},Q^{M_2})&\geq E(\lambda_2).
\end{align}

\section{Conclusion}
\label{sec:conc}
We revisited the problem of statistical sequence matching and proposed low-complexity exponentially consistent tests for both discrete and continuous observed sequences. We considered both cases when the number of matches is known and unknown. For each case, we proposed non-parametric fixed-length tests and sequential tests, characterized the exponential decay rates of error probabilities and demonstrated the benefit of sequentiality by showing that our proposed sequential tests achieve better performance than our proposed fixed-length tests. Compared with existing exhaustive search tests~\cite{unnikrishnan2015asymptotically,zhou2024tit,zhou2025csm,zhou2025seq} that have exponential complexity with respect to the number of sequences and the number of matches, our tests have polynomial complexity with respect to the number of sequences in each database regardless of the number of matches, which strike a better tradeoff between detection performance and computational complexity. In particular, when the number of sequences of either database and the number of matches is large, the tests in~\cite{unnikrishnan2015asymptotically,zhou2024tit,zhou2025csm,zhou2025seq} cannot be used due to prohibitively high complexity but our tests are feasible.

We next discuss future research directions. Firstly, we assumed that a pair of matched sequences are generated from exactly the same distribution. However, in practice, even matched sequences could be generated from distributions that deviate slightly. Thus, towards a further step to practical applications, it would be worthwhile to model distribution uncertainty and analyze the impact of distribution uncertainty on the performance of tests, potentially borrowing ideas from \cite{hsu2020binary,zhu2025report}. Secondly, we derived achievability results in this paper. Without a matching converse result, it is unclear whether our results are optimal and our claimed benefit of sequentiality is general. Thus, it is valuable to derive converse results for the settings in this paper. Finally, for sequential tests, we assumed that all sequences of both databases are available at each time point. However, in certain cases, the set of available sequences could be determined in an online manner by the test designer to further improve sample efficiency. Thus, it is beneficial to generalize our results for sequential tests with an additional sampling step, potentially borrowing ideas from~\cite{yavas2025bandit}.

\bibliographystyle{IEEEtran}
\bibliography{IEEEfull_lin}
\end{document}